\documentclass[prb,showpacs,twocolumn,showpacs]{revtex4}
\usepackage{graphicx}
\newcommand{\figwidth}{3.375in}

\newcommand{\eq}[1]{(\ref{#1})}
\newcommand{\kb}{k_{\rm B}}
\newcommand{\tw}{t_{\rm w}}
\newcommand{\Lo}{L_{0}}

\newcommand \be {\begin{equation}}
\newcommand \ee {\end{equation}}
\newcommand \beq {\begin{equation}}
\newcommand \eeq {\end{equation}}
\newcommand \bmat {\begin{displaymath}}
\newcommand \emat {\end{displaymath}}

\newcommand \qea {q_{\rm EA}}
\newcommand \qd {q_{\rm D}}
\newcommand \chiea {\chi_{\rm EA}}
\newcommand \chid {\chi_{\rm D}}
\newcommand \chifc {\chi_{\rm FC}}
\newcommand \chizfc {\chi_{\rm ZFC}}

\begin{document}
\preprint{ISSP}
\title[Short Title]{
Extended droplet theory for aging in short-ranged spin glasses and
a numerical examination
} 
\author{Hajime Yoshino}
\email{yoshino@ess.sci.osaka-u.ac.jp}
\affiliation{
Department  of Earth and Space Science, Faculty of Science, 
Osaka University, Toyonaka, Osaka 560-0043, Japan 
}
\author{Koji Hukushima}
\email{hukusima@issp.u-tokyo.ac.jp}
\affiliation{
Institute for Solid State Physics, University~of~Tokyo, 5-1-5 Kashiwa-no-ha,
Kashiwa, Chiba 277-8581, Japan}
%\email{hukusima@issp.u-tokyo.ac.jp}
\author{Hajime Takayama}
\email{takayama@issp.u-tokyo.ac.jp}
\affiliation{
Institute for Solid State Physics, University~of~Tokyo, 5-1-5 Kashiwa-no-ha,
Kashiwa, Chiba 277-8581, Japan} 
\date{March 10, 2002: Version 6.2}
\begin{abstract}
We analyze isothermal aging of a four dimensional Edwards-Anderson model in
detail by Monte Carlo simulations. We analyze the
data in the view of an extended version of the droplet theory 
proposed recently (cond-mat/0202110) which is based
on the original droplet theory plus conjectures on the anomalously 
soft droplets in the presence of domain walls.
We found that the scaling laws including some fundamental predictions
of the original droplet theory explain well our results.
The results of our simulation strongly suggest the separation of the
 breaking of the time translational invariance and the fluctuation
 dissipation theorem in agreement with our scenario.
\end{abstract}
\maketitle
%%%%%%%%%%%%%%
%\tableofcontents
%%%%%%%%%%

%Introduction
%\input{sec-intro-v6}

\section{Introduction}

Spin glasses exhibit characteristic slow dynamics below 
the spin-glass (SG) transition temperature $T_{\rm c}$.
Recently aging phenomena, which have been known for a long time
in glassy systems \cite{Struik}, has attracted renewed interest 
both in experimental and theoretical studies
\cite{APY-review,Uppsala,VHO,Saclay,Review-Rieger,BCKM}.
A major theoretical progress was the development of the dynamical mean-field
theories (MFT) of spin glasses and related systems \cite{MFT}.
Most remarkably it found that the Field Cooled (FC) susceptibility $\chifc$
is larger than the equilibrium susceptibility $\chiea$
which is related to the Edwards-Anderson (EA) order parameter 
$q_{\rm EA}$ by the fluctuation dissipation theorem (FDT)
as $\kb T \chi_{\rm EA}=1-q_{\rm EA}$.
This finding corresponds to the well known  experimental 
observation \cite{Nagata79} that  $\chifc$
is {\it larger} than the Zero Field Cooled (ZFC) susceptibility  $\chizfc$.
In addition, many new view points for the glassy dynamics were discovered
subsequently such as the concept of effective temperature \cite{CKP97,CK99}. 
However, the MFT does not provide insights into what will 
become important in realistic finite dimensional systems. Most seriously,
thermally activated nucleation processes which are presumably important
in finite dimensional glassy systems cannot be captured at the
mean-field level.  

Recently we proposed a refined scenario \cite{our-letter} for the isothermal 
aging based on the droplet theory for spin glasses \cite{BM,FH1,FH2}.
We conjectured that the original idea of effective stiffness of 
droplets in the presence of frozen-in domain wall, introduced 
by Fisher and Huse \cite{FH2}, can be extended to take into 
account anomalously soft droplet excitations
which are as large as frozen-in extended defects, i.~e. domain walls.
This conjecture is partly motivated by 
the results of recent active studies of spin-glass models at $T=0$
\cite{LEE} which revealed existence of 
anomalous low energy and large scale excitations.
The anomalously soft droplets allow  emergence of a new dynamical order parameter 
$q_{\rm D}$ and the dynamical susceptibility $\chid$
associated with the former by FDT
$\kb T \chid=1-\qd$.
The dynamical order parameter $q_{\rm D}$
is expected to be {\it smaller} than the equilibrium 
EA order parameter 
$q_{\rm EA}$ which means that the dynamical susceptibility
$\chi_{\rm D}$ is {\it } {\it larger} than the equilibrium susceptibility
$\chi_{\rm EA}$.
Consequently, our scenario implies a novel feature that
breaking of the time translational invariance (TTI) 
and FDT 
separates asymptotically at large length (time) scales:
the breaking point of TTI  converges to the equilibrium EA 
order parameter $q_{\rm EA}$ while that of FDT 
converges to the new dynamical order parameter $q_{\rm D}$ smaller 
than $q_{\rm EA}$. We will find that $\chifc = \chid$
so that our scenario also suggests  $\chifc > \chiea$.

Both the original droplet theory \cite{FH2} and our refined scenario \cite{our-letter}
predict scaling laws for the time dependent quantities 
measured in aging such as the magnetic autocorrelation function 
and the dynamical susceptibilities 
in terms of a time-dependent length scale $L(t)$
which presumably grows extremely slowly 
in a logarithmic fashion due to thermally activated processes.
By now it is well understood that length scale that can be
explored in practice is very much limited not only in numerical
simulations (typically $1-10$ lattice spacings)
but also in real experiments (typically $ \sim 100$ lattice spacings). 
The latter implies one must seriously take care possible pre-asymptotic behaviors
to elucidate the desired asymptotic behavior associated with 
the putative $T=0$ glassy fixed point. To cope with such a complicated 
situation still in a controlled way, we examine the scaling theory 
by Monte Carlo simulations in two strokes.

First, we examine the growth law of the dynamical length 
scale $L(t)$ itself by directly measuring the 
spatial coherence using two real replicas.
As realized in recent studies \cite{HYT,Saclay-AC,Uppsala01,BB02}, 
the problem of crossover from critical to activated dynamics 
is the central issue here.
Second, we examine the scaling properties of the time dependent
quantities of our interest by parameterizing the times using
the data of the time dependent length scale obtained by 
the separate simulation. The original droplet theory and  
our extended version provide some useful information of
finite {\it length} correction terms to the asymptotic 
limit $L \to \infty$. The two-strokes (or parametric) 
strategy of the present paper, which is already employed partially in
the previous studies \cite{Huse,KYT,KYT2,TYK,HYT}, allows
us to cope with the mixture pre-asymptotic behaviors of different
origins in a controlled way and far more advantages
than usual approaches which try to examine the scaling laws 
in one stroke directly as a function of times blindly 
with many uncontrolled fitting parameters.

In this paper we present a detailed study on the isothermal aging
by Monte Carlo (MC) simulations on a four-dimensional ($4d$)
EA Ising SG model. 
While the model system in $4d$ is somewhat non realistic, 
the advantage of studying the $4d$ Ising EA model is that
important equilibrium properties concerning both the critical
phenomena at $T_c$ and some essential scaling properties associated
with the $T=0$ glassy fixed point,  both of which will turn out to provide
extremely useful information to study off-equilibrium dynamics, 
are far better known in $4d$ than in three dimensions ($3d$).
In oder to take care of the critical fluctuations, we will use 
the well established information provided by the previous studies.
Furthermore the value of stiffness exponent $\theta$ is found to be
considerably larger than that of $3d$ which implies easier access 
to low temperature properties in $4d$ than in $3d$ within limited
length scales. Indeed, a recent analysis of defect free-energy in $4d$ 
could clarify the anticipated crossover from critical regime 
to low temperature regime \cite{KH}. In our analysis
we employ the values of these parameters and  fix them
so that we are left with a few free parameters in our scaling analysis.

The present paper is organized as follows.  In the next section 
we introduce our model system studied. 
In section \ref{sec:def-two-time}, we introduce the two time quantities used in
this paper and summarize some of their
basic properties for the convenience of later sections.
In section \ref{sec:theory} we explain our extended
droplet scaling theory \cite{our-letter}
in a more self-contained and comprehensive manner
recalling also the fundamental results of the original 
droplet theory \cite{BM,FH1,FH2}.
In section \ref{sec:growthlaw} we examine the growth law of the
dynamical length $L(t)$
by MC simulations.  In section \ref{sec:analysis-one-time} and
\ref{sec:analysis-two-time}, we examine time dependent physical
quantities by MC simulations and perform scaling analysis using the
growth law $L(t)$ obtained in  section \ref{sec:growthlaw}.
A part of the results was already reported  in Ref.~\onlinecite{HYT}.
Finally in section \ref{sec:discussion}, we present some discussions and
conclude this paper.

%Model and Methods
%\input{sec02-v6}

\section{Model and simulation method}
\label{sec:model-simulation-method}

We study the $4d$ Ising EA SG model, defined
by the Hamiltonian 
\begin{equation}
 H = -\sum_{\langle ij\rangle} J_{ij}S_iS_j -h\sum_iS_i, 
\end{equation}
where the sum runs over pairs of nearest neighbor sites. Ising variables are
defined on a hypercubic lattice with periodic
boundary conditions in all directions. 
The interactions are quenched random variables drawn with equal
probability among $\pm J$ with $J > 0$. We will use $J$ as the energy unit. 
The last term in the Hamiltonian 
represents the Zeeman energy  with $h$ being the
strength of the external uniform magnetic field. In this representation,
the magnetic field $h$ has the dimension of energy so that we will
also measure it in the unit of $J$.
In the simulations, we use $\kb T/J$ for the temperature scale
and we set the Boltzmann constant $\kb=1$ for simplicity. 

It has been well-established that a SG phase transition does occur at
a finite temperature with strong ordering, namely a finite 
amplitude of SG  order parameter  in the ordered phase. 
Recent extensive MC studies\cite{KH,Marinari99} have estimated the
critical temperature to be $2.0J$.
The critical exponent $\nu$ of the diverging coherence length is also
obtained as $\nu\sim 0.9-1.0$\cite{Marinari99,KH,BernardiCampbell}.
Another important exponent associated with $T=0$ glassy fixed point is
the stiffness exponent $\theta$ whose value is also obtained around
$0.7$ by MC simulation\cite{KH} and ground state
calculation\cite{Hartmann}. We note that the  
value of $\theta\sim 0.7$ in $4d$ is significantly larger than that of
$3d$ Ising  EA model, $\theta_{3d}\sim 0.2$\cite{theta3D}. 
This fact allows us to analyze the asymptotic behaviors rather easily
than $3d$ case, which  
is one of our main reasons for investigating the $4d$ EA model. 

The simulation method is a standard single-spin-flip MC
method using the two-sublattice dynamics with heat-bath transition
probability. We define one Monte Carlo step (MCS) as $N$ spin trials. 
We also use the multi-spin coding technique, which simulates 32
different systems independently at the same time on a 32-bit computer. 
The system sizes studied are $L=8$, $16$, $24$ and $32$ at $T_{\rm c}$. 
Below $T_{\rm c}$, we mainly study $L=24$ systems, and $L=32$ in order
to check finite-size effects. 
There is no significant difference between data of $L=24$ and $32$ at
least within our time window ($10^{5}$ MCS).

\section{Two-Time Quantities}
\label{sec:def-two-time}

Experimentally, isothermal aging of spin glasses is
investigated by observing response of the system to 
an applied external magnetic field. In the present paper, we 
study dynamical DC linear magnetic susceptibilities and their conjugate 
magnetic (spin) autocorrelation function during isothermal aging 
by Monte Carlo simulations. In the present section, we introduce
the two time quantities and summarize some basic properties.

To mimic the experimental protocol of isothermal aging, 
we consider that the configuration of the system is completely 
random at time $t=0$ and then start to relax in touch with a 
heat-bath at temperature $T$ which is lower than the critical
temperature $T_{c}$. Thus cooling rate is infinitely fast.

There are two standard protocols used in DC magnetization
measurements. In the so-called zero field cooling (ZFC) procedure, 
the system first evolves for a waiting time $\tw$ without
an applied magnetic field then a small probing magnetic field 
of strength $h$ is switched on. The growth of the induced magnetization 
is measured afterwards. In the measurement of the so-called thermoremanent 
magnetization (TRM), the system evolves under the applied magnetic field 
of strength $h$ 
for the waiting time $\tw$ and then the field is cut-off.
The decay of the magnetization induced during the waiting time 
is measured afterwards. In our simulations, we measure
the linear susceptibility at time $t > \tw$ as,
\be
\chi(t,\tw)=\frac{1}{N} \frac{\langle  M(t) \rangle }{h}
\label{eq-def-chi}
\ee
where $M(t)$ is the total magnetization measured at time $t$
with $N$ being the number of spins. The total magnetization is given
by sum
\be
M(t)=\sum_{i}S_i(t)
\ee
where $S_i(t)$ is the Ising spin variable at site $i$ at time $t$
after the quench.
Correspondingly,  we measure the magnetic (spin) autocorrelation function,
\be
C(t,\tw)=\frac{1}{N}\langle M(t)M(\tw)\rangle =\frac{1}{N} \sum_{i}\langle S_i(t)S_i(\tw)\rangle 
\label{eq-def-c}.
\ee
The last equation holds for our model with no ferromagnetic or
anti-ferromagnetic bias so that
the sum of the cross terms of $i \neq j$ vanishes as $1/\sqrt{N} \to 0$ in the
thermodynamic limit $N \to \infty$.
Furthermore in our Ising model, the spins are
normalized  $S_{i}^{2}=1$ which yields,
\be
C(t,t)=\frac{1}{N} \sum_{i}\langle S_i(t)^{2}\rangle  =1
\label{eq-spin-normalization}
\ee
for any $t$.

In the above equations, bracket $\langle \ldots \rangle$
means to take an average over different realizations 
of initial conditions, thermal noises and realization of random 
exchange couplings of the system.  However, the above quantities 
are presumably self-averaging for thermodynamically large systems 
$N \to \infty$. 

%In glassy systems, the above quantities show waiting time 
%effects, namely they are not only functions of the time difference 
%$\tau=t-\tw$ but the waiting time $\tw$. 
%A primary interest is then to find
%universal scaling properties of these quantities.

If linear response holds, the dynamical linear susceptibilities
measured in the ZFC and TRM procedure can be written as,
\be
\chi_{\rm TRM}(t,\tw)=\int_{0}^{\tw} dt' R(t,t') 
\label{eq:trm-linear}
\ee
\be
\chi_{\rm ZFC}(t,\tw)=\int_{\tw}^{t} dt' R(t,t')
\label{eq:zfc-linear}
\ee
where $R(t,t')$ is the magnetic linear response function. The latter is
define as,
\be
R(t,t')=\frac{1}{N}\lim_{\delta h \to 0}\frac{\delta \langle M(t) \rangle}{\delta h(t')} \qquad t > t'
\label{eq-def-response}
\ee
where $\delta \langle M(t) \rangle$ is the induced magnetization at time
$t$ by an infinitesimal probing pulse field $\delta h(t')$ 
applied only at time $t'( < t)$.
From these, it follows that the sum of the two susceptibilities
\be
\chi_{\rm ZFC}(t,\tw)+\chi_{\rm TRM}(t,\tw)
=\int_{0}^{t}dt'R(t,t')=\chi_{\rm ZFC}(t,0)
\label{eq:sum-rule}
\ee
becomes independent of the waiting time $\tw$ and only a function
of the total time $t$ elapsed after the temperature-quench
\cite{Uppsala-superposition,Uppsala}.
One can use this {\it sum rule} as a criterion
to check linearity of measurements 
\cite{Uppsala-superposition,VHO,MC-sumrule}.

Let us briefly discuss the implication of the sum rule 
\eq{eq:sum-rule} combined
with the following very mild assumptions.
First, the ZFC linear susceptibility 
$\lim_{\tw \to 0}\chi_{\rm ZFC}(t,\tw)$ increases with
$t$ but saturates since it is bounded from above. 
Let us define in particular the limit, 
\be
\chi_{\rm FC} \equiv \lim_{t \to \infty}\chi_{\rm ZFC}(t,0).
\label{eq-def-chifc}
\ee
We call the latter as Field Cooled (FC) susceptibility because
it is more or less similar to what is called FC susceptibility
as we discuss below. 
Second, it is natural to assume {\it weak long term memory} \cite{MFT,BCKM},
\be
\lim_{t \to \infty}\chi_{\rm TRM}(t,\tw) =0
\label{eq-weak-long-term-memory}
\ee  
for any large but finite  $\tw$. The latter simply means that
TRM should relax down to zero for any large but finite waiting time $\tw$
during which the magnetic field is applied.
Then we find using above assumptions in the sum rule \eq{eq:sum-rule},
\be
\lim_{t \to \infty}\chi_{\rm ZFC}(t,\tw) 
= \lim_{t \to \infty}\chi_{\rm ZFC}(t,0) = \chi_{\rm FC}
\ee
for any large but finite $\tw$. 

Let us explain why we call \eq{eq-def-chifc} as the FC susceptibility.
Usually the FC magnetization is measured by cooling down the temperature 
with a certain cooling rate from above $T_{c}$ down to 
a target temperature $T$ below $T_{c}$ with the magnetic field 
$h$ being applied. Suppose that it takes time $\epsilon$ to 
cool down the temperature (typically of order 30 sec)
and the target temperature is reached at time $t=0$. 
Then the linear susceptibility
measured in this protocol can be expressed as,
\be
M_{\rm FC}(t)/h=\int_{-\epsilon}^{0} dt' R(t,t') + \chi_{\rm ZFC}(t,0).
\label{eq-fc-exp}
\ee
The above expression is formally valid as far as linear response holds. 
Note that the response function in the first term is defined 
with respect to the particular schedule of the temperature changes.
The contribution of the first term decreases with time $t$ because
of the weak long term memory property \eq{eq-weak-long-term-memory}.
Thus in the limit $t \to \infty$ the susceptibility $M_{\rm FC}(t)/h$
converges to the FC susceptibility of our definition in \eq{eq-def-chifc},
$\lim_{t \to \infty} M_{\rm FC}(t)/h= \lim_{t \to \infty}
\chi_{\rm ZFC}(t,0)=\chi_{\rm FC}$.  The approach to the limit
may well be slow. Experimental observations (See for instance
Fig 13. of \onlinecite{Uppsala}) show that the correction 
term to the asymptotic limit relaxes slowly but the amplitude
is very small such that it can be made 
much less than $1\%$ of the asymptotic value 
well within the experimental time window.

In the present paper, we do not discuss the possible effects of finite
cooling rates and assume the idealized temperature quench
$\epsilon=0$. In this idealized situation, the first term in \eq{eq-fc-exp}
is absent and the TRM becomes equivalent to the 
so-called isothermal remanent magnetization (IRM).

%As we discuss later, the droplet theory predicts {\it even weaker 
%long term memory} that $\lim_{\tw \to \infty}\chi_{\rm TRM}(t,\tw) =0$
%with the ratio $x=L(t)/L(\tw)$ being fixed to a constant larger
%than $1$. Here $L(t)$ and $L(\tw)$ are dynamical length scales 
%given by the growth law \eq{eqn:growth}. Then the sum rule 
%implies $\lim_{t \to \infty}\chi_{\rm ZFC}(t,\tw)= \chi_{\rm FC}$
%swith the ratio $x=L(t)/L(\tw) > 1$ being fixed.

Another interesting limit is to consider $\tw \to \infty$ first 
with fixed time separation $\tau=t-\tw$. 
In this limit, one expects to find equilibrium (stationary) response,
\be
\chi_{\rm eq}(\tau) 
\equiv \lim_{\tw \to \infty} \chi_{\rm ZFC}(\tau+\tw,\tw),
\label{eq-def-chieq}
\ee
which only depends on the time separation $\tau$.
The equilibrium susceptibility $\chi_{\rm EA}$ is defined as,
\be
\chi_{\rm EA} \equiv \lim_{\tau \to \infty}\chi_{\rm eq}(\tau).
\label{eq-def-chiea}
\ee
The last static susceptibility $\chiea$ is more or less close to what is
called  the ZFC magnetization (divided by $h$). 
% The ZFC magnetization is usually measured 
% as the following. First the system is cooled down 
% from above $T_{c}$ down to a temperature well below $T_{c}$.
% Then a small field $h$ is switched on and the induced magnetization
% is measured by heating up the system with a certain heating rate.

A very important issue is then {\it whether the two susceptibilities
$\chi_{\rm EA}$ and $\chi_{\rm FC}$ are the same 
or different}. As we noted in the introduction this is
intimately related with the fundamental 
experimental observation in spin-glass systems \cite{Nagata79}, namely  
$\chifc > \chizfc$. One of the most remarkable finding 
of the dynamical mean-field theory\cite{MFT} is that indeed an inequality
\be
\chi_{\rm FC} > \chi_{\rm EA}
\label{eq-fc-ea}
\ee
holds with  $\chi_{\rm FC}$  defined in \eq{eq-def-chifc} and  
$\chi_{\rm EA}$  defined in \eq{eq-def-chiea}.
The difference is due to anomalous contribution of the slowly relaxing, aging part 
of the response function $R(t,t')$. On the other hand, the conventional
droplet theory \cite{BM,FH1,FH2} was understood \cite{BCKM} to predict 
$\chifc=\chiea$, i.~e. no anomaly. As we explain later in section
\ref{subsec:domainwall-softdroplets}, our extended droplet theory \cite{our-letter}
predicts \eq{eq-fc-ea} with $\chifc$ being identified with the dynamical
susceptibility $\chid$ associated with the noble dynamical order
parameter $\qd (< \qea)$ as $\kb T\chid=1-\qd$.

Another important issue is to what extent FDT holds in aging systems. In
our present context FDT reads as,
\be 
R(t,t')=\frac{1}{\kb T}\partial_{t'} C(t,t')
\label{eq-fdt-differential}
\ee where $C(t,t')$ is the  autocorrelation function.  
It becomes after integration over time $\int^{t}_{\tw} dt'\cdots$,
\be
\mbox{FDT} \qquad  1-C(t,\tw)=\kb T\chi_{\rm ZFC}(t,\tw). 
\label{eq-fdt}
\ee
Since this {\it must} hold precisely in equilibrium, 
the equilibrium limit of ZFC linear susceptibility 
\eq{eq-def-chieq} must be
related to that of the spin autocorrelation function as,
\be
T \chi_{\rm eq}(\tau) 
 = 1- C_{\rm eq}(\tau)
\label{eq-fdt-eq}
\ee
where 
\be
C_{\rm eq}(\tau) 
\equiv \lim_{\tw \to \infty} C(\tau+\tw,\tw),
\label{eq-def-ceq}
\ee
is the spin autocorrelation function in the equilibrium.
In the {\it static limit} ${\tau \to \infty}$
\eq{eq-fdt-eq} becomes the {\it static} FDT,
\be
\kb T\chi_{\rm EA}=1-q_{\rm EA},
\label{eq-qea-chiea}
\ee
where $q_{\rm EA}$ is the static EA order parameter
defined as,
\be
q_{\rm EA} = \lim_{\tau \to \infty}C_{\rm eq}(\tau).
\label{eq-def-qea}
\ee

Except for the ideal equilibrium limit, FDT \eq{eq-fdt} 
is not guaranteed in general. However, it was realized
recently by Dean, Cugliandolo and Kurchan that possible 
amplitude of the violation of the  FDT \eq{eq-fdt},
\begin{eqnarray}
I(t,\tw)  \equiv  1-C(t,\tw)-\kb T\chi_{\rm ZFC}(t,\tw) 
\label{eq-integral-violation-fdt}
\end{eqnarray}
is bounded from above by the entropy production rate \cite{FDT-bound}. 
The bound implies even for very slowly relaxing systems in which entropy 
production rate becomes small, 
the FDT \eq{eq-fdt} should hold 
between spontaneous thermal fluctuations and linear 
responses at least for short enough time scales $\tau$.

Let us consider asymptotic limit $\tw \to \infty$
of the two time quantities with fixed value of the autocorrelation 
function $C=C(\tau+\tw,\tw)$.
First let us note that $\lim_{\tau \to \infty}C_{\rm eq}(\tau)=\qea$ implies
the time translational invariance (TTI) is strongly broken at $0 < C < \qea$,
i.~e. the two time quantities cannot be a function of only the time
separation in this regime.
Concerning the integral FDT violation \eq{eq-integral-violation-fdt},
it becomes a function of $C$, i. e. $\lim_{\tw\to \infty} I(t,\tw)=I(C)$.
In conventional cases $I(\qea < C < 1)=0$
while $I(0 < C < \qea) > 0$ being a non-trivial function of $C$
by the dynamical MFT \cite{MFT,FMGPP} and $I(0 < C < \qea)=\qea-C$
in usual coarsening systems \cite{B94,HF87,spherical-SK}.
On the other hand, our scenario \cite{our-letter} suggests $I(\qd < C < 1)=0$ and
$I(0 < C < \qd)=\qd-C$. Here $\qd$ is the new dynamical order parameter
which is smaller than $\qea$. Thus the breaking points of TTI and FDT take
place separately at $\qea$ and $\qd$ respectively in our scenario while
both of them take place simultaneously at $\qea$ in the dynamical MFT
and usual coarsening systems \cite{B94,HF87,spherical-SK}.

%Droplet Theory (modified version)
%\input{sec03-v6}

%%%%%%%%% COMMENT here
%%%%%%%%%%

\section{Theoretical Background}
\label{sec:theory}

\newcommand{\Lmax}{L_{\rm m}}
\renewcommand{\Lo}{L_{0}}
\newcommand{\rhoo}{\tilde{\rho}(0)}
\newcommand \FtypLR {F_{L,R}^{\rm typ}}
\newcommand \FtypL {F_{L}^{\rm typ}}
\newcommand \Ueff {\Upsilon_{\rm eff}}

In this section we discuss our extended 
droplet theory sketched in Ref.~\onlinecite{our-letter}
concerning isothermal aging  which will be our basis to
analyze the data of Monte Carlo simulations in later sections.
As we noted in the introduction,
we pay a special attention to the idea of so-called 
{\it effective stiffness} of droplet excitations in the presence
of domain walls which are present as extended defects during
isothermal aging. 
For clarity, we will try to present this section in a 
self-contained fashion including summaries of the results of 
the original droplet theory \cite{FH1,FH2}
which are almost fully included in our scenario.
To simplify notations, we consider systems 
of $N$ Ising spins $S_{i}=\pm 1$ ($i=1,\ldots,N$) 
in a $d$-dimensional space coupled by short-ranged interactions
of energy scale $J$ with random signs with no ferromagnetic or 
anti-ferromagnetic bias.

\subsection{Basics}

Let us recall briefly the starting point of the 
droplet theory.\cite{FH1}
It assumes that thermodynamic states of (Ising) spin-glass phases 
consist of a pair of pure states of an infinite system
which are related by global spin inversion. 
At low but finite temperature in the spin-glass 
phase, an equilibrium state can be considered as made of a ground 
state, say $\Gamma$ or its global spin inversion $\bar{\Gamma}$,
plus thermally activated droplet excitations of various
sizes taking place on top of $\Gamma$. 
In simple systems such as ferromagnets droplet excitations exist 
but play a rather limited role \cite{HF87}. 
An essential finding of the original droplet 
theory \cite{BM,FH1} is that the temperature 
is {\it dangerously irrelevant} in spin-glass phases 
because of strong impacts of thermally activated droplets.

A droplet at a given length scale $L$ is supposed to be
a compact cluster of spins with a volume $L^{d}$ with $d$ being
the dimension of the space and a surface volume
$L^{d_f}$ with $d_f$ being the (fractal) surface dimension.
The typical value of its excitation gap $F_L^{\rm typ}$ 
with respect to the ground state is supposed to scales as  
\begin{equation}
 F_L^{\rm typ}\sim \Upsilon (L/\Lo)^\theta, 
\label{eqn:fgap}
\end{equation}
where the exponent $\theta >0$ is the stiffness exponent,
$\Upsilon$ is the stiffness constant and $\Lo$ is a microscopic
length scale. 
The excitation gap is however broadly distributed with
the typical value given above. The probability 
distribution of the free-energy gap
is expected to follow a universal scaling form,
\be
\rho(F_{L})dF_{L} = 
\tilde{\rho}(F_{L}/\FtypL) dF_{L}/\FtypL
\label{eq-scaling-pf}
\ee
with non-vanishing amplitude at the origin 
\be
\rhoo > 0
\label{eq-p0}
\ee
which allows marginal droplets.
This property $\rhoo > 0$  allows {\it dangerously active droplets} 
which will respond to arbitrarily weak perturbations so that they
play extremely important roles as the Goldstone modes: 
they dominate spontaneous thermal fluctuations and linear responses.

Dynamically, the excitation of 
such a cluster of spins is supposed to happen only by thermal 
activated process. The typical value of free-energy barrier 
$B_L^{\rm typ}$ to flip the cluster of spins is supposed to scale as
\begin{equation}
 B_L^{\rm typ}\sim \Delta (L/L_0)^\psi,
\label{eqn:fbarrier}
\end{equation}
where $\Delta$ is a characteristic free-energy scale of the barriers.
The Arrhenius law implies that a droplet of length scale $L(t)$,
\begin{equation}
L(t) \sim \left(\frac{\kb T}{\Delta}\ln(t/\tau_0)\right)^{1/\psi}, 
\label{eqn:growth}
\end{equation}
can be activated within a time scale of $t$. Here $\tau_0$ is 
a certain unit time scale for the activated processes. Let us call
this time dependent length scale as the dynamical length scale.

\subsection{Crossover from Critical to Low Temperature Regime}
\label{subsec:crossover-critical-activated}

In practice, it is necessary to take into account of critical 
fluctuations near $T_{\rm c}$. Even at $T < T_{\rm c}$, the length
scales shorter than the coherence length of the critical
fluctuation 
\be
\xi_{-} \sim L_{0}|1-T/T_{\rm c}|^{-\nu}
\label{eq-xi}
\ee
should be dominated by critical fluctuations. Asymptotic low
temperature properties should appear only at larger length scales.

\begin{figure}[]
\resizebox{\figwidth}{!}{\includegraphics{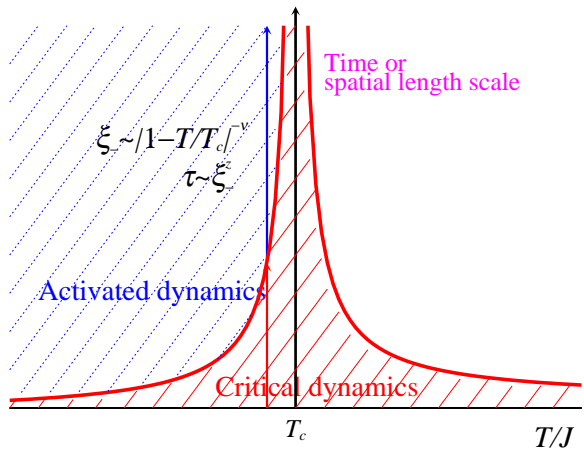}}
 \caption{
Schematic picture of length scale.
}
\label{fig:sketch-xi}
\end{figure}

Correspondingly, we expect two typical stages in the dynamical
length scale $L(t)$ as shown in Fig.~\ref{fig:sketch-xi}.
One is a critical dynamics associated with critical slowing
down in time range $\tau_0(T)\gg t \gg t_0$, where $L(t)$ follows a
power law with the dynamical critical exponent $z$, 
\begin{equation}
 L(t)=l_0(t/t_0)^{1/z}.
\label{eqn:LTcritical}
\end{equation}
Here $l_0$ is a microscopic length scale which is of
order $1$ lattice distance in EA models and $t_0$ is a microscopic time
scale which is  
typically $t_0 \sim 10^{-12}-10^{-13}$
(sec) in real spin systems while it is 1 Monte Carlo Step (MCS) 
in usual heat-bath Monte Carlo simulations.
This formula means that even at $T < T_{\rm c}$, $L(t)$ behaves
like the critical power-law at short time-length scales. 
On the other hand, the intrinsic low-temperature dynamics associated
with the $T=0$ glassy fixed point
(\ref{eqn:growth}) should occur  only at larger time scales beyond a 
certain crossover time $t\gg\tau_0(T)$. 

The crossover length $L_0(T)$ would be determined by a comparison
between free-energy barrier and thermal energy, 
\begin{equation}
 k_{\rm B}T\sim \Delta(T)(L_0(T)/l_0)^\psi,
\end{equation}
where the characteristic free-energy scale $\Delta(T)$ behaves like
$\Delta(T) = J|1-T/T_{\rm c}|^{\psi\nu}$ near $T_{\rm c}$\cite{FH2}. 
It leads to the temperature dependence of $L_0(T)$, 
\begin{equation}
 L_{0}(T)=l_0(T/J)^{1/\psi}|1-T/T_{\rm c}|^{-\nu},
\end{equation}
which is essentially equivalent to \eq{eq-xi}.
The corresponding crossover time $\tau_0(T)$ is given by 
\begin{equation}
 \tau_{0}(T)=t_0(L_0(T)/l_0)^z.
\end{equation}
We obtain the singular part of the crossover time at $T$,
$\tau_0(T)\sim |1-T/T_{\rm c}|^{-z\nu}$, as expected from a critical
scaling theory. 
Consequently, the scaling formula of the growth law which describes the
whole crossover from the critical dynamics at $t\ll\tau_0(T)$ and the
activated dynamics at $t\gg\tau_0(T)$ is given by   
\begin{equation}
 L(t)/L_0(T)=\tilde{L}(t/\tau_0(T)),
\label{eqn:LTcross}
\end{equation}
where
\begin{eqnarray}
\tilde{L}(x) \sim \left\{ 
\begin{array}{cc}
 x^{1/z}  & (x \ll 1), \\
 \log^{1/\psi}(x) & (x\gg 1). 
\end{array}
\right.
\label{eqn:scal-g}
\end{eqnarray}
One should note that the crossover could be very gradual and
functional form of the intermediate regime (which will dominate
realistic time ranges in simulations and experiments) 
can have very complicated expression which is not obvious.
This crossover in the growth law of the dynamical length scale is
numerically examined in section \ref{sec:growthlaw}. 

The importance of the crossover from critical to low temperature behavior
has been pointed out by Bokil et al \cite{Bokil00} concerning some static 
properties of low-temperature SG phase.
Let us note that the above analysis of the crossover from the critical 
to activated dynamics is a direct (dynamical) analogue of the
analysis of defect free-energy in $4d$ EA model\cite{KH} by one of us (KH).
In the latter study, the defect free-energy was found to become
a universal constant in the limit $L \ll \xi_{-}$ and grows 
as $\Upsilon(T)(L/\Lo)^{\theta}$ 
and $\Upsilon(T) \sim J |1-T/T_{\rm c}|^{\theta\nu}$ 
with $\theta=0.82$,  $\nu=0.93$ at $L \gg \xi_{-}$.

\subsection{Domain Growth}
\label{subsec:one-time}

During isothermal aging up to a time $t$ after quench,   
domains with the mean size $L(t)$ separating different pure states
grow up by coarsening domain walls of smaller length scales.
The droplet theory proposed scaling properties of time-dependent
physical quantities in terms of time-dependent mean domain size $L(t)$. 
We discuss the scaling by $L(t)$ of one-time quantities with time elapse
after the quench, which examine later by MC simulations. One example is
time development of energy per spin defined by 
\begin{equation}
e(t)=-\frac{1}{N}\sum_{\langle ij\rangle}\langle J_{ij}S_i(t)S_j(t)\rangle.
\label{eq:energy}
\end{equation}
Relaxation of the energy per spin is expected to be
due to relaxation of excessive energy associated with domain walls.
Thus it is expected to decrease as \cite{KYT},
\begin{equation}
e(t)-e_{\rm eq} =  \Upsilon'\left(\frac{L(t)}{l_{0}}\right)^{\theta-d},
\label{eq:ene-decay}
\end{equation}
where $\Upsilon'$ is a temperature dependent parameter. 

Another interesting one-time quantity is domain-wall density $\rho_s(t)$
in which the morphology of the domain with the fractal surface dimension
$d_f\geq d-1$ appears. In coarsening dynamics the density decreases with
time during isothermal aging. In simple systems such as a ferromagnet
where the domain wall becomes flat at sufficiently low temperature, the
density of domain wall is proportional to the inverse of the mean size
$1/L(t)$.  However, it could be rough with the fractal dimension in spin
glasses because of the disorder and frustration so that we expect 
\begin{equation}
 \rho_s (t) \sim \left(\frac{L (t)}{l_0}\right)^{d_f-d}.
\label{eq:density-decay}
\end{equation}
In MC simulations, two replicas with identical interaction bonds are
updated independently and the domain-wall density is calculated as\cite{Huse}
\begin{equation}
 \rho_s (t) = \frac{1}{2}\left(1-
\frac{1}{N_B}\sum_{\langle ij\rangle}
\sigma_i^{(\alpha)}(t)
\sigma_j^{(\alpha)}(t)
\sigma_i^{(\beta)}(t)\sigma_j^{(\beta)}(t)
\right), 
\label{eq:dwdensity}
\end{equation}
where $N_B$ is the number of bonds and the suffixes $\alpha$ and $\beta$
denote replica indices. Here, following Ref.~\onlinecite{Huse} we take
short time average of each spin as 
\begin{equation}
 \sigma_i^{(\alpha)}(t)=\mbox{sgn}\left(\sum_{\tau=t/2}^{t}
S_i^{(\alpha)}(\tau)\right). 
\label{eqn:sigma}
\end{equation}
This procedure in (\ref{eqn:sigma})  means that smaller fluctuations
associated with small droplets are eliminated and the coarsening domain
walls are  emphasized. 
Without taking such average over time, the density is similar to
so-called link overlap function, which does not vanish  in the long time
limit.

\subsection{Domain Walls and Soft Droplets}
\label{subsec:domainwall-softdroplets}

Following Ref.~\onlinecite{FH2}, the whole process of isothermal 
aging may be divided into {\it epoch}s such that 
the typical separation between domain walls (hereafter
simply denoted as domain size) 
$\Lo,a\Lo,a^{2}\Lo,\ldots, a^{n}\Lo,\ldots$
where $a>1$. At each epoch, droplets of various sizes 
up to that of the domain size can be thermally activated or polarized 
by the magnetic field. 
The two time quantities we introduced in section \ref{sec:def-two-time}
namely the autocorrelation function $C(\tau+\tw,\tw)$ 
defined in \eq{eq-def-c} and the linear 
susceptibility $\chi(\tau+\tw,\tw)$ defined in \eq{eq-def-chi} 
probe such thermal fluctuations 
and linear responses of droplets smaller than the size of the domain 
if the time separation is limited such that $L(\tau) \leq  L(\tw)$.

In the previous work \cite{our-letter}, we extended the 
idea of Fisher and Huse \cite{FH2} who noticed that droplet excitations
can be {\it softened} in the presence of a {\it frozen-in} defect 
or domain wall compared with ideal equilibrium with no
extended defects. This is because droplets which touch 
the defects can reduce the excitation gap compared with those 
in equilibrium. This effect will have very important impacts
on the two time quantities.

Let us consider a system with a frozen-in defect of size $R$:  
a large  droplet of size $R$ is flipped with respect to $\Gamma$,
which is a ground state of an infinite system, and then it is frozen. 
The typical free-energy gap $\FtypLR$ of a smaller 
droplet of size $L$ in the interior of the frozen-in 
defect is expected to scale as
\be
\FtypLR = \Upsilon_{\rm eff}[L/R](L/L_0)^\theta   \qquad L < R
\label{eqn:effU-general}
\ee
with $\Ueff[L/R]$ being an {\it effective stiffness} 
which is only a function of the ratio $y=L/R$.

For $y \ll 1$, $\Ueff[y]$ will decrease with $y$ as~\cite{FH2}
\begin{equation}
\Upsilon_{\rm eff}[y]/ \Upsilon=1-c_{v} y^{d-\theta} \qquad 
\mbox{for} \qquad   y \ll 1.
\label{eqn:effU-small}
\end{equation}
Here $\Upsilon$ is the original stiffness constant $\Upsilon=\Ueff(0)$. 
A basic conjecture, on which our new scenario based, is that 
 at the other limit $y \sim 1$ the effective stiffness vanishes as,
\begin{equation}
\Upsilon_{\rm eff}[y]/\Upsilon \sim (1-y)^{\alpha}
\qquad  y \sim 1
\label{eqn:effU-large}
\end{equation}
with  $ 0 < \alpha < 1$ being an unknown exponent, and that the lower
bound for $\FtypLR$ should be of order $J$, say $ F_{0}$. 

We assume that the probability distribution of the free-energy gap
$F_{L,R}$ is broad and obey the same functional form as
\eq{eq-scaling-pf} where $\FtypL$ should be replaced by $\FtypLR$.
Then the scaling form of the distribution of the gap \eq{eq-scaling-pf}
and non-zero amplitude of gap-less droplets $\rhoo >0$ \eq{eq-p0}
implies that the probability that a gap is smaller
than a certain threshold $\delta U  (\ll \FtypL)$ scales as,
\be
\mbox{Prob}(F_{L,R} \langle  \delta U) \sim \rhoo \delta U/\FtypLR.
\label{eq-gap-deg}
\ee
Since the probability is {\it linear} with $\delta U$, these active droplet
are dangerously irrelevant \cite{BM,FH1,FH2}: 
arbitrary small perturbation $\delta U$ may trigger a droplet excitation.

\subsubsection{Two Length Quantities}

To explain the consequence of our conjecture above introduced, 
let us consider thermal fluctuations and magnetic linear responses of
droplet excitations of size $L$ enclosed in a compact region made by
a frozen-in defect at scale $R$ which has $N_{d} \propto (R/L_{0})^{d}$ spins.
To this end, let us construct a toy droplet model 
\cite{FH1} defined on logarithmically separated shells of length scales 
$L_k/\Lo=a^{k} < R/\Lo$ with $a>1$ and $0 \leq k \leq n_{2} \leq n_{1}$ 
where $R/\Lo=a^{n_1}$ and $\Lmax/\Lo=a^{n_2}$.
For each shell an {\it optimal } droplet is assigned whose 
free-energy gap is minimized within the shell. 
Then $\FtypLR$ will be of order $F_{0}$ at the shell $k=n_{1}$
so that we have 
\be
\tilde{F}_{L,R}^{\rm typ}=\FtypLR(1-\delta_{k,n_1})+F_{0}\delta_{k,n_1}.
\label{eq-ftyp-model}
\ee 
For simplicity, droplets at
different scales are assumed to be independent from each other.

At the shell $L/\Lo=a^{k}$, the system is decomposed into 
compact cells of volume $L^{d}$ 
such that each cells represents a droplet of size $L$.
The number density of droplets per spin associated with the shell
scales as,
\be
N_{d} \left(\frac{L}{L_{0}}\right)^{-d} \ln a.
\label{eq-droplet-density}
\ee
Each droplet excitation 
will induce a random change of the magnetization of order
\be
M_{L} \sim m\sqrt{(L/\Lo)^{d}},
\label{eq-def-ml}
\ee
where $m$ is the average magnetic moment within a volume of $\Lo$. 

The thermal fluctuation can be measured by an
order parameter 
\be
q=\overline{N_{d}^{-1}\sum_{i}\langle S_{i}\rangle^{2}}, 
\ee
where the sum
runs over sites in the interior of the frozen-in defect.
Here we have put the overline $\overline{\cdots}$ which means
to take average over different realization of the randomness.
In the absence of  any droplet excitations,  $q=1$ holds due to the
normalization of the Ising spins. 
A spontaneous thermal fluctuation of a droplet will take place
if its excitation gap happens to be smaller than the thermal energy $\kb T$.  
The probability of the latter is found to be 
proportional to $\rhoo \kb T/\FtypLR$ due to  \eq{eq-gap-deg}.
So the reduction from $1$ due to a droplet excitation at scale 
$L$ is of order $M^{2}_{L} \rhoo(\kb T/\FtypLR)$.
Then the reduction of the spin autocorrelation  by droplet excitations
at scale $L$ is estimated as,
\be
\delta q_{L} \sim  M_{L}^{2} \rhoo  \frac{\kb T}{\FtypL}
\left(\frac{L}{L_{0}}\right)^{-d} \ln a 
\label{eq-c-dl}
\ee
where the last factor is due to the number of droplets
per spin given in \eq{eq-droplet-density}.

The magnetic linear response by weak external magnetic field $h$ 
is measured by a linear susceptibility 
\be
\chi=\lim_{h \to 0}\overline{N_{d}^{-1}\sum_{i}\langle S_{i}\rangle_h/h},
\ee
where $\sum_{i}\langle
S_{i}\rangle_h$ is the induced magnetization by the field.
Since a droplet excitation can induce a magnetization 
of order $M_{L}$ given in \eq{eq-def-ml} with random signs, 
it can gain a Zeeman energy of order $\delta U_{L} \sim  M_{L} \delta h$
by responding to the field.
The probability that the droplet excitation takes place is then proportional
to the probability that the gain by the Zeeman energy is larger 
than the excitation gap of the droplet which can be found
using \eq{eq-gap-deg}. Then the expectation value of the 
magnetic moment induced by droplet excitations
at scale $L$ is estimated as,
\be
h \delta \chi_{L} \sim M_{L} \rhoo \frac{h M_{L}}{\FtypL} \left(\frac{L}{L_{0}}\right)^{-d} \ln a.
\label{eq-chi-dl}
\ee

Now summing over contributions at different length scales 
$0 < k < n_{2}$, we obtain
the total reduction of the order parameter from $1$ and the
linear susceptibility as using \eq{eq-ftyp-model}, \eq{eq-c-dl}
and  \eq{eq-chi-dl} by,
\begin{eqnarray}
&& 1-q (\Lmax,R)  =  \kb T\chi(\Lmax,R)  \nonumber \\
&&   =    \rhoo m^{2}  \sum_{k=0}^{n_2}\frac{\kb T}{\FtypLR(1-\delta_{k,n_1})+F_{0}\delta_{k,n_1}} \nonumber \\
&& = \rhoo m^{2} \kb T \times \nonumber \\
&&  \int_{L_{0}}^{\Lmax}   \frac{d L}{L} \left[  
  \frac{1-\Delta_{a} (\ln (L/R)) }{\FtypLR}
 + 
\frac{\Delta_{a} (\ln (L/R))}{F_{0}}
\right].     \hspace*{.5cm}
\label{eq-q-chi-lr}.
\end{eqnarray}
In the last equation, the sum  $\sum_{k=0}^{n_2}$ is replaced by an integral 
$\int_{\Lo}^{\Lmax}dL/L$ and 
$\Delta_{a}(z)$ is a {\it pseudo} $\delta$-function of 
width $\ln a$ \cite{note}.
Note that FDT is satisfied between (\ref{eq-c-dl}) and (\ref{eq-chi-dl}).

In the following let us call the two length quantities 
$q(L,R)$ and $\chi(L,R)$ as the 
generalized order parameter and the generalized 
linear susceptibility,  respectively.
We will associate these two-{\it length} quantities 
with the two-{\it time} quantities measured
in aging experiments.

\subsubsection{Edwards-Anderson Order Parameter}

For clarity, let us consider the {\it equilibrium limit}
where there is no extended defects which can be realized
by taking $ R \to \infty$ first. In the latter limit
we recover the result of the original droplet theory \cite{FH1},
\be
\lim_{R \to \infty}q(L,R)=\qea
+ c \frac{\rhoo m^{2}\kb T}{\Upsilon(L/\Lo)^{\theta}},
\ee
\label{eq-nodefect-limit}
with 
\be
c=\int_{1}^{\infty}dyy^{-1-\theta}. 
\label{eq-def-const-c}
\ee
and the Edwards-Anderson (EA) order parameter 
defined in \eq{eq-def-qea} evaluated as,
\be
q_{\rm EA} =\lim_{L \to \infty}\lim_{R \to \infty}q(L,R)
=1-c \rhoo m^{2}\frac{\kb T}{\Upsilon}.
\label{eq-qea-eval}
\ee 
The associated {\it equilibrium}  susceptibility $\chiea$ is
defined as \eq{eq-qea-chiea} $\kb T\chiea \equiv 1-\qea$. 
The above expressions become useful when we consider equilibrium
dynamics.

\subsubsection{Dynamical Order Parameter}

It is useful to consider asymptotic behavior at large sizes 
$R/\Lo \gg 1$ with the ratio 
$x=\Lmax/R$ being fixed. We obtain for $0 < x \leq 1$,
\be 
 q (xR,R) = \qea 
+ \frac{\rhoo m^{2}\kb T}{\Upsilon(R/\Lo)^{\theta}} A(x)
-\frac{\rhoo m^{2}\kb T}{F_{0}}\Theta_{a} (\ln x)
\label{eq-qxRR}
\ee
with $\Theta_{a}(z)$ being a  {\it pseudo} step-function of width $\ln a$
 \cite{note} and 
\begin{eqnarray}
&& A(x)=
\int_{x}^{\infty}dy y^{-1-\theta}  \nonumber \\
&& -\int_{0}^{x}dyy^{-1-{\theta}} 
(\Upsilon/\Upsilon_{\rm eff}[y](1-\Delta_{a} (\ln y))-1).
\label{eq-def-a}
\end{eqnarray}
Note that the second integral  converges 
because of \eq{eqn:effU-small} and 
the inequality $\theta < (d-1)/2$ \cite{FH1}.
Thus as far as $0 < x < 1$, the order parameter converges to
the {\it equilibrium} EA order parameter $\qea$ evaluated
in \eq{eq-qea-eval}.

In the intriguing case $x \sim 1$, $A(x)$ will remain finite
as far as $0 < \alpha < 1$. At $x \sim 1$ 
the last term of  \eq{eq-qxRR}, which is due to the anomalously
soft droplets, contributes and we obtain,
\begin{eqnarray}
q(R,R) = \qd+   A(1)\frac{\rhoo m^{2} \kb T}{\Upsilon(R/\Lo)^{\theta}},
\label{eq-q-cross}
\end{eqnarray}
where we have defined the {\it dynamical order parameter} 
\be
q_{D}\equiv\lim_{R \to \infty}q(R,R) =\qea - 
\rhoo m^{2}\frac{\kb T}{F_{0}}.
 \label{eq-def-qd}
\ee
Naturally, we can define 
the associated dynamical linear susceptibility $\chid$ as
\be
\kb T\chid \equiv 1-\qd.
\label{eq-def-chid}
\ee 

The above results imply
\be
\qd < \qea
\label{eq-qd-qea}
\ee
and
\be
\chid > \chiea
\label{eq-chid-chiea}
\ee
As we discuss below $\qd$ and $\chid$ play significantly important roles
in the dynamical observables of aging. We will see that
the field cooled susceptibility $\chi_{\rm FC}$ defined in
\eq{eq-def-chifc} is equal to the dynamical susceptibility
$\chid$ defined above. Thus the inequality \eq{eq-chid-chiea}
implies the anticipated inequality $\chi_{\rm FC} > \chiea$ 
given in \eq{eq-fc-ea}.

\subsubsection{Two-time and Two-length Quantities}
\label{subsubsec:twotime-twolength}

We can now associate the two length quantities
discussed above with the two time quantities
for short time separations $L(\tau)\leq L(\tw)$.
In the latter regime, the autocorrelation function
$C(\tau+\tw,\tw)$ and the ZFC linear susceptibility
$\chi_{\rm zfc}(\tau+\tw,\tw)$ measure respectively   
thermal fluctuations and linear responses of 
droplets smaller than the size of the domain $L(\tw)$.

The autocorrelation function at a given time separation
$\tau$ probes thermal fluctuations
of droplets as large as $L(\tau)$ in the presence of domain
walls of size $R(\tw)$ so that we expect
\be
C(\tau+\tw,\tw)=q(L(\tau),L(\tw)), 
\label{eq-c-q}
\ee 
where $q(L,L')$  is the generalized order parameter
given in \eq{eq-q-chi-lr}.
Similarly, the ZFC susceptibility probes the polarization
of droplets as large as $L(\tau)$ in the presence of
domain walls of size $R(\tw)$ we expect,
\begin{eqnarray}
\kb T \chi_{\rm ZFC}(\tau+\tw,\tw) &=& \kb T \chi(L(\tau),L(\tw)) \nonumber \\
&=& 1-C(\tau+\tw,\tw), 
\label{eq-chi-chi}
\end{eqnarray}
where $\chi(L,L')$  is the generalized susceptibility given in
\eq{eq-q-chi-lr}. 

Let us note that the generalized order parameter $q(L,L')$ 
and the generalized susceptibility $\chi(L,L')$  are defined via {\it
disorder averaging}  
of many different realization of small systems of size $L'$.
Here it is important to recall that
at any finite time, a macroscopic system will contain macroscopic 
number of domains no matter how large their size $L'=L(\tw)$ is. 
Thus we can safely evaluate these two time quantities, which 
are macroscopic quantities,
by the disorder-averaged two-length quantities.

Let us emphasize that the FDT \eq{eq-fdt} is satisfied while 
TTI (Time Translational Invariance) is broken in the above
two length/time quantities for $L(\tau) \leq L(\tw)$.
The latter is due to the fact that the autocorrelation function
and the susceptibility depends on two length/time, i.e.
not only $L(\tau)$ but also on $L(\tw)$. 
% Thus our scenario
% is significantly different from the dynamical MFT \cite{MFT}
% and usual coarsening system \cite{B94,spherical-SK,response-coarsening}
% where  breaking of FDT and TTI takes place simultaneously.
% This novel feature is entirely due to the anomalously soft 
% droplets which satisfy FDT but disappear in equilibrium
% where there are no extended defects.

For larger time separations $L(\tau) > L(\tw)$, we take into account
decay of memory due to coarsening of  domain walls 
following the original droplet theory \cite{FH2}.

\subsection{Scaling Properties of Two-Time Quantities}
\label{subsec:theory-two-time}

Using the results of the previous section, we now 
discuss scaling properties of the two time quantities
at different regimes in detail. Basically we expect three distinct
regimes i) quasi-equilibrium regime $L(\tau) < L(\tw)$,
ii) crossover regime $L(\tau) \sim L(\tw)$ and 
iii) aging regime $L(\tau) > L(\tw)$. In the following
we first consider the ideal equilibrium limit for clarity 
and then the three regimes subsequently.

\subsubsection{Equilibrium Limit}
\label{subsubsec:theory-eq}

Let us consider first the equilibrium limits of the autocorrelation
function and the ZFC susceptibility
which are obtained by taking the limit $L(\tw) \to \infty$ with fixed $L(\tau)$
in  \eq{eq-q-chi-lr},\eq{eq-c-q} and \eq{eq-chi-chi}.
From \eq{eq-nodefect-limit} one immediately finds \cite{FH2},
\be
C_{\rm eq}(\tau) 
=q_{\rm EA} +  c \rhoo m^{2}(\kb T/\Upsilon)(L(\tau)/L_{0})^{-\theta}
\label{eq-c-eq}
\ee
and
\be
\chi_{\rm eq}(\tau) \equiv \lim_{\tw \to \infty} \chi_{\rm ZFC}(\tau+\tw,\tw)
=\chi_{\rm EA} -  c \frac{\rhoo m^{2}}{\Upsilon(L(\tau)/L_{0})^{\theta}}.
\label{eq-chi-eq}
\ee
In (\ref{eq-c-eq}) $\qea$ is the EA order parameter 
defined in \eq{eq-def-qea} and evaluated in \eq{eq-qea-eval}.
The numerical constant $c$ is given in \eq{eq-def-const-c}.
Correspondingly $\chiea$ is the equilibrium susceptibility related to
$\qea$ as $\kb T\chi_{\rm EA}=1-q_{\rm EA}$ as in \eq{eq-qea-chiea}.

\subsubsection{Quasi-equilibrium regime}
\label{subsubsec:theory-quasi-eq}

Now we discuss the quasi-equilibrium regime $L(\tau) < L(\tw)$. 
The autocorrelation function and ZFC linear susceptibility
can be evaluated using \eq{eq-q-chi-lr}, \eq{eq-c-q} and \eq{eq-chi-chi}.
For simplicity, here we assume to stay deep in the quasi-equilibrium 
regime $L(\tau) \ll L(\tw)$.  Then using the scaling property
of the effective stiffness \eq{eqn:effU-small},
one obtains \cite{KYT2},
\begin{eqnarray}
&& C(t=\tau+\tw,\tw)   = C_{\rm eq}(\tau)   \nonumber \\
&-&  c'\rhoo m^{2} \frac{\kb T}{\Upsilon(L(\tau)/L_0)^{\theta}} 
\left(\frac{L(\tau)}{L(t_{\rm w})}\right)^{d-\theta} 
 +\cdots , 
\label{eq-c-quasieq}
\end{eqnarray}
with $C_{\rm eq}(\tau)$ being the equilibrium part given in \eq{eq-c-eq} and 
$c'=c_{v}\int_{0}^{1}dy y^{2((d-1/2)-\theta)}$.
The 2nd term is the weak non-equilibrium correction 
term due to the weak softening 
of small droplets which gives rise to some weak waiting time dependences.

The above formula combined with the formula for the equilibrium part
\eq{eq-c-eq} yields a systematic extrapolation scheme 
to determine the EA order parameter $q_{\rm EA}$. Such an extrapolation
was already demonstrated in our previous paper \cite{HYT}.
In section \ref{subsec:c-quasieq}, we perform such an analysis numerically.

Similarly, the ZFC susceptibility
$\chi_{\rm ZFC}(\tau+\tw,\tw)$ is evaluated as,
\begin{eqnarray}
&& \kb T\chi_{\rm ZFC}(t=\tau+\tw,t_{\rm w}) = \kb T\chi_{\rm eq}(\tau) \nonumber \\
&&  
+c' \rhoo m^{2} \frac{\kb T}{\Upsilon(L(\tau)/L_0)^{\theta}} \left(\frac{L(\tau)}{L(t_{\rm w})}\right)^{d-\theta} 
 +\cdots. 
\label{eq-chi-quasieq}
\end{eqnarray}
with $\chi_{\rm eq}(\tau)$ being the equilibrium part given in
\eq{eq-chi-eq}  and  $\chi_{\rm EA}$ is the equilibrium susceptibility.

The quasi-equilibrium regime is most relevant for 
the measurements of the relaxation of AC susceptibilities 
during isothermal aging. Recently a related scaling 
analysis was performed in an experiment \cite{Uppsala01}.
Let us note that previous experimental analysis 
of spontaneous thermal fluctuation of the magnetization
and AC susceptibility have already confirmed that FDT holds 
for quasi-equilibrium regime while
non-stationarity is observed clearly \cite{SGNoise,Saclay-AC-Noise,VHO}.

\subsubsection{Crossover Regime}
\label{subsubsec:theory-crossover}

Next let us consider the crossover regime $L(\tau) \sim L(\tw)$.
Here we have to consider also the contribution of anomalously
soft droplets of sizes as large as that of the domain. 

From \eq{eq-q-cross} the spin autocorrelation function is obtained 
immediately as,
\begin{eqnarray}
&&  C(\tau+\tw,\tw)|_{L(\tau)/L(\tw) \sim 1} \nonumber \\
&& \sim  q_{\rm D} + \rhoo m^{2} A(1) \frac{\kb T}{\Upsilon (L(\tw)/\Lo)^{\rm \theta}} 
\label{eq-c-cross}
\end{eqnarray}
where $q_{\rm D}$ is the {\it dynamical order parameter} defined 
in  \eq{eq-def-qd}.
Similarly the ZFC susceptibility is obtained as,
\begin{eqnarray}
&& \kb T\chi_{\rm ZFC}(\tw+\tw,\tw) |_{L(\tau)/L(\tw) \sim 1} \nonumber \\
&& 
\sim \kb T\chi_{\rm D}
- \rhoo m^{2} A(1)\frac{\kb T}{\Upsilon (L(\tw)/\Lo)^{\rm \theta}} 
\label{eq-chi-crossover}
\end{eqnarray}
where $\chi_{\rm D}$ is the {\it dynamical susceptibility}
defined in \eq{eq-def-chid}.

As we discuss in section \ref{subsubsec:theory-classification},
the change from the quasi-equilibrium regime and the crossover
regime becomes very abrupt as function of $x=L(\tau)/L(\tw)$
in the asymptotic limit $L(\tw) \to \infty$
 (See Fig. \ref{fig:c-conjecture}). This novel feature
deserves to be studied in simulations and experiments. 
A convenient measure is the modified relaxation rate function
$S_{\rm mod}(x,\tw)$ we introduce in \eq{eq-def-s-mod}.

\subsubsection{Aging regime : autocorrelation function}
\label{subsubsec:theory-aging-c}

Let us now consider aging regime $L(t) \sim L(\tau)  > L(\tw)$
where we need to consider growth of the domains explicitly.
For simplicity, we used the notion of {\it epochs} mentioned previously
\cite{FH2}. The $n$-th epoch spans logarithmically separated 
time scales between $t_{n-1}$ and $t_{n}$ such that 
$L(t_{n-1})=a^{n-1}L_{0}$ and $L(t_{n})=a^{n}L_{0}$ with $a>1$.
At each epoch some of the smaller domains are eliminated so that
domain walls are coarsened. Thus a given site may or may not
belong to the same domain ($\Gamma$ or $\bar{\Gamma}$) at two different
epochs.  
The probability $P_{\rm s}(R_{1},R_{2})$ 
that a given site belongs to the same domain 
at the two different epochs characterized by the sizes of the domains
$R_{1}$ and $R_{2}$ presumably becomes a function of the ratio 
of the size of the domains, i.~e. 
\be
P_{\rm s}(R_{1},R_{2})=P_{\rm s}\left(\frac{R_{1}}{R_{2}}\right)
\ee
reflecting the self-similar nature of the domain growth 
process \cite{FH2,B94}. This function describes the slow decay of 
memory by the domain growth and plays the central role in the following.

For simplicity, let us neglect thermally active 
droplet excitations in the interior of the
domains and consider only the domain growth itself for the moment.
Then the auto-correlation function between two 
epochs such that size of domains are  $R_{1}=L(t)$ and $R_{2}=L(\tw)$ 
becomes
\be
C(t=\tau+\tw,\tw)=\tilde{C} \left(\frac{L(t)}{L(\tw)}\right)
\ee
with
\be
\tilde{C}(x)= 2P_{\rm s}(x)-1.
\label{eq-c-p}
\ee
In the limit $x \to 1^{+}$, the scaling function should converge as
\be
\lim_{x \to 1^{+}}\tilde{C}(x)=1,
\label{eq-scale-aging-zero}
\ee
because the normalization of the Ising spins.
In the other limit, the scaling function is 
expected to behave asymptotically as
\be
\tilde{C}(x)\sim x^{-\lambda}\qquad (x\gg 1)
\label{eq-scale-aging-lambda}
\ee
with $\lambda$ being a non-equilibrium exponent \cite{FH2,B94}
\be
d/2 < \lambda < d.
\label{eq-bound-lambda}
\ee

By taking into account thermally active droplet excitations,
we expect the scaling form of the spin autocorrelation function 
as
\be
C(t,\tw) \sim  q_{\rm D}\tilde{C}\left( \frac{L(t)}{L(\tw)} \right). 
\label{eq-scale-c-aging}
\ee
where $q_{\rm D}$ is the dynamical order parameter introduced 
in \eq{eq-def-qd}. Note that in usual coarsening process 
\cite{B94,spherical-SK,response-coarsening},
the amplitude is given by EA order parameter 
$\qea$ which was also assumed in 
the original droplet theory \cite{FH2}. However, we expect
the dynamical order parameter $\qd$ is more natural 
because of the anomalously soft droplets which are as large as domains.
The above scaling form \eq{eq-scale-c-aging} should be manifested
in the asymptotic limit $L(\tw) \to \infty$ with the 
ratio $x=L(t)/L(\tw)$ fixed to a certain value larger than $1$. 
Note that the normalization 
\eq{eq-scale-aging-zero} allows matching with
the crossover regime discussed in the previous section
in the asymptotic limit $L(\tw) \to \infty$.

At finite time scales, we expect some correction terms because of 
the following reasons. First let us recall that the asymptotic 
amplitude of the oder parameter is attained only by integrating 
out contributions of droplet excitations up to infinitely large ones. 
At finite length scales, the amplitude of the order parameter 
should be larger than $q_{\rm D}$ due to the algebraic 
correction term such as the second term in \eq{eq-q-cross} or
\eq{eq-c-cross}.  
Then the factor $q_{\rm D}$ in the scaling form 
\eq{eq-scale-c-aging} should be replaced by some time-dependent 
factor at finite time scales. Second, we also expect 
some additive correction terms since with some probability 
there will be some region where domain walls don't pass through 
so that the dynamics due to droplets in the interior of the domains
continues. Such quasi-equilibrium corrections will be additive as
considered in dynamical MFT \cite{MFT}.
However, we don't know how to collaborate both 
the multiplicative and additive correction 
terms simultaneously and we will leave the problem of correction terms 
for the autocorrelation function in the aging regime for future studies.

\subsubsection{Aging regime : linear susceptibilities}
\label{subsubsec:theory-aging-chi}

Finally let us consider the linear response to magnetic field during 
the domain growth (aging). Suppose that magnetic field $h$ is
applied only during a certain {\it epoch} where the
size of the domain is $L_{n}/L_{0}=a^{n}$ 
with $n$ being a certain positive integer. Then let us
consider the magnetization 
$\delta M_{n}(t)$ measured at some time $t>t_{n-1}$ during
and after the epoch, 
\be
\delta M_{n}(t)= h \int_{t_{n-1}}^{{\rm min}(t,t_{n})} dt' R(t,t') 
\ee
where $R(t,t')$ is the response function defined in \eq{eq-def-response}.  

During the epoch, droplets over the length scales from $L_{0}$ 
up to $L(\tau=t-t_{n-1})$ can be polarized by the field. Importantly 
the size of the domain can be considered as frozen in time
to the value $L_{n}$ during this epoch so that
the linear response is the same as that in the quasi-equilibrium/crossover
regimes.
Thus the magnetization measured {\it within the same epoch} will be 
\begin{eqnarray}
\delta M_{n}(t) &\sim &h \chi(L(\tau),L_{n}) \nonumber \\
&\sim & h \rhoo \int_{L_{0}}^{L(\tau)}\frac{dL}{L}
\frac{m^{2}}{F^{\rm typ}_{L,L_{n}}} \nonumber \\
&& \mbox{for} \qquad  t_{n-1} < t < t_{n}
\label{eq-def-mn-quasi-stationary}
\end{eqnarray}
where we used the generalized susceptibility given in \eq{eq-q-chi-lr}.
This part due to the droplets satisfies the FDT \eq{eq-fdt}.

When the field is cut off at time $t_{n}$, de-polarization 
of the droplets will start. Let us now follow the
argument of Fisher and Huse \cite{FH2},  
and  first consider a certain early stage of the $n+1$-th epoch, 
say up to time $t_{n}^{+}$ slightly after $t_{n}$, 
such that further domain growth still does not proceed appreciably.
Note that the switching-off of the field is equivalent to adding 
additional field of the opposite sign $-h$. The latter will
induce additional {\it negative} magnetization whose 
amplitude grow like \eq{eq-def-mn-quasi-stationary} 
with $\tau$ being understood as the time after 
the field change. Up to the time $t_{n}^{+}$ most of the
magnetizations will be canceled.
However some residual magnetization of order,
\be
\delta M_{n}(t_n^{+}) 
\sim c'' \rhoo  \frac{hm^{2}}{\Upsilon(L_{n}/L_{0})^{\theta}}
\ee
may be left behind. Here $c''$ is a certain numerical constant.
As domain growth proceeds further, the residual magnetization will
remain only if the domain to which the polarized droplet belongs
is not eliminated. Then the remanent magnetization will decay 
by the further domain growth as,
\begin{eqnarray}
\delta M_{n}(t) &\sim& \delta M_{n}(t^{+}_n) 
\left(2P_{s}\left(\frac{L(t)}{L_{n}}\right)-1 \right)\nonumber \\
&\sim&  c'' \rhoo \frac{hm^{2}}{\Upsilon(L_{n}/L_{0})^{\theta}}
\tilde{C}\left(\frac{L(t)}{L_{n}}\right) \nonumber \\
\mbox{for} &&  t \gg  t_{n}
\label{eq-def-mn-non-stationary}
\end{eqnarray}
Here $\tilde{C}(L(t)/L)$ is related to the probability $P_{s}(L(t)/L)$ 
that a given site belongs to the same 
domain at the two different epochs $L_{n}$ and $L(t)(> L_n)$
as given in \eq{eq-c-p}. This is the aging part of the response
which violates the FDT \eq{eq-fdt} strongly.

Combining the above results we can now evaluate the ZFC and TRM
susceptibilities defined in \eq{eq:zfc-linear} and \eq{eq:trm-linear}
by summing over the responses at different epochs 
given by \eq{eq-def-mn-quasi-stationary} and \eq{eq-def-mn-non-stationary}.
Suppose that the observation time $t$ belongs to the $n$-th epoch and
the waiting time $\tw$ belongs to the $m$-th epoch such that $m < n$. 
Then the ZFC susceptibility is obtained as,
\begin{eqnarray}
\chi_{\rm ZFC}(t,\tw) &=& \int_{\tw}^{t}dt'R(t,t') 
= \frac{1}{h} \sum_{k=m}^{n} \delta M_{k}(t) \nonumber \\
&& \sim  \chi(L(\tau),L(t)) \nonumber \\
&& +c'' \int_{L(\tw)}^{L(t)}\frac{dL}{L}
\frac{\rhoo m^{2}}{\Upsilon(L/L_{0})^{\theta}}
\tilde{C}\left(\frac{L(t)}{L}\right).
\end{eqnarray}
The first term in the last equation is the response of droplets
\eq{eq-def-mn-quasi-stationary} during the last epoch
where the observation is being done
and second term is the aging part \eq{eq-def-mn-non-stationary} 
due to the remanence of the response made at previous epochs. 
Note that the expression is valid also
for the quasi-equilibrium and crossover regimes because there
$L(t)\sim L(\tw)$ holds and the second term is absent so that
the expression becomes the same as \eq{eq-chi-chi}.

In the special case of $\tw=0$, the above result becomes
\begin{eqnarray}
 \chi_{\rm ZFC}(t,0)
&\sim  &  \chi(L(t),L(t)) \nonumber \\
& +& c'' \int_{L_{0}}^{L(t)}\frac{dL}{L}
\frac{\rhoo m^{2}}{\Upsilon(L/L_{0})^{\theta}}
\tilde{C}\left(\frac{L(t)}{L}\right) \nonumber \\
& \sim & \chi_{\rm D} - c'''
\frac{\rhoo m^{2}}{\Upsilon(L(t)/L_{0})^{\theta}}
\label{eq-scaling-chizfc0}
\end{eqnarray}
where 
\be
c'''=A(1)-c_{\rm nst}  
\label{eq-c'''}
\ee
with
\be
c_{\rm nst}=c\int_{0}^{1} dy y^{-1-\theta+\lambda} >0.
\label{eq-cnst}
\ee
To obtain the last equation, we used the dynamical susceptibility
$\chi_{\rm D}$ defined in \eq{eq-def-chid} and the scaling form 
$\tilde{C}(x) \sim x^{-\lambda}$ given in \eq{eq-scale-aging-lambda}
and assumed that $t$ is large enough so that $L(t)/L_{0} \ll 1$  holds.
Note that $c_{\rm nst}$  represents the strength of the 
aging part of the response whose contribution to 
$\chi_{\rm ZFC}(t,0)$ scales as $c_{\rm nst}m^{2}/(L(t)/L_{0})^{\theta}$.
Now comparing the result \eq{eq-scaling-chizfc0}
with \eq{eq-def-chifc} which reads as
\bmat
\chi_{\rm FC} \equiv \lim_{t \to \infty}\chi_{\rm ZFC}(t,0)
\emat
we find the dynamical 
susceptibility $\chid$ defined in \eq{eq-def-chid} is nothing 
but the desired FC susceptibility $\chifc$. Thus using 
the inequality \eq{eq-chid-chiea} we obtain,
\be
\chi_{\rm FC}=\chid > \chiea
\label{eq-chifc-chid}
\ee
which is nothing but the anticipated inequality \eq{eq-fc-ea}.

Lastly let us evaluate the TRM susceptibility in the aging regime.
Suppose the waiting time $\tw$ belongs to the $m$-th epoch
and observation is done {\it well after the waiting time}
such that the domain growth proceeds appreciably $L(t) > L(\tw)$.
Then  summing over the aging part of the response
\eq{eq-def-mn-non-stationary} we obtain,
\begin{eqnarray}
&& \chi_{\rm TRM}(t=\tau+\tw,\tw) =\int_{0}^{\tw}dt'R(t,t')
\sim
\sum_{k=1}^{m} \delta M_{k} (t) \nonumber \\
&& \sim  c'' \int_{L_{0}}^{L(\tw)}\frac{dL}{L}
\frac{\rhoo m^{2}}{\Upsilon(L/L_{0})^{\theta}}
\tilde{C}\left(\frac{L(t)}{L}\right) \nonumber \\
&& \sim  c_{\rm nst} \frac{m^{2}}{\Upsilon(L(\tw)/L_{0})^{\theta}}
\left(\frac{L(t)}{L(\tw)}\right)^{-\lambda}\nonumber \\
\mbox{for} &&  L(t)\sim  L(\tau) > L(\tw).
\label{eq-chitrm-aging}
\end{eqnarray}
Here $c_{\rm nst}$ is defined above in \eq{eq-cnst} which represents
the strength of the integral of the aging part of the response.
To obtain the last equation, we used the scaling form 
$\tilde{C}(x) \sim x^{-\lambda}$ given in \eq{eq-scale-aging-lambda}
and assumed that $\tw$ is large enough so that $L_{0}/L(\tw) \ll 1$  holds.
The functional form agrees with what was anticipated by Fisher
and Huse \cite{FH2}.

Thanks to the sum rule \eq{eq:sum-rule}, the relaxation of TRM
susceptibility \eq{eq:trm-linear} and ZFC susceptibility \eq{eq:zfc-linear}
can be obtained from each other using $\chi_{\rm ZFC}(t,0)$ 
obtained as \eq{eq-scaling-chizfc0}. Here one must pay attention
to the fact that $\chi_{\rm ZFC}(t,0)$ contains {\it both} 
the response due to droplets (which satisfy the FDT)
and aging part (which violates the FDT)
as can be seen in \eq{eq-scaling-chizfc0}. 

In the previous subsections \ref{subsubsec:theory-quasi-eq}
and \ref{subsubsec:theory-crossover}, we obtained scaling properties
of the ZFC susceptibility in the quasi-equilibrium/crossover regime. The
ZFC susceptibility in these 
regimes contain only response of droplets which satisfy the FDT. 
However one can check that the TRM susceptibility in these regimes, 
which can be readily obtained via the sum rule, becomes a 
mixture of response due to droplets and aging part. 
Thus the ZFC susceptibility in the quasi-equilibrium/crossover regimes 
is better suited to examine responses of droplets than 
the TRM susceptibility.

Conversely the TRM susceptibility in the aging regime 
is better suited to examine aging part of the response 
than the ZFC susceptibility. As can be seen in 
\eq{eq-chitrm-aging}, the TRM susceptibility in the aging regime
contains only the aging part of the response. On the other hand 
one can check that the ZFC susceptibility
which can be readily obtained by the sum rule
becomes a mixture of response due to droplets and aging part.

\subsubsection{Summary}
\label{subsubsec:theory-classification}

\begin{figure}[h]
\resizebox{\figwidth}{!}{\includegraphics{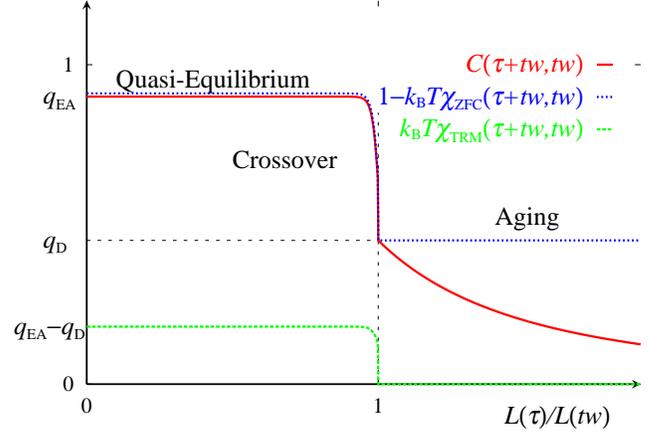}}
 \caption{Different asymptotic regimes of the two-time quantities.
See text for details.}
\label{fig:c-conjecture}
\end{figure}

\begin{figure}[h]
\resizebox{\figwidth}{!}{\includegraphics{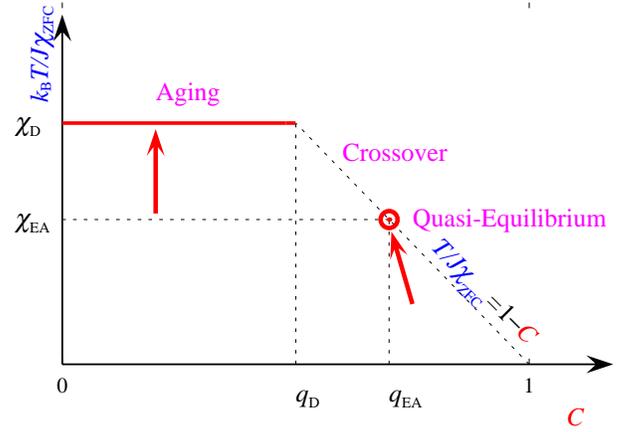}}
 \caption{Different asymptotic regimes in the $C-T\chi_{\rm ZFC}$ 
plane. The dotted line labeled $\kb T\chi=1-C$ represents the
FDT line. See text for details.}
\label{fig:ck-conjecture}
\end{figure}

Let us consider large time limit
such that $L(\tw) \to \infty$ is taken with the ratio 
\be
x=\frac{L(\tau)}{L(\tw)}
\ee
being fixed to certain values.  Here it is useful to consider 
the large time limit of the sum rule \eq{eq:sum-rule}. For 
{\it any  $\tau$ that may be allowed to grow with $\tw$} we find,
\be
 \lim_{\tw \to \infty} \left[\chi_{\rm TRM}(\tau+\tw,\tw)+ \chi_{\rm ZFC}\chi(\tau+\tw,\tw) \right]
 =  \chi_{\rm D} 
\label{eq-sum-rule-largetime}
\ee
In the last equation we used the definition of the field cooled (FC)
susceptibility defined in \eq{eq-def-chifc} and our result \eq{eq-chifc-chid}.
The asymptotic behaviors discussed below are
displayed in Fig. \ref{fig:c-conjecture} and Fig. \ref{fig:ck-conjecture}.

First $x<1$ corresponds to the {\it quasi-equilibrium regime}
where the spin autocorrelation function
slowly decays from $1$ to static order parameter $\qea$
accompanying some {\it weak} waiting time dependence or
weak violation of time translational invariance (TTI). 
Here if one fix $L(\tau)$ and let $L(\tw) \to \infty$ one 
obtains the ideal equilibrium limit behavior where the weak
violation of TTI is removed. On the other hand, the FDT \eq{eq-fdt} 
$1-C(\tau+\tw,\tw)=\kb T\chi_{\rm ZFC}(\tau+\tw,\tw)$
is satisfied even in the presence of the weak violation of TTI. 
In the large time limit $L(\tw) \to \infty$ with fixed $x<1$,
the time dependences (including the weak violation of TTI)
disappears such that the spin autocorrelation converges to
the static EA order parameter $q_{\rm EA}$ 
and the ZFC susceptibility converges to the equilibrium susceptibility
$\kb T\chi_{\rm EA}$.
Correspondingly the TRM susceptibility converges to 
$\kb T\chi_{\rm D}-\kb T\chi_{\rm EA}=q_{\rm D}-q_{\rm EA}$ 
because of the sum rule \eq{eq-sum-rule-largetime}. 
In the last equation we used $\kb T\chi_{\rm D}=1-q_{\rm D}$
given in \eq{eq-def-chid}.
To summarize, the spin autocorrelation function
and the susceptibilities asymptotically become flat lines in the 
quasi-equilibrium regime ($x<1$) as displayed in Fig. \ref{fig:c-conjecture}.
In the parametric plot of Fig. \ref{fig:ck-conjecture}, the whole
quasi-equilibrium regime converges to a single point 
$(q_{\rm EA}, \kb T\chi_{\rm EA})$. 
Importantly, the  scaling theory has provided
not only such asymptotic limits but details of finite-time 
$(\tau,\tw)$ correction terms by which the asymptotic limit 
is approached as discussed in section 
\ref{subsubsec:theory-quasi-eq}. The latter are well amenable 
to be examined seriously by experiments and simulations.
In section \ref{sec:analysis-two-time} we present such a detailed 
analysis for the 4D EA model based on  MC simulations.

Second $x \sim 1$ corresponds to the {\it crossover regime}.
From \eq{eq-c-cross}, we expect
the spin autocorrelation function drops vertically 
(against $L(\tau)/L(\tw)$) 
from the EA order parameter $q_{\rm EA}$
down to the dynamical order parameter  $q_{\rm D}$ 
which is smaller than $q_{\rm EA}$
due to the anomalously soft droplets.
In this regime we expect the FDT \eq{eq-fdt} is still satisfied
in spite of the strong non-stationary character 
and we expect the ZFC susceptibility 
increases vertically from the
static susceptibility $\kb T\chi_{\rm EA}$ to a larger value
$\kb T\chi_{\rm D}=1-q_{\rm D}$.
Correspondingly the TRM susceptibility drops off vertically 
from $\kb T\chi_{\rm D}-\kb T\chi_{\rm EA}=q_{\rm EA}-q_{\rm D}$ 
to zero because of the sum rule \eq{eq-sum-rule-largetime}.
In the parametric plot of Fig. \ref{fig:ck-conjecture}, the crossover
regime converges to the line of points in the section of the FDT line between
$(q_{\rm EA}, \kb T\chi_{\rm EA})$ and $(q_{\rm D}, \kb T\chi_{\rm D})$.

The abruptness of the changes of the two time quantities at $x \sim 1$
is very surprising. Here let us recall the well known experimental
\cite{Uppsala-S} and numerical \cite{MC-S} observations
that the so called relaxation rate 
\be
S(\tau,\tw) \equiv  \frac{d  \chi_{\rm ZFC}(\tau+\tw,\tw)}{d \ln (\tau)}
\label{eq-def-s}
\ee
has a broad peak centered at around $\tau \sim \tw$.
Our scenario naturally suggests a {\it modified}
relaxation rate function,
\be
S_{\rm mod}(x, \tw) \equiv  \left. \frac{d  \chi_{\rm ZFC}(\tau+\tw,\tw)}{d x}\right|
_{x=L(\tau)/L(\tw)}.
\label{eq-def-s-mod}
\ee
which should have a sharper peak at $x\sim 1$ with increasing $L(\tw)$.
This modified relaxation rate will be useful for further numerical 
simulations and experiments.

In order to describe the interior of the crossover regime
more closely, 
different scaling variables other than $L(\tau)$ and $L(\tw)$
are certainly needed. Unfortunately, the droplet theory which is based
on the dynamical length scales cannot provide information for
proper scaling variable to describe the interior of the crossover regime. 
A possible scaling variable would be 
$\alpha=\tau/\tw$ as proposed by Fisher and Huse \cite{FH2} 
with which the abruptness will be absent.
Note that possible large time 
limits $\tw \to \infty$ classified by different values of $\alpha$ 
are {\it smashed to a  point} $x=L(\tau)/L(\tw)=1$  in the x-axis
for all $\alpha$. 

Finally $x > 1$ corresponds to the {\it aging regime}
where the spin autocorrelation function takes a continuous
value between $q_{\rm D}$ and $0$ as deceasing function of $x$
as given in \eq{eq-scale-c-aging}.
Here the result \eq{eq-chitrm-aging} implies the TRM susceptibility is
asymptotically zero. 
The latter means the ZFC susceptibility
converges to the FC susceptibility $\kb T\chi_{\rm FC}$ because of
the sum rule \eq{eq-sum-rule-largetime}. The FDT \eq{eq-fdt} is thus
strongly violated.
In the parametric plot of Fig. \ref{fig:ck-conjecture}, asymptotic
limit for each $x$ converges to a point on the flat horizontal line 
connecting $(0,T\chi_{\rm D})$ and  $(q_{\rm D}, \kb T\chi_{\rm D})$.

In conventional understanding \cite{BCKM} which includes
usual coarsening systems 
\cite{B94,spherical-SK,response-coarsening} and mean-field 
spin-glass models \cite{MFT}, both the violation 
of TTI and FDT happens
asymptotically at the same static oder parameter $q_{\rm EA}$.
A remarkable feature of our present scenario is that the break points
of TTI and FDT are separated: the break point of TTI 
is located at $q_{\rm EA}$ while that of FDT is located at
the dynamic oder parameter  $q_{\rm D}$.

So far it is implicitly assumed that FDT \eq{eq-fdt} is 
valid in the quasi-equilibrium regime 
and also in the crossover regime in spite of 
the fact that there are weak and strong waiting time effect.
A supplementary argument for the validity of FDT 
can be made by considering the bound on the
possible violation of FDT found in \cite{FDT-bound}.
The rigorous bound \cite{FDT-bound} on the
integral violation $I(t,\tw)$ defined in \eq{eq-integral-violation-fdt}
is put in terms of the entropy production rate 
which presumably has the same scaling form
as the energy relaxation rate. Then using the scaling form 
of the energy relaxation \eq{eq:ene-decay} one finds,
\begin{eqnarray}
|I(t,\tw)| & \leq & K \int_{\tw}^{t}ds \sqrt{\frac{d e(s)}{ds}}\nonumber \\
&\leq & K' \left((t/t_{0})^{1/2+\alpha}-(\tw/t_{0})^{1/2+\alpha}\right)
\end{eqnarray}
with $K$ and $K'$ being certain finite constants and $ 0< \alpha \ll 1$ 
being an arbitrary small positive non-zero number.
Now let us consider the bound in the large time limit $\tw \to \infty$
in the quasi-equilibrium regime $x=L(\tau)/L(\tw) <  1$. 
For convenience let us introduce $y$ such that,
\begin{equation}
\tau=t-\tw=t_{0}(\tw/t_{0})^{y}
\end{equation}
Then we find
\begin{eqnarray}
 \lim_{\tw \to \infty} I(t,\tw) && 
\leq   (1/2+\alpha) K' \lim_{\tw \to \infty} (\tw/t_{0})^{-1/2+y+\alpha}
\nonumber \\
&& =0
\end{eqnarray}
The last equations holds for $0 \leq y < 1/2$ since $\alpha$ is
an arbitrary small positive number. This observation implies
that FDT is satisfied not only in the equilibrium limit but also 
in the quasi-equilibrium regime. In order to verify the validity 
of FDT in the crossover regime $y \sim 1$, apparently improved bounds
are needed.

\subsubsection{Comparison with Conventional Pictures}
\label{subsubsec:comarison}

Finally, let us compare our scenario with the conventional picture
for isothermal aging \cite{BCKM} which applies for
the dynamical mean-field theory (MFT) of spin-glasses  \cite{MFT}
and usual coarsening systems \cite{B94,spherical-SK,response-coarsening}.
Here let us consider the asymptotic limit  $\tw \to \infty$
of the two time quantities  by fixing the value of the autocorrelation 
function to a certain value $C$ in the range $0 < C < 1$. 
In that limit, the ZFC susceptibility becomes a function of 
$C$, $\chi=\chi(C)$. This definition of asymptotic limit can 
be considered in general and our scenario implies 
the simple structure of $\chi(C)$ shown in Fig. \ref{fig:ck-conjecture}.
In the usual coarsening systems and in the dynamical MFT,
the FDT line terminates at $(\qea, T\chiea)$ while it extends up to
$(\qd, T\chid)$ in our scenario. In the usual coarsening systems,
a horizontal line connects $(0,\kb T\chiea)$ and $(\qea,\kb T\chiea)$
where the FDT is violated strongly. On the other hand,
the dynamical MFT predicts a {\it  curved line} between 
$(0,\chid)$ and $(\qea,\chiea)$. Our picture is different from both of them.

The difference between usual coarsening systems and our scenario
is the presence of the anomalously soft droplets. Droplet excitations
exist also in simple coarsening systems like ferromagnets \cite{HF87}
but with extremely small probability. In our scenario, the anomalously 
soft droplet can exist with probability of order $O(1)$ in a given 
domain irrespective of the size of the domain 
so that the thermal fluctuations in 
each domain are anomalously large compared with equilibrium where
there are no extended defects. 

For clarity, let us note \cite{thanks-Ludovic}  
that there are also some excessive response in usual coarsening systems 
due to thermalized domain walls \cite{response-coarsening}
at wavenumbers $k$ such that $kL(t) \gg 1$. They are similar
to the anomalously soft droplets in our scenario in the sense that
1) they satisfy FDT but 2) disappear in the ideal equilibrium. However
their integral contribution to the response decreases with the 
growth of the domain $L(t)$ so that their contribution 
vanishes asymptotically. On the other hand the excessive response 
of the anomalously soft droplets in our scenario give $O(1)$ 
contribution irrespective of the size of the domain so that
their contribution to the response do not disappear 
as far as $\tw \to \infty$ limit (ideal equilrbium) is not took first.

Although both the dynamical MFT and our scenario concludes
the inequality \eq{eq-fc-ea} $\chifc > \chiea$, the origins
are different. The difference between the two, called  {\it anomaly}, 
is attributed to contributions of aging part of the response function
which violates the FDT in the case of the dynamical MFT 
while it is rather attributed to the responses due to the anomalously 
soft droplets which still satisfy the FDT in our scenario. 
For example, one can see in \eq{eq-scaling-chizfc0} 
that $\chifc(=\chi_{\rm D})$ is generated not from the 
aging part but from the part 
due to the anomalously soft droplets which satisfies the FDT.

There is a conjecture \cite{FMGPP} that there is a connection between
the statics and dynamics such that the function $\chi=\chi(C)$
in the dynamics should be related to the overlap distribution 
function $P(q)$ in {\it equilibrium},
as 
\be
\kb T\chi(C)=\int_{C}^{1}dC' \int_{0}^{C'} dC''P(C'').
\label{eq-c-chi-pofq}
\ee
This is known to hold exactly in some (not all) mean-field models \cite{MFT}
and usual coarsening systems \cite{spherical-SK,response-coarsening}.
Moreover, an interesting conjecture \cite{BB01} was proposed 
recently that at finite time (length) scales, the  above formula holds
for the two time quantities at a finite waiting time $\tw$
and a $P(q)$ measured in equilibrium of a finite system of size 
$L(\tw)$. We agree with this proposal partially but not completely 
as we discuss in the following.

In order to fully reproduce the parametric plot Fig. \ref{fig:ck-conjecture},
we would need $P(q)$ with a delta peak at $q=\qd$ rather than
at  $q=\qea$. On the other hand, we expect 
a $P(q)$ with delta peak at $q=\qea$ in the absence of any extended 
defects as the original droplet theory has predicted \cite{FH1}. 
Thus the correspondence between the dynamics and statics does not hold
at all in this sense in our scenario.

However, one can explicitly consider a rather special static situation 
in the presence of extended defects of size, say $R$, 
which is actually what we considered in section \ref{subsec:domainwall-softdroplets}.
In any finite size systems, it is likely that the existence of 
boundaries (periodic/free e.t.c.) will intrinsically induce 
certain defects as compared with infinite systems \cite{NS98,M99,KYT}. 
Thus we expect actually the circumstance we are considering is relevant
in practice. 
Our scenario implies the {\it average} 
overlap $\bar{q}=\int_{0}^{1}dq q P(q)$ measured in such an equilibrium 
is equivalent to
$q(R,R) = \qd+  \rhoo m^{2} \kb T/\Upsilon(R/\Lo)^{\theta} A(1)$ 
of \eq{eq-q-cross}.
Thus our scenario suggests 
$\chifc=\lim_{\tau \to \infty}\chi_{\rm ZFC}(\tau+\tw,\tw)
=1-\bar{q}=1-\int_{0}^{1}dq q P(q)$  for any $\tw$ while 
$\lim_{\tau \to \infty} C(\tau+\tw,\tw) \to 0$ for any $\tw$.
Thus our scenario agrees with the conjecture of Ref.~\onlinecite{BB01} at the special point $C=0$
plus the usual FDT part between $C=\qea$ and $C=1$ but not in 
the section $0 < C < \qea$.
Here $P(q)$ should be understood as measured in equilibrium but with 
the extended defects.  It will be natural to expect that
$P(q)$ in such a situation becomes a non-trivial, non-self-averaging 
function of $q$ with finite amplitude at $q=0$ as observed in many
numerical simulations of finite size systems \cite{low-temp-MC,Rome-review}.

In the above arguments, we considered asymptotic limits with fixed $C$
while it is more natural to consider asymptotic limits
with fixed  ratio of the two length $x=L(\tau)/L(\tw)$  in the droplet theory.
In the dynamical MFT, one needs infinitely many kinds of time 
reparametrization functions $h(t)$ to span the whole correlation 
range $0 < C < 1$ but the droplet theory has only one natural 
variable $L(t)$. However it should be remarked that the scaling 
variable to describe the interior of the crossover regime 
$\qd < C < \qea$ is not known at present within the droplet theory.

%Growth law of the correlation length in off-equilibrium
%\input{sec04-v6}

\section{Growth law of the correlation length in off-equilibrium}
\label{sec:growthlaw}

As discussed in Sec.~\ref{sec:theory}, the dynamical length scale $L(t)$
and its growth law with time are important to understand nature of aging
in SG systems from the view point of the scaling theory. 
In the present section, we discuss our numerical results of the length
scale and its growth law during isothermal aging. 
A plausible definition of the length scale $L(t)$ is given by 
a decay constant of a correlation function $G(r,t)$. 
We measure the equal-time replica correlation function in
off-equilibrium under zero magnetic field, defined by 
\begin{equation}
 G(r,t) = \sum_i \langle S_i^{(\alpha)} (t)S_i^{(\beta)} (t) 
S_{i+r}^{(\alpha)} (t) S_{i+r}^{(\beta)} (t) \rangle, 
\label{eqn:gofr}
\end{equation}
where $\alpha$ and $\beta$ denote the replica indices which are updated
independently from different initial random spin configurations. 
In our simulation, only one MC sequence is performed for each random
bond.
Typical number of samples averaged over bond realizations is about
$128$. 
As reported previously\cite{Kisker,KYT,BB02}, the correlation function
$G(r,t)$ exhibits a complicated functional form which is not a simple
exponential and depends on time $t$. This may be because a
characteristic distance above which $G(r,t)$ approximately follows an
exponential form depends on time considerably.  In order to avoid the
artificial effect, we do not employ the so-called ratio
method\cite{Kisker,KYT} where the full data of $G(r,t)$ are used
independent of $t$. Instead, we estimate $L(t)$ by fitting directly the
tail part of  $G(r,t)$ to an exponential formula for each time $t$. 
In the fitting procedure, we focus our attention only on the large
distance tail of $G(r,t)$ and carefully choose the range depending on
time $t$. 
%%%%%%%%%%%%%%%
\begin{figure}[]
 \resizebox{\figwidth}{!}{\includegraphics{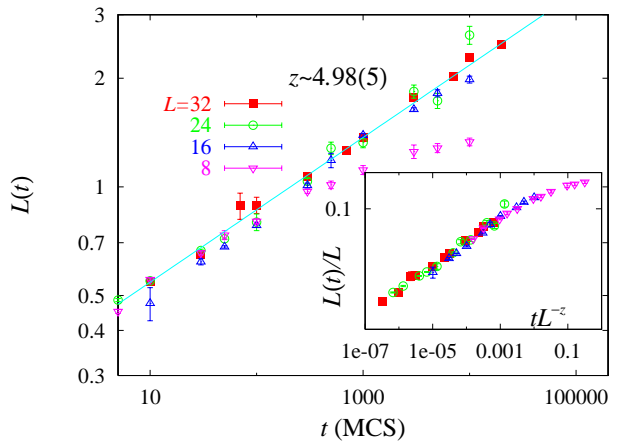}}
 \caption{$R(t)$ of the $4d$ Ising SG model at $T/J=2.0$ with different
 system sizes. The dashed line represents a power law with exponent
 $1/z$. In the inset, an expected finite-size scaling plot is shown. 
 }
\label{fig:xi-tc}
\end{figure}

In Fig.~\ref{fig:xi-tc} we show our results of the length scale
$L(t)$ at the critical temperature $T_{\rm c}=2.0J$.  
The data of the length scale $L(t)$ follows expected critical power law
(\ref{eqn:LTcritical}) except for the data with $L=8$, which is due to
finite-size effect.  
The estimate of the exponent $z$, $4.98(5)$, is roughly consistent  
with that of the previous work\cite{BernardiCampbell}, which quoted
$z=4.45(1)$.  
The data,  including the size $L=8$ deviated from the power law,  scale well
using a standard finite-size scaling form, shown in the inset of
Fig.~\ref{fig:xi-tc}.  
It is seen that the length 
scale $L(t)$ manages to reach at most few lattice spacing even at about
$10^5$ MCS. 
Nonetheless it should be noted that $L(t)$ already 
captures macroscopic behavior
in the sense that the critical exponent $z$ is successfully estimated
from $L(t)$ and that a strong finite-size effect is observed already in  
$L=8$ data.

\begin{figure}[]
\resizebox{\figwidth}{!}{\includegraphics{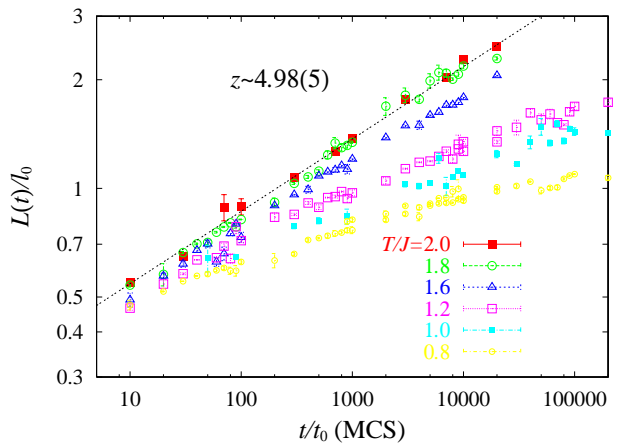}}
 \caption{Time evolution of $R(t)$ of  the $4d$ Ising SG model at
 $T_{\rm c}$ and below  $T_{\rm c}$ with $L=24$.}
\label{fig:xi-raw}
\end{figure}

\begin{figure}[h]
 \resizebox{\figwidth}{!}{\includegraphics{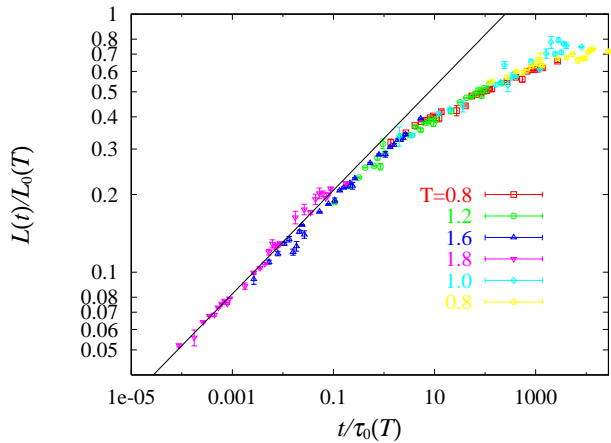}}
 \caption{Scaling plot of $R(t)$ of  the $4d$ Ising SG model, where the SG 
 transition temperature $T_{\rm c}=2.0$ and the critical exponent $\nu=0.93$ 
 are fixed,  but the dynamical exponent $z$ is obtained to be $4.98(5)$ 
 from the best scaling.   }
\label{fig:xi-scal}
\end{figure}

Let us now examine the crossover from critical to activated dynamics
discussed in section \ref{subsec:crossover-critical-activated}.
In Fig.~\ref{fig:xi-raw}, we display the raw data of the length scale
$L(t)$ below $T_{\rm c}$ and at $T_{\rm c}$. 
The critical dynamics is expected to be dominant near
$T_{\rm c}$.  In fact, the length scale $L(t)$ at $T/J=1.8$ is not
distinguishable from that at $T=T_{\rm c}$ within our time regime.  
Let us examine the crossover scaling defined in (\ref{eqn:LTcross}) and
(\ref{eqn:scal-g}). 
We present the scaling plot in Fig.~{\ref{fig:xi-scal}} by using the
same data  as shown in Fig.~{\ref{fig:xi-raw}}.  In this plot, for two
scaling parameters, $T_{\rm c}$ and $\nu$, we use the known values 
obtained previously\cite{KH}, while only one remaining parameter $\psi$
is appropriately chosen for the data with different temperatures to
merge into a universal curve. 
The best scaling plot is obtained by $\psi\sim 2.5-3.0$. 
The proposed scaling pretty well  works in the observed time regime. It is
clearly found that the scaling function exhibits a crossover from the
critical power law to slower growth law associated with low temperature
dynamics  which is compatible with the logarithmic growth law predicted
by the droplet theory. 

%One time quantities
%\input{sec05-v6}

\section{Relaxation of Energy and Density of domain wall}
\label{sec:analysis-one-time}

In the present section, we examine scaling properties of 
one-time quantities under isothermal
aging after quench based on the view point presented in section
\ref{subsec:one-time}.
In the following sections, we perform simulations mainly at  two low
temperatures $T/J=1.2$ and $0.8$ which amount to $0.6T_{c}$  and 
$0.4T_{c}$,  respectively.  
It is found that the effects of critical fluctuations do not 
dominate our time window at these temperatures
but can be taken into account in a renormalized way.

\begin{figure}[]
 \resizebox{\figwidth}{!}{\includegraphics{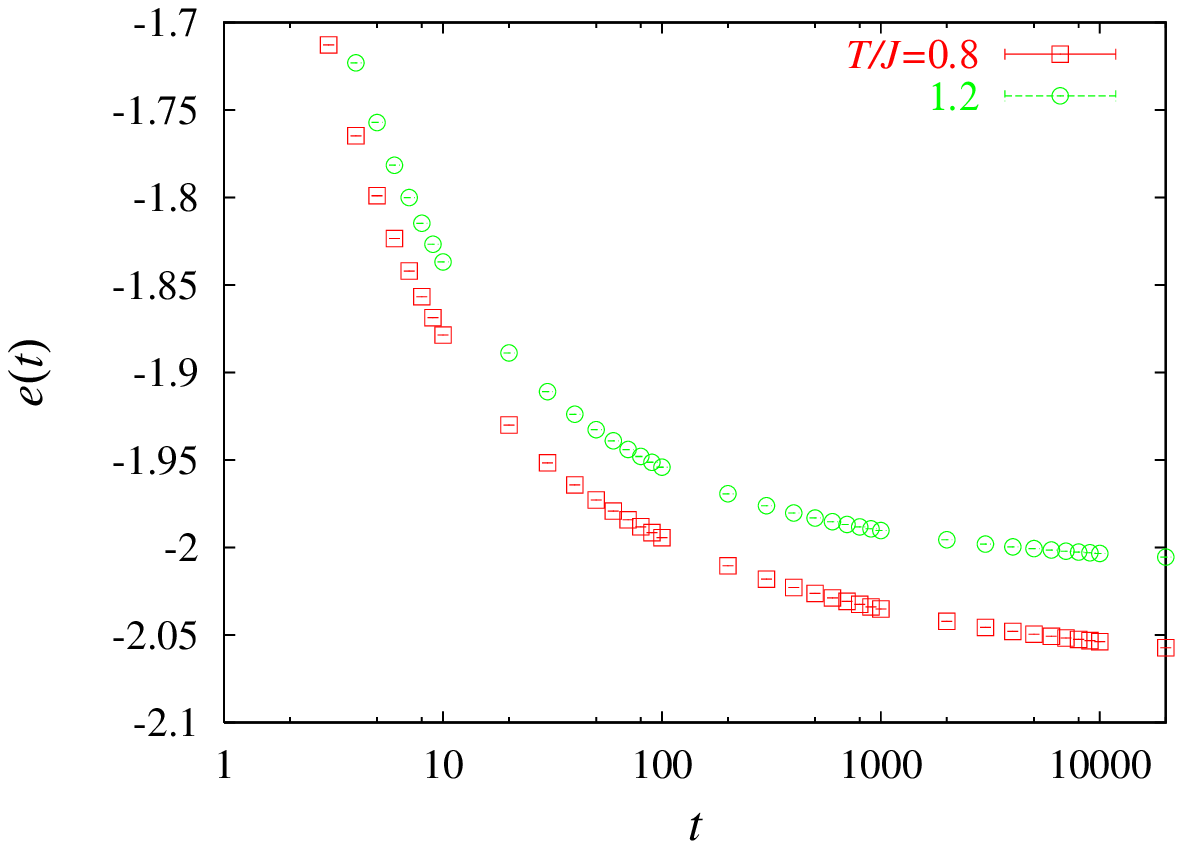}}
 \resizebox{\figwidth}{!}{\includegraphics{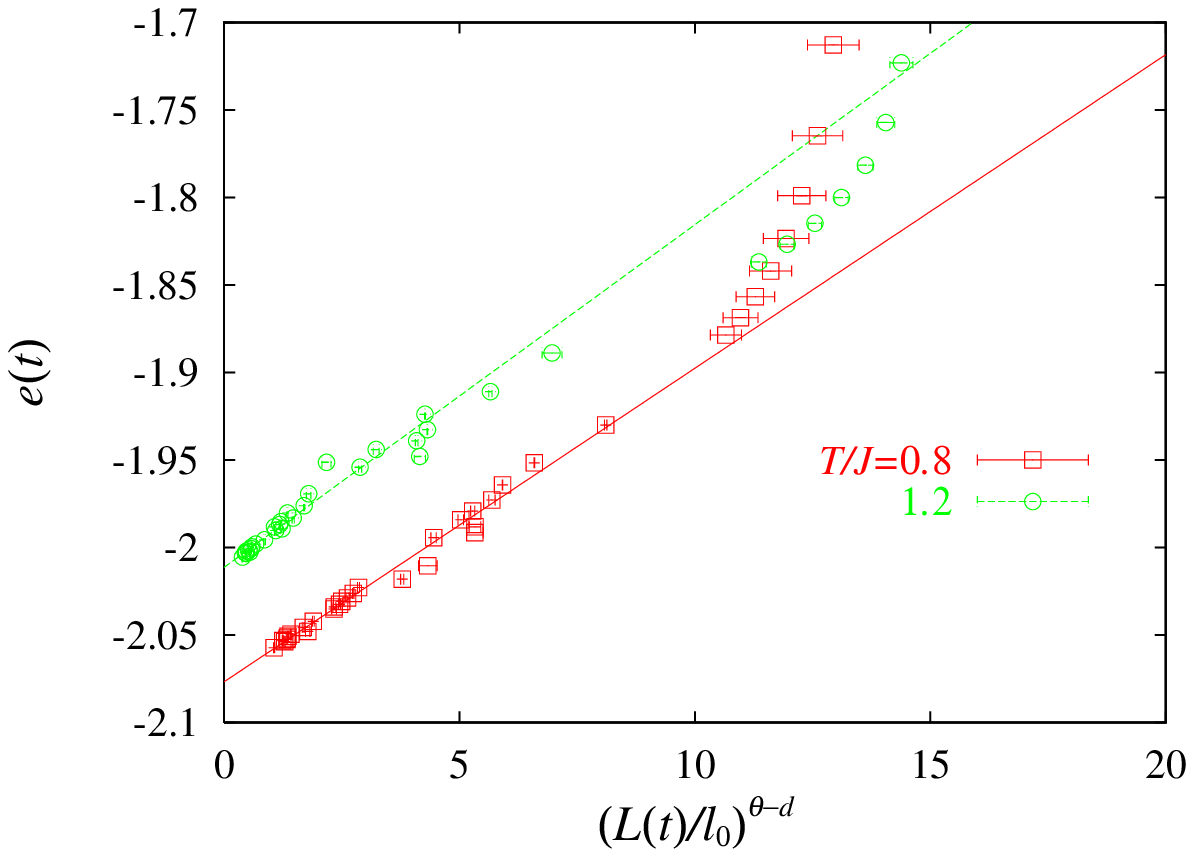}}
 \caption{
Energy relaxation at $T/J=0.8$ and $1.2$ as a function of elapse time
 (upper figure) and the length scale $L(t)$ (lower figure). 
 The straight lines in the lower figure represent fitting result to a
 linear function function of $(L(t)/l_0)^{\theta-d}$. 
}
\label{fig:ene}
\end{figure}

\subsection{Energy}
In Fig.~\ref{fig:ene}, we present the data of the energy per spin $e(t)$
defined in (\ref{eq:energy}) at $T/J=0.8$ and $1.2$. 
As seen in the upper figure, the energy function relaxes with time to an
equilibrium value at each temperature. However, it is rather hard to
extract the equilibrium value from the figure because of the extremely
slow dynamics. On the other hand, the scaling formula (\ref{eq:ene-decay})
says that the energy approaches its equilibrium value linearly as a
function of $L^{\theta-d}(t)$. We thus plot the same data set against
$L^{\theta-d}(t)$ in the lower one by using the
length scale $L(t)$ estimated independently in the previous section and
the known value of the stiffness exponent $\theta=0.82$\cite{KH}. 
We see the linear behavior as a function of $L^{\theta-d}(t)$ for large
$L(t)$ which supports the validity of the scaling formula
(\ref{eq:ene-decay}). One can take the long time limit of the energy
relaxation using the scaling formula. 
A similar scaling analysis of the energy has been confirmed in the $3d$
Ising EA SG model including finite size effects\cite{KYT}. 
This two stroke strategy has an advantage over the direct analysis with
time and gives us a more powerful tool in the analysis of two-time
quantities discussed in the following sections.

\subsection{Density of domain wall}
\begin{figure}[]
 \resizebox{\figwidth}{!}{\includegraphics{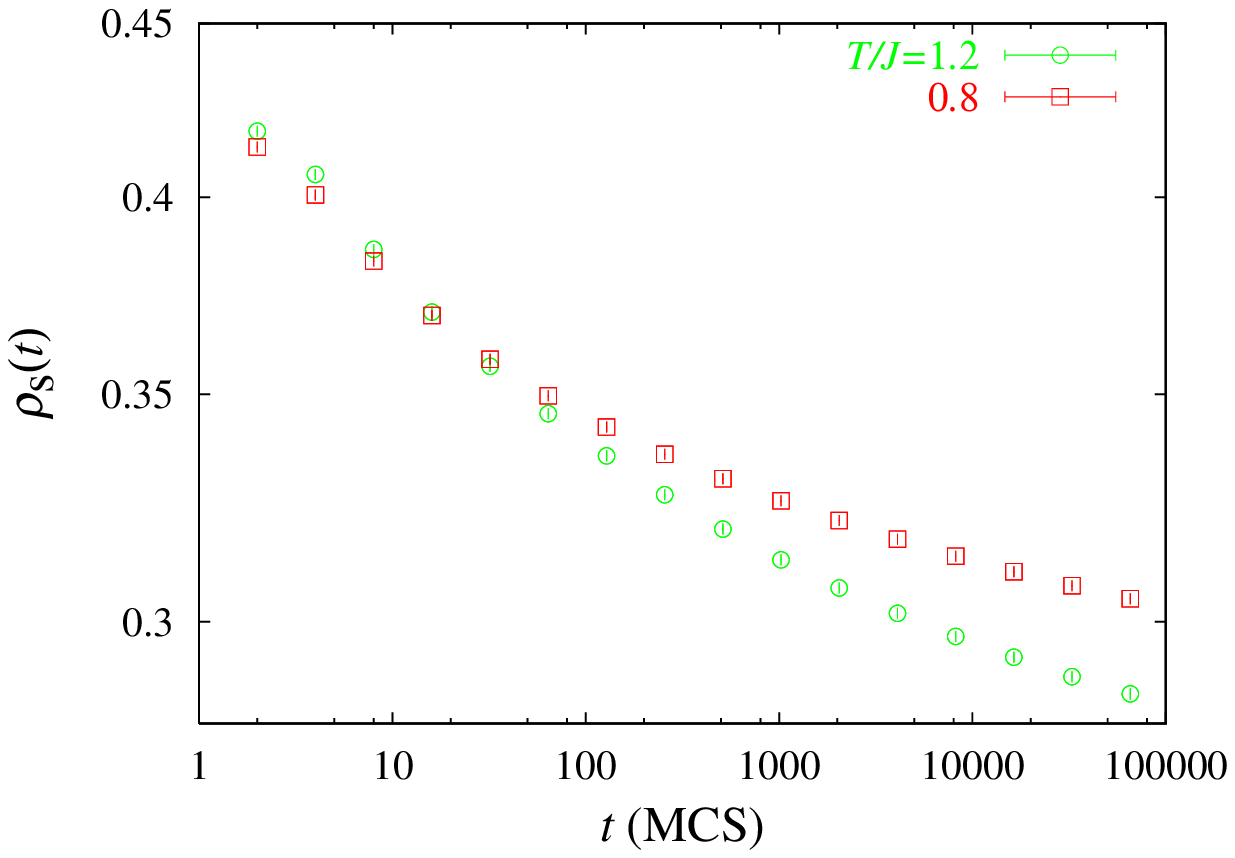}}
 \resizebox{\figwidth}{!}{\includegraphics{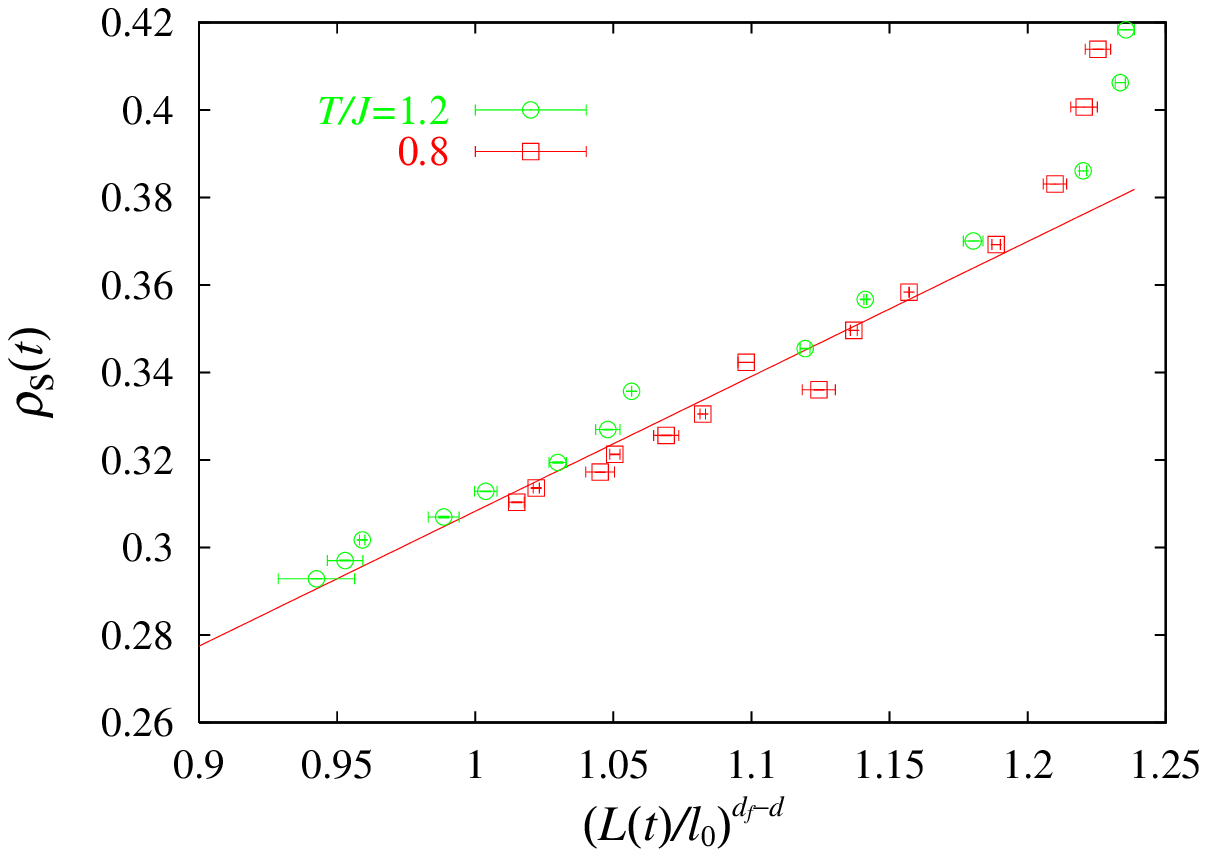}}
 \caption{Density of the domain wall at $T/J=0.8$ and $1.2$ with $L=24$
 as a function of time $t$(upper figure) and $L(t)$ (lower figure).
 The straight line in the lower figure represents fitting result to a
 linear function of $L^{d_f-d}(t)$  with $d_f=3.75$\cite{Palassini00}. 
}
\label{fig:rho-t12}
\end{figure}

In Fig.~\ref{fig:rho-t12}, we  show the domain-wall density $\rho_s(t)$
defined in (\ref{eq:dwdensity}). 
The average over bond realizations is taken over 256 samples with $L=24$. 
It is found that the density monotonically decays and there is no
tendency of saturation, which is similar behavior observed in the $3d$
Ising EA model\cite{Huse}. This is clearly seen in the lower figure,
where following (\ref{eq:density-decay}) we plot $\rho_s(t)$ as a function of
$L^{d_f-d}(t)$ estimated in the previous section. Here we use the value
of the fractal dimension $d_f=3.75$ recently evaluated in
Ref.~\onlinecite{Palassini00}. 
As shown in the lower figure, $\rho_s(t)$ is well fitted by a linear
function of $L^{d_f-d}(t)$  and the fitting function is down to zero in
the large time/length limit.  
We stress again the validity of the scaling formula with the length
scale $L(t)$.

%Two time quantities
%\input{sec06-v6}

\newcommand{\figwidthsmall}{8cm}

\section{Relaxation of Correlation function
and Linear susceptibilities}
\label{sec:analysis-two-time}

We now turn to the two time quantities. First we display the
data in the next section \ref{subsec:data-two-time} 
and discuss the overall features qualitatively in section \ref{subsec:overall}
from the point of view of the scaling theory 
explained in section \ref{subsec:theory-two-time}.
Then we go to more detailed examinations of the
scaling properties in the subsequent sections.
In order to test the scaling ansatz explained in 
section \ref{subsec:theory-two-time} which are
expressed in terms of the dynamical length scale $L(t)$,
we use the data of $L(t)$ discussed in the previous 
section \ref{sec:growthlaw}.

\subsection{Measurements of Spin Autocorrelation Function and 
Linear Susceptibilities}
\label{subsec:data-two-time}

\begin{figure}[t]
\resizebox{\figwidthsmall}{!}{\includegraphics{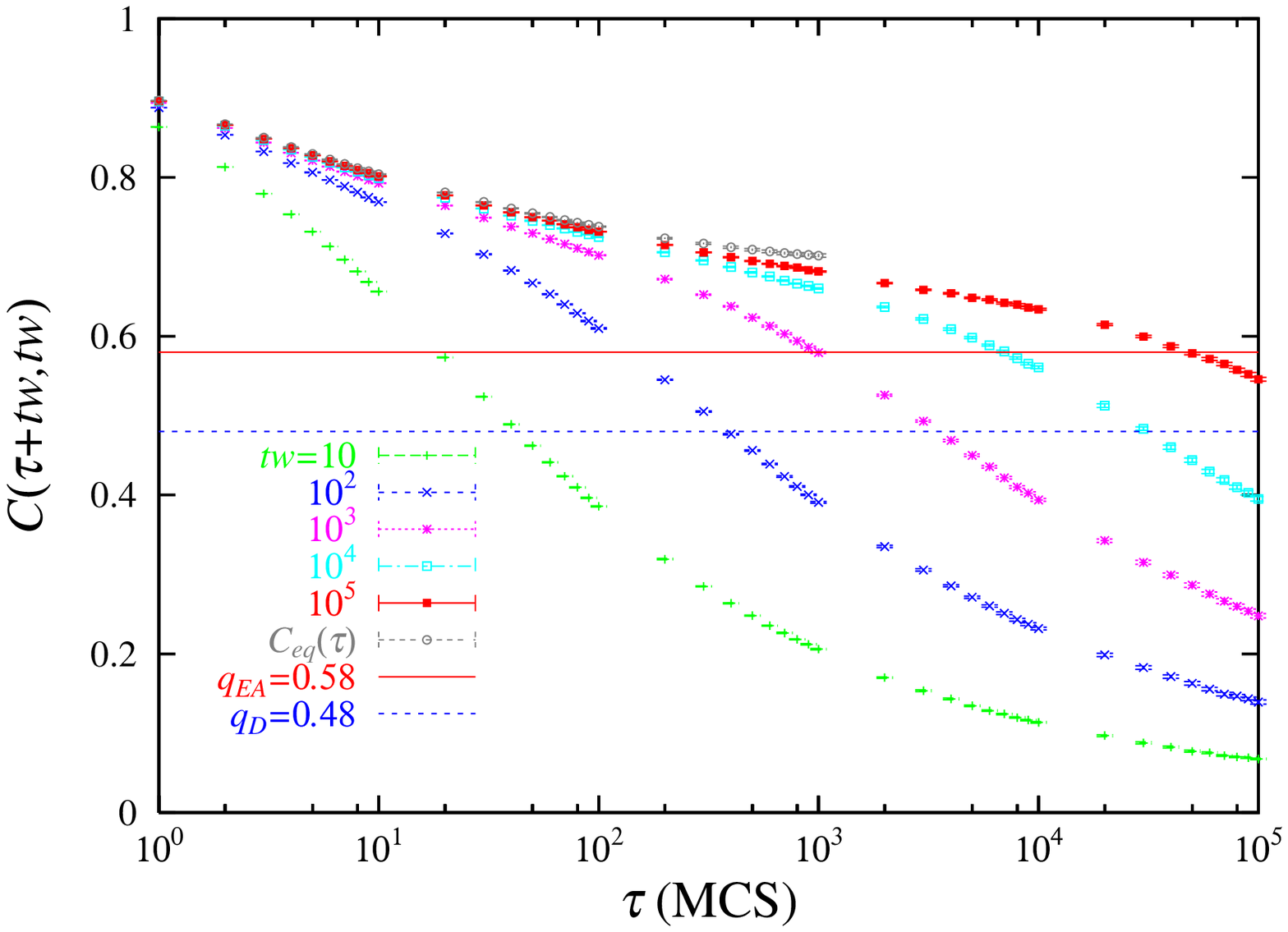}}
\resizebox{\figwidthsmall}{!}{\includegraphics{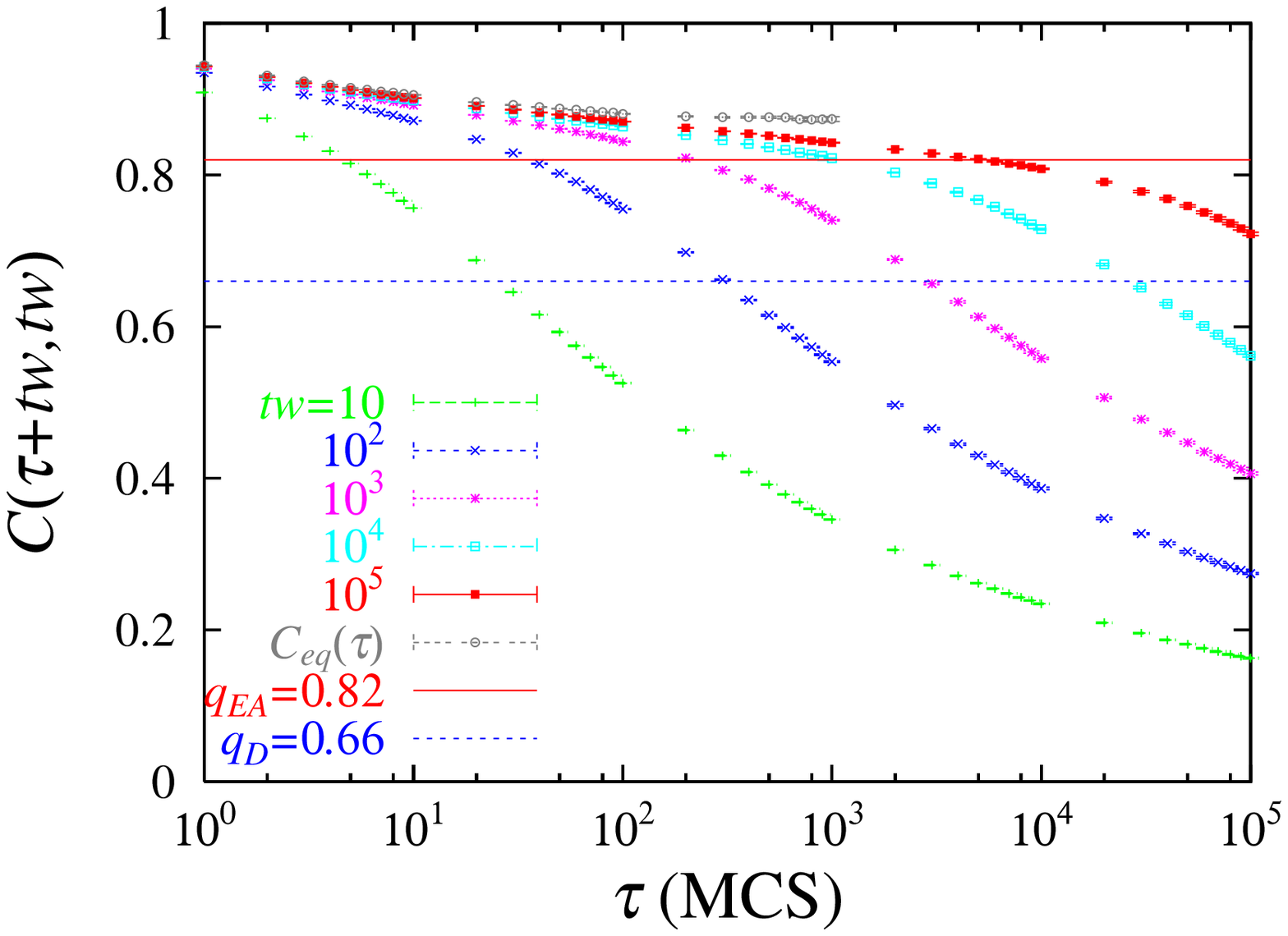}}
 \caption{Spin autocorrelation function of the $4D$ Ising SG model at
 $T/J=1.2$ (upper figure) and $0.8$ (lower figure)
with different waiting time. 
The top data curves are the
equilibrium curves obtained in section \ref{subsec:c-quasieq}.
}
\label{fig:cinv-t}
\end{figure}

In Fig.~\ref{fig:cinv-t}, we present the data of the spin autocorrelation
function $C(\tau+\tw,\tw)$ measured up to $10^5$ MCS 
for various waiting times 
$\tw=10,10^{2},10^{3},10^{4},10^{5}$ at $T/J=1.2$ and $0.8$ . 
The system size is $L=24$. The data are
obtained by performing MC simulations starting from random initial
conditions. The average over realizations of randomness 
is taken over 32 samples. The data show clear waiting time dependences.
We also show the curve of equilibrium relaxation $C_{\rm eq}(\tau)$ 
extracted in the analysis which will be 
explained later in section \ref{subsec:c-quasieq}.
The latter yields the value of static EA order parameter 
$q_{\rm EA}$ whose value is also indicated in the figure.
In addition we also indicate
in the figures the values of the dynamical order parameter
$\qd$ which will be obtained in the analysis explained later
in section \ref{subsec:scaling-sus}.
As discussed in section \ref{subsec:theory-two-time}, we expect
that the spin autocorrelation function develops a plateau at
$q_{\rm EA}$ in the quasi-equilibrium regime and decays down to
zero in the aging regime. The decay is expected to be most 
steep at around the dynamical order parameter $\qd$.
The data indeed appears qualitatively 
compatible with these expectations. 

\begin{figure}[t]
\resizebox{\figwidthsmall}{!}{\includegraphics{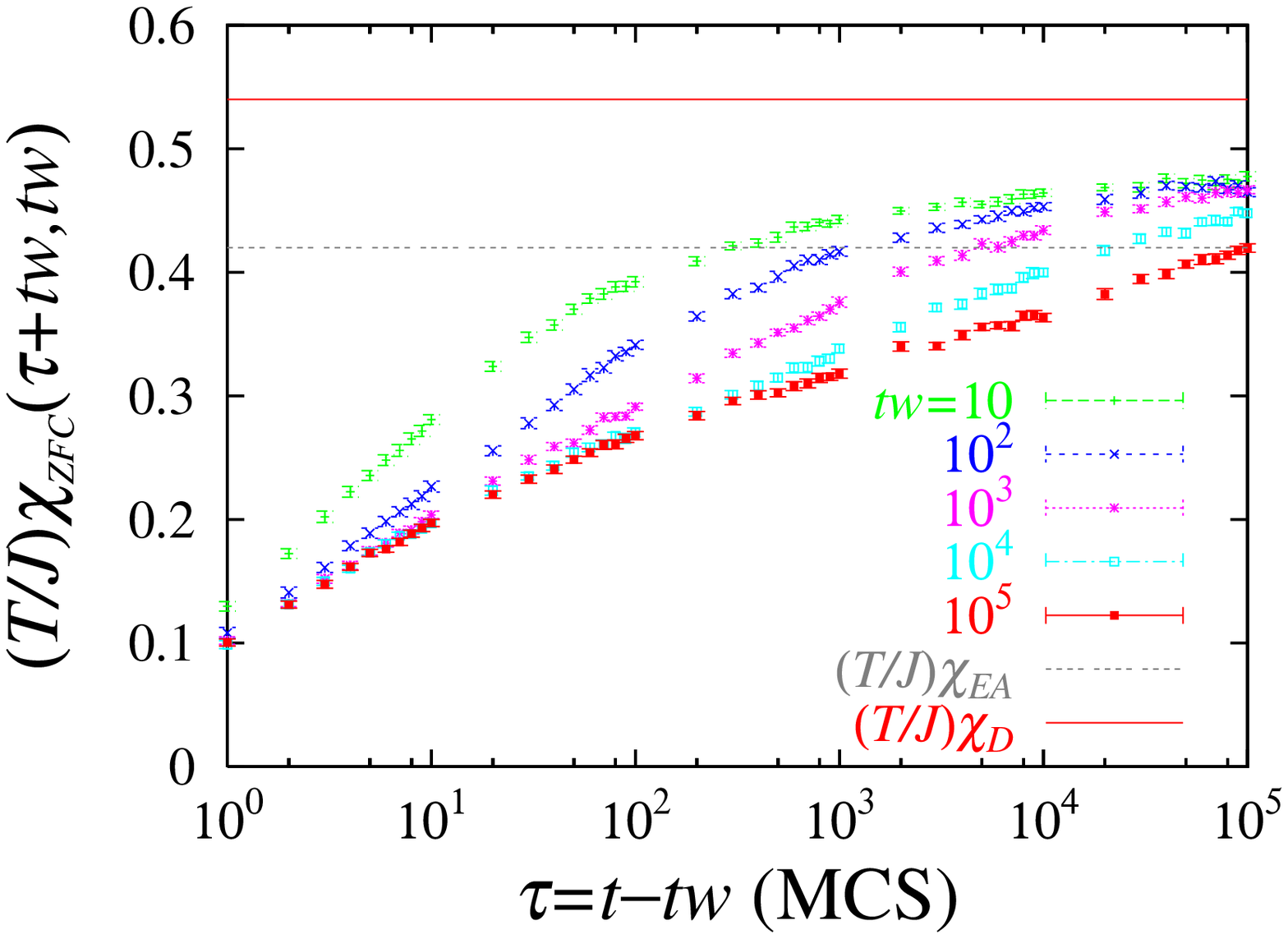}}
\resizebox{\figwidthsmall}{!}{\includegraphics{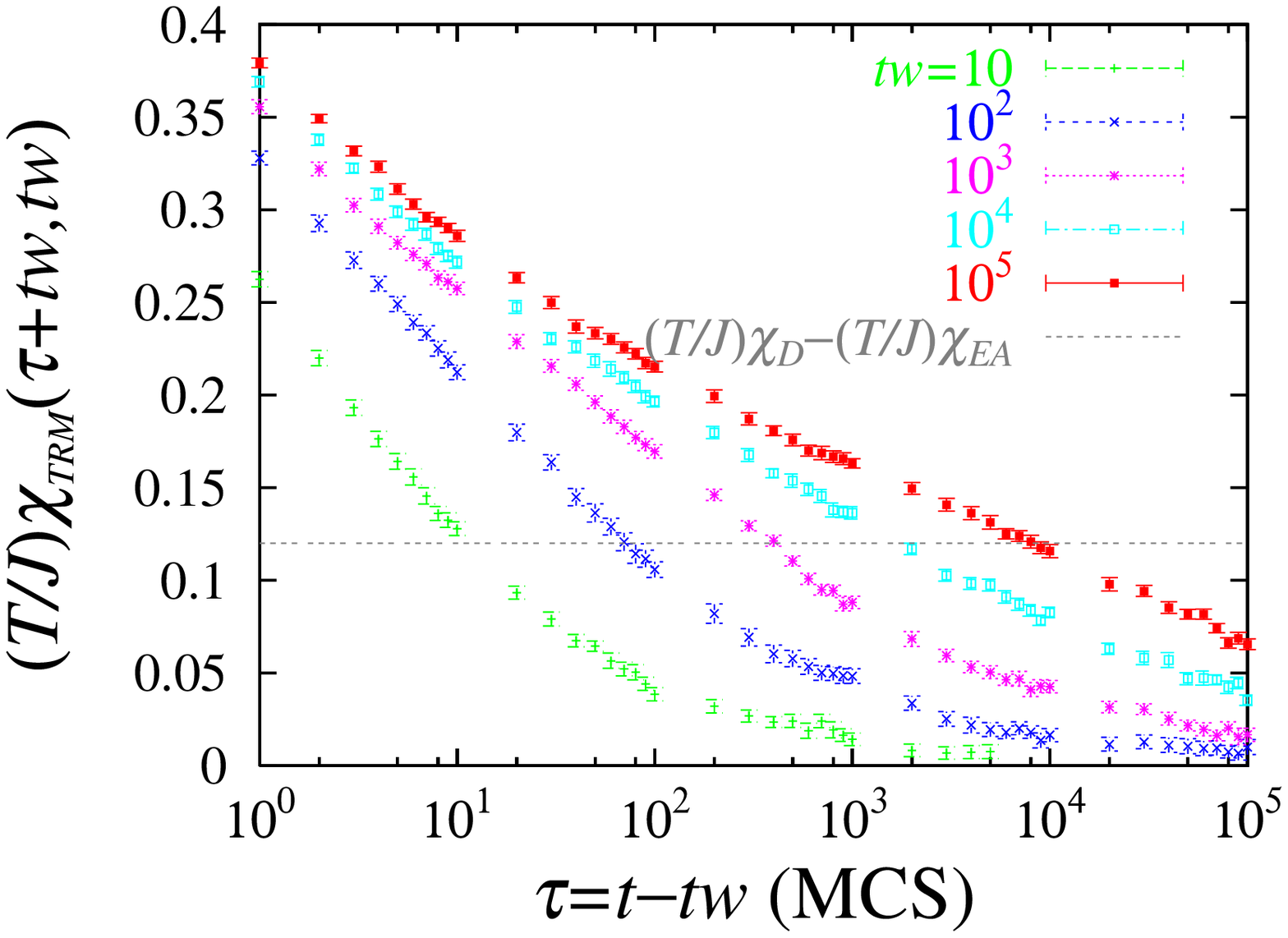}}
\caption{Linear susceptibilities at $T/J=1.2$:
ZFC susceptibility (upper figure) and TRM susceptibility (lower figure).
Here $(T/J) \chi_{\rm EA}=0.42$ 
and  $(T/J) \chi_{\rm D}=0.52$.}
\label{fig:plot-m-12}
\end{figure}

\begin{figure}[t]
\resizebox{\figwidthsmall}{!}{\includegraphics{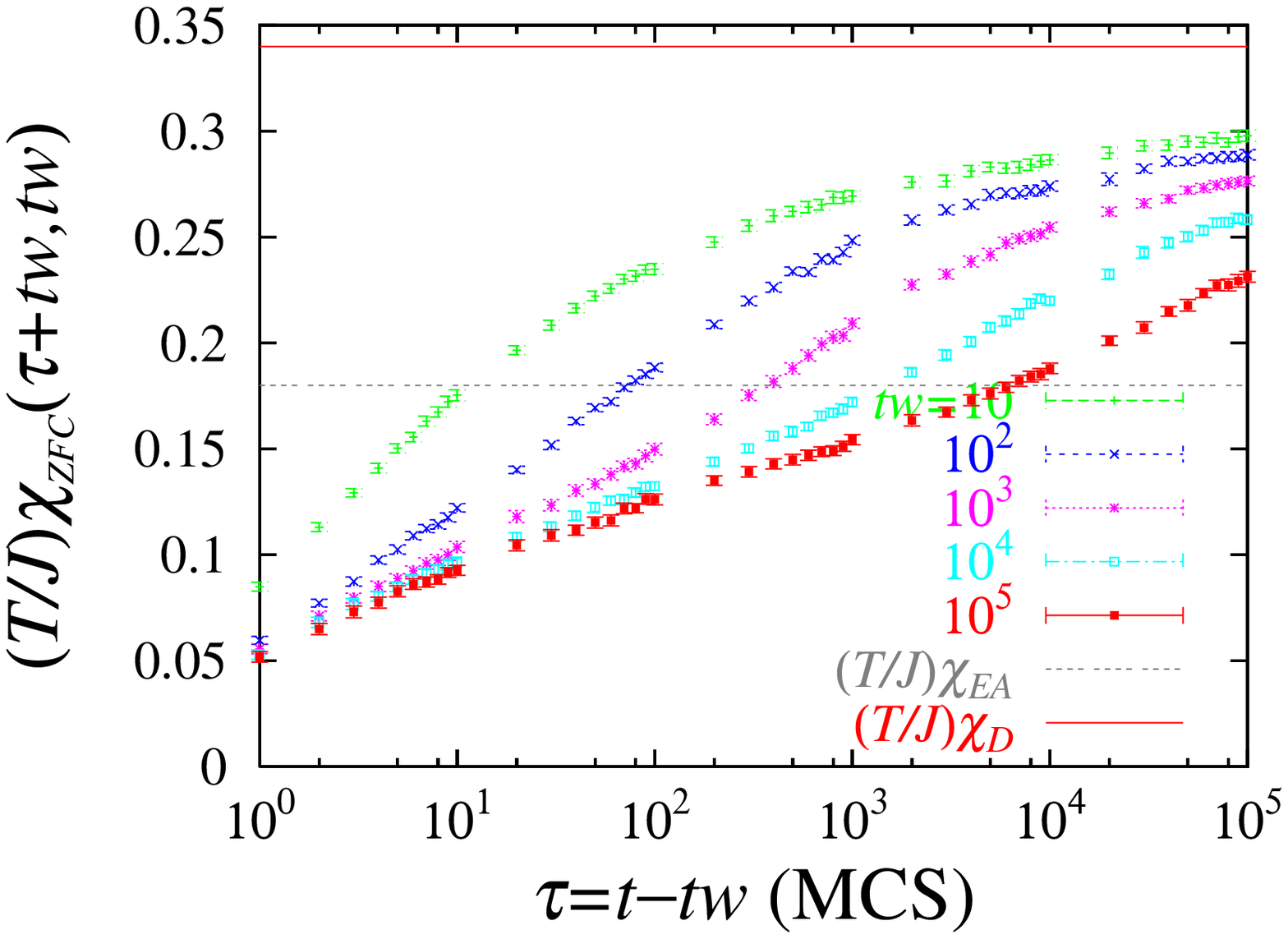}}
\resizebox{\figwidthsmall}{!}{\includegraphics{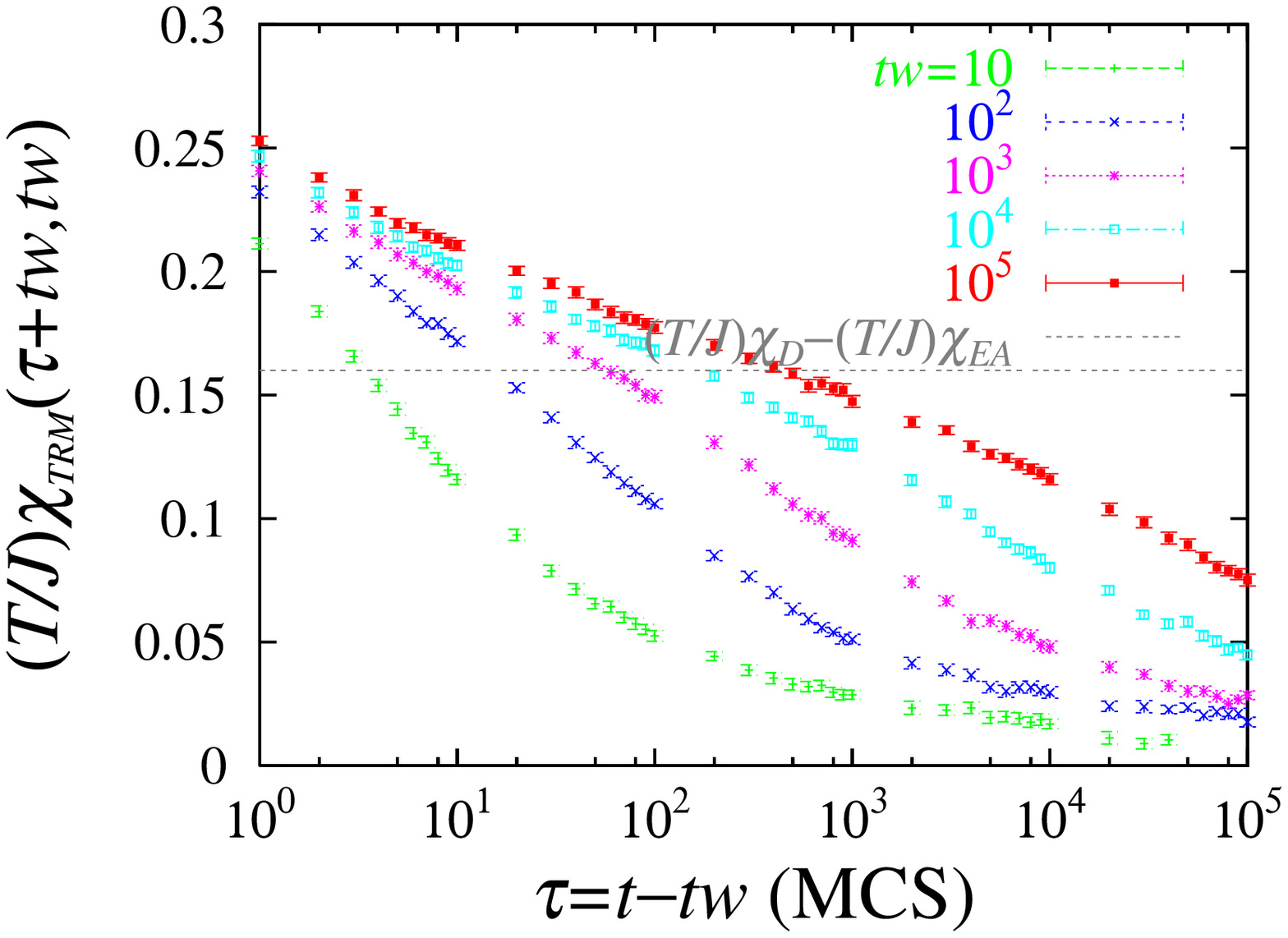}}
\caption{Linear susceptibilities at $T/J=0.8$.
%ZFC susceptibility (upper figure) and TRM susceptibility (lower figure).
Here $(T/J) \chi_{\rm EA}=0.2$ and $(T/J) \chi_{\rm D}=0.33$.
}
\label{fig:plot-m-08}
\end{figure}

\begin{figure}[h]
\resizebox{\figwidthsmall}{!}{\includegraphics{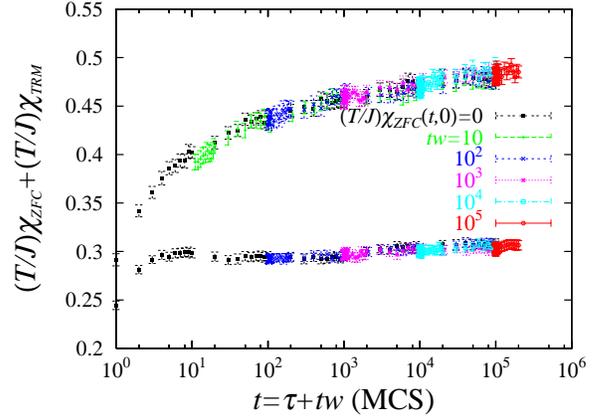}}
\caption{Check of the sum rule \eq{eq:sum-rule} 
%Here the same 
%data presented in Fig. \ref{fig:plot-m-12} and \ref{fig:plot-m-08} 
%are used. 
at $T/J=1.2$ (upper figure) and $0.8$ (lower figure).
The data of the ZFC susceptibility $(T/J)\chi_{\rm ZFC}(t,0)$
is also included.}
\label{fig:check-linear}
\end{figure}

In Figs.\ref{fig:plot-m-12} and \ref{fig:plot-m-08}, 
we display the data of the ZFC susceptibility
$(T/J)\chi_{\rm ZFC}(\tau+\tw,\tw)$
and the TRM susceptibility $(T/J)\chi_{\rm TRM}(\tau+\tw,\tw)$
measured  up to $10^5$ MCS for various waiting times 
$\tw=10,10^{2},10^{3},10^{4},10^{5}$
at $T/J=1.2$ and $T/J=0.8$ using field of strength 
$h/J=0.1$. Again the system size is $L=24$. 
The average over realizations of randomness is taken over $32$ samples. 
Here the susceptibilities are defined by
dividing the measured magnetization per spin 
$m(t)=(1/N)\sum_{i=1}^{N}S_{i}(t)$ at time $t$ by the strength 
of the external magnetic field $h/J$. 
For the measurement of the TRM susceptibility 
$\chi_{\rm TRM}(t=\tau+\tw,\tw)$, some waiting time $\tw$
is elapsed under the field and then the field is switched off
at $t=\tw$. Conversely, the ZFC  susceptibility 
$\chi_{\rm ZFC}(t=\tau+\tw,\tw)$ 
is measured by first elapsing the waiting time $\tw$ with no applied 
field and then field is switched on at $t=\tw$.

To check the linearity of the response, 
we examined the sum rule \eq{eq:sum-rule}: the sum of the
two susceptibilities become $(T/J)\chi_{ZFC}(t,0)$.
As shown in Fig.\ref{fig:check-linear}, the sum rule is satisfied
over our time window within the statistical accuracy.

In Figs.\ref{fig:plot-m-12} and \ref{fig:plot-m-08}, we can see
clear waiting time dependences of the susceptibilities. 
As discussed in section \ref{subsec:theory-two-time}, we
expect that the ZFC susceptibility first develops a plateau at the static
susceptibility $\chi_{\rm EA}$ within the quasi-equilibrium regime
and then grows further up to the dynamical susceptibility 
$\chi_{\rm D}$ later in the aging regime. Correspondingly, we
expect that the TRM susceptibility develops a plateau at
$\chi_{\rm D}-\chi_{\rm EA}$ within the quasi-equilibrium regime
and decays down to zero in the aging regime.
For the references, we indicated the values of the static 
susceptibility $(T/J) \chi_{\rm EA}$ , 
the dynamical susceptibility $(T/J) \chi_{\rm D}$
and their difference in the figures using the values which will be
obtained in section 
\ref{subsec:c-quasieq} and \ref{subsec:scaling-sus}.
The data indeed appears qualitatively compatible with the expected behavior.

\subsection{Overall Features}
\label{subsec:overall}

\begin{figure}[h]
 \resizebox{\figwidth}{!}{\includegraphics{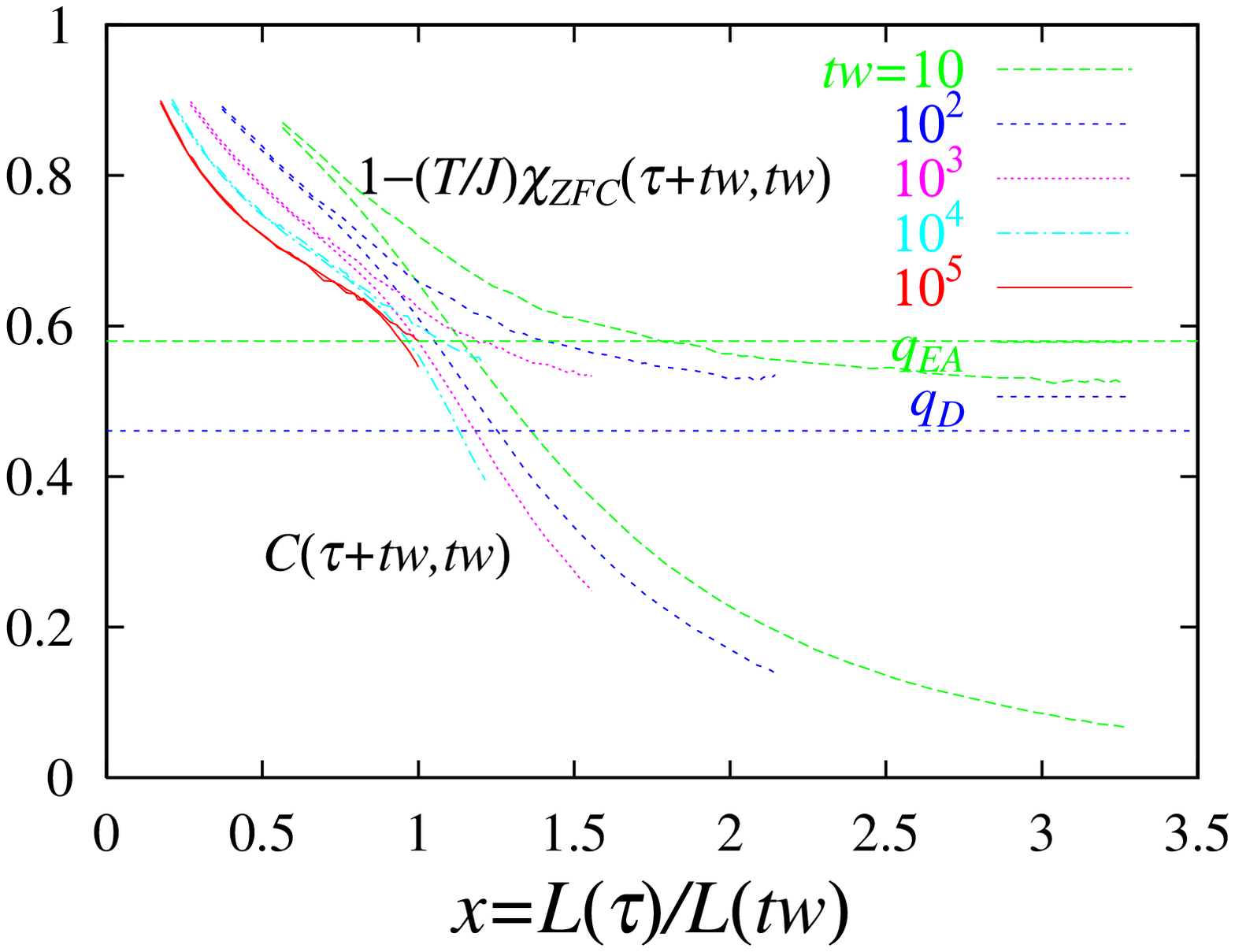}}
 \resizebox{\figwidth}{!}{\includegraphics{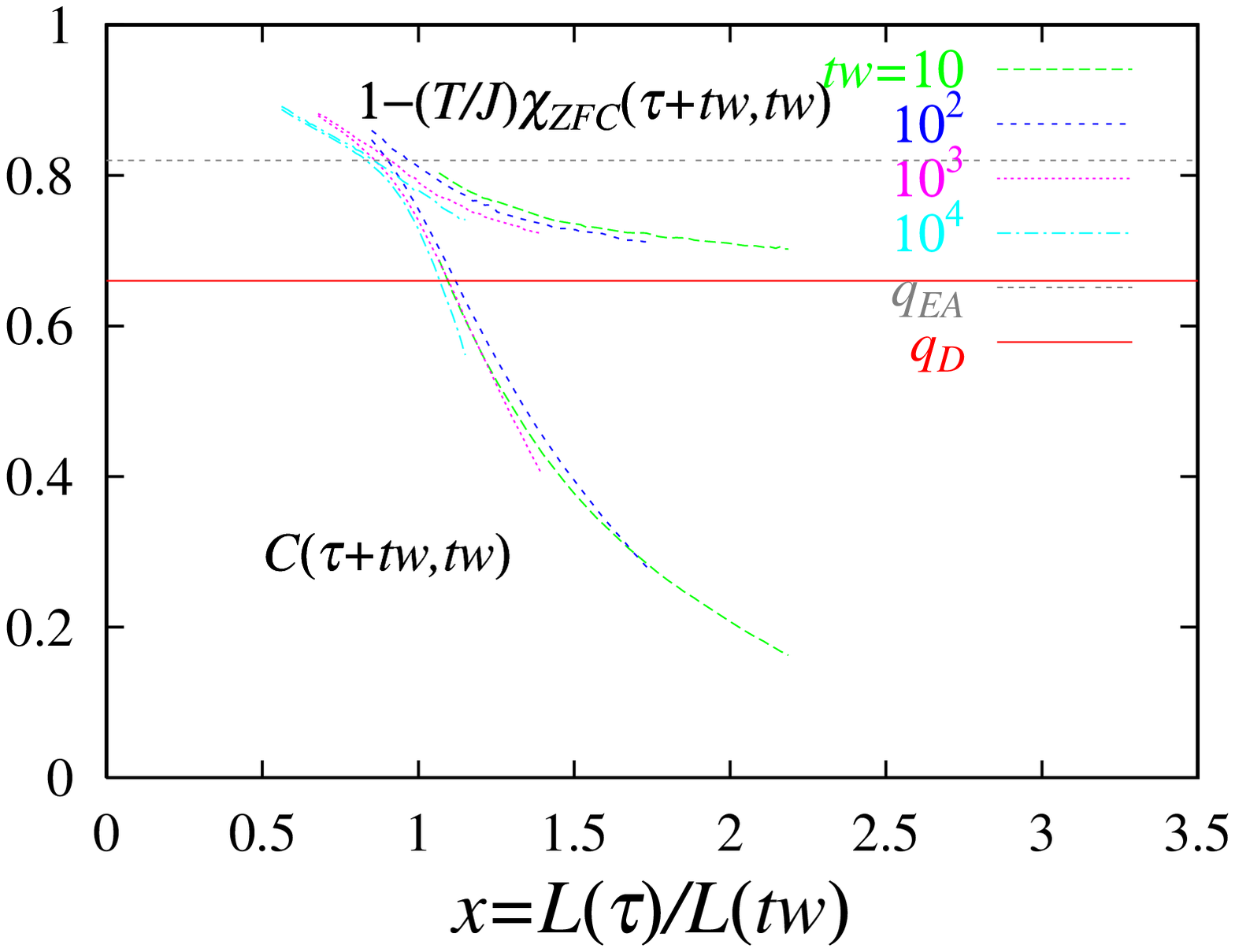}}
   \caption{Plot of spin autocorrelation function $C(\tau+\tw,\tw)$ and
$1-(T/J)\chi_{ZFC}(\tau+\tw,\tw)$ against $x=L(\tau)/L(\tw)$ at 
$T/J=1.2$ (upper figure) and $0.8$ (lower figure).
For each waiting time $\tw$, the lower curve is the 
spin autocorrelation function and the upper curve is the susceptibility.
}
\label{fig:scale-c-m}
\end{figure}

In Fig. \ref{fig:scale-c-m}, we show the spin autocorrelation 
function $C(\tau+\tw,\tw)$ 
and the ZFC susceptibility $1-(T/J)\chi(\tau+\tw,\tw)$ 
plotted against $x=L(\tau)/L(\tw)$. 

Here we parametrized the times $\tau$ and $\tw$ using 
the time dependent length $L(t)$ obtained in section \ref{sec:growthlaw}.
More precisely, we first fitted the data of the domain size $L(t)$ 
as $L(t)/\Lo=a_{0}+a_{1}\log(t)+a_{2}\log^{2}(t)+a_{3}\log^{2}(t)$
with $t$ measured in the unit of MCS and $\Lo=1$ (lattice distance).
This fitting was enough to model the data of $L(t)$ within our
time window. Then we used the resultant fitting functions
to do the parameterization here and in the following analysis.
When we do the parameterization, we discard short time data $t < 10$
because $L(t)$ is not available there.

\begin{figure}[t]
 \resizebox{\figwidth}{!}{\includegraphics{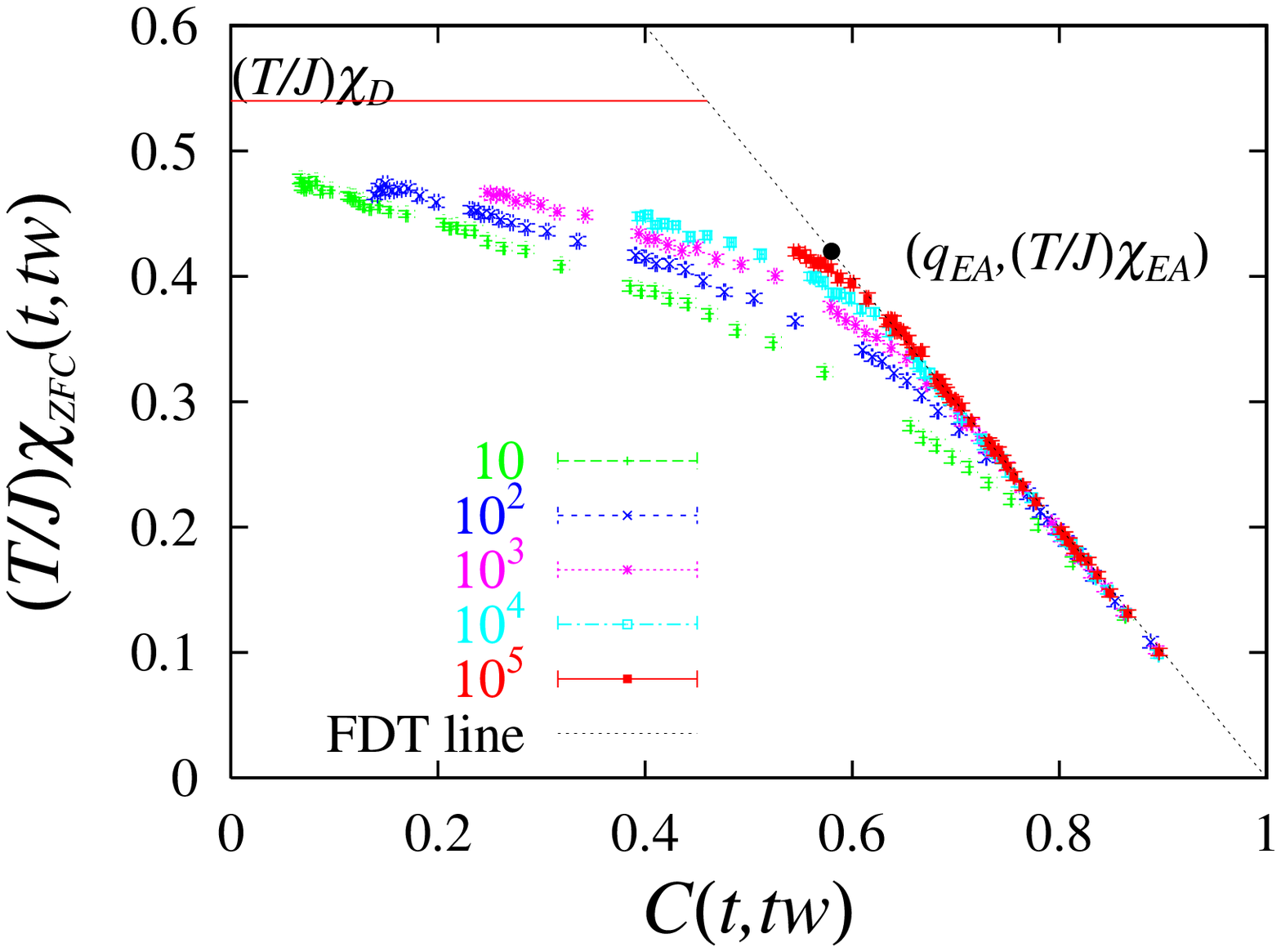}}
 \resizebox{\figwidth}{!}{\includegraphics{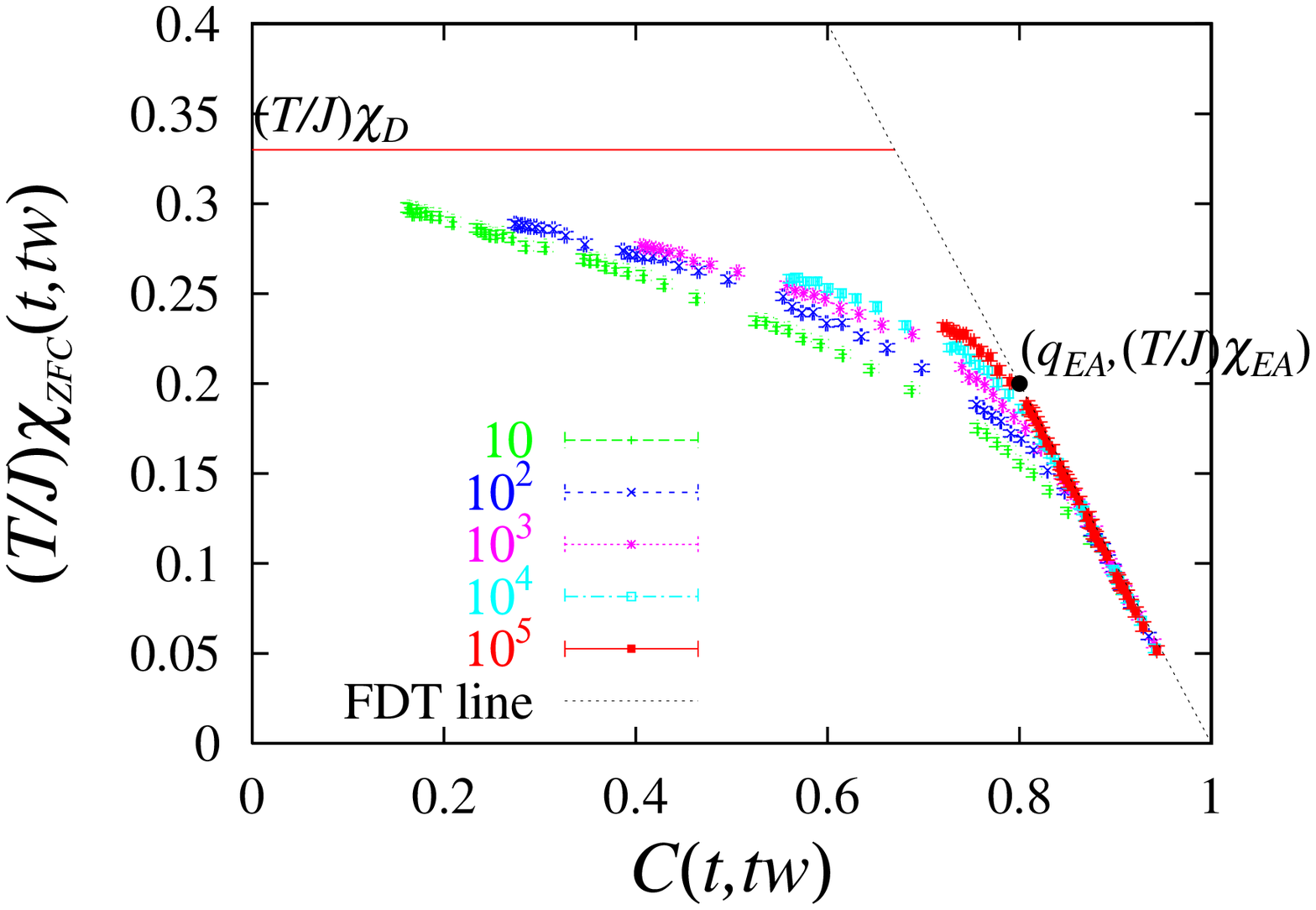}}
 \caption{Parametric plot  $(T/J)\chi_{\rm ZFC}(t,\tw)$ vs $C(t,\tw)$ 
at $T/J=1.2$ (upper figure) and $0.8$ (lower figure).
The straight tangent lines represents the FDT \eq{eq-fdt}.
The convergence points of the
break  points of TTI $(q_{\rm EA}, (T/J)\chi_{\rm EA})$ 
and the break points of FDT $(q_{\rm D} ,(T/J)\chi_{\rm D})$ 
which will be obtained later in section \ref{subsec:c-quasieq} and 
\ref{subsec:scaling-sus} respectively are indicated.}
\label{fig:ck}
\end{figure}

The figure should be compared with  Fig. \ref{fig:c-conjecture}
which explains the expected asymptotic behavior of the two time
quantities in the large time limit $L(\tw) \to \infty$ with the ratio
$x=L(\tau)/L(\tw)$  
fixed to certain values.  We expect
three distinct regimes depending on the value of the fixed ratio 
i) quasi-equilibrium regime $x < 1$ where the autocorrelation functions
spans values in the range $q_{\rm EA} < C < 1$
ii) crossover regime $x \sim 1$ for  $q_{\rm D} < C < q_{\rm EA}$
and iii) aging regime $x >1$ for $0  < C < q_{\rm D}$. 
Here $q_{\rm EA}$ and $q_{\rm D}$ are convergence points of
the break points of TTI and FDT respectively 
in the limit $L(\tw) \to \infty$.

In the quasi-equilibrium regime $x < 1$, we expect that the
spin autocorrelation function convergence to a plateau 
$q_{\rm EA}$ in the large time limit $L(\tw) \to \infty$.
Correspondingly the  ZFC susceptibility $(T/J)\chi_{\rm ZFC}(\tau+\tw,\tw)$ 
is expected to converge to the static susceptibility
$(T/J)\chi_{\rm EA}=1-q_{\rm EA}$. 
In the figure,  we  indicated the value of $q_{\rm EA}$ 
which will be obtained later in section \ref{subsec:c-quasieq}.
The data indeed appears descending down toward the plateau from above
with increasing $\tw$ (at $x<1$) but still far from it. 
Fortunately, the scaling theory provides prediction on the correction terms
to the asymptotic limit as explained 
in section \ref{subsubsec:theory-quasi-eq}.
In the section \ref{subsec:c-quasieq}, we will examine the correction terms
in detail which actually yields the value of $q_{\rm EA}$.
Second important observation is that
the FDT $1-C(t,\tw)=(T/J)\chi_{\rm ZFC}(t,\tw)$  is well satisfied
in the quasi-equilibrium regime $x < 1$ as expected.

In the crossover regime  $x \sim 1$, we expect a vertical drop from the
plateau at the static EA order parameter 
$q_{\rm EA}$ down to the dynamical oder parameter $q_{\rm D}$ 
for the spin autocorrelation function. Correspondingly
the ZFC susceptibility $(T/J)\chi_{\rm ZFC}(\tau+\tw,\tw)$ 
is expected to jump from the static susceptibility 
$(T/J)\chi_{\rm EA}=1-q_{\rm EA}$
up to the dynamical susceptibility $(T/J) \chi_{\rm D}=1-q_{\rm D}$
still keeping the FDT $1-C(t,\tw)= (T/J)\chi_{\rm ZFC}(t,\tw)$
as explained in section \ref{subsubsec:theory-quasi-eq}.
In the figure,  we  indicated the value of $\qd$ 
which will be obtained later in section \ref{subsec:scaling-sus}.
We will examine the slope of the curves 
around $x\sim 1$ in section  \ref{subsec:scaling-sus}
and find indeed that the suitably refined relaxation rate
function $S_{\rm mod}(x,\tw)$  defined in 
\eq{eq-def-s-mod} have a sharply pronounced peak at around $x \sim 1$.
Furthermore, we will examine the violation of FDT 
$I(\tau+\tw,\tw)=1-C(\tau+\tw,\tw)-(T/J)\chi_{\rm ZFC}(\tau+\tw,\tw)$
defined in \eq{eq-integral-violation-fdt}
in section \ref{subsec:integralFDT}
and find indeed that it is decreasing with increasing $\tw$ 
in the crossover regime $x \sim 1$. The result is compatible with our expectation that
the FDT is asymptotically valid even in the crossover regime.

In the aging regime  $x>1$, we expect the 
ZFC susceptibility $1-(T/J)\chi_{\rm ZFC}(\tau+\tw,\tw)$
converges to the dynamical susceptibility
$(T/J)\chi_{\rm D}=1-q_{\rm D}$ 
while spin autocorrelation function
$C(\tau+\tw,\tw)$ becomes a scaling function of $L(\tau)/L(\tw)$.
Indeed the data appears slow converges to such limits.
For the susceptibility, the scaling theory provides prediction 
on the scaling forms of the correction terms to the asymptotic 
limit as explained 
in section \ref{subsubsec:theory-aging-chi}. We will examine and confirm them
in section \ref{subsec:scaling-sus}. 
The value of $q_{\rm D}$ indicated in Fig. \ref{fig:scale-c-m}
will be actually obtained as the result of such an  analysis.

In Fig. \ref{fig:ck}, the susceptibility $(T/J)\chi_{\rm ZFC}(\tau+\tw,\tw)$
and the spin autocorrelation function $C(\tau+\tw,\tw)$ are
plotted in a parametric way. It should be compared with 
Fig. \ref{fig:ck-conjecture}, which explains the expected 
asymptotic regimes in the large time limit 
$L(\tw) \to \infty$  with the fixed ratio $x=L(\tau)/L(\tw)$.
The curves apparently continue to move upwards with increasing 
waiting time $\tw$ and the break point of FDT does not appear
to converge to $(q_{\rm EA},(T/J)\chi_{\rm EA})$ which will be
obtained in section \ref{subsec:c-quasieq}. This observation is
consistent with our picture presented in Fig. \ref{fig:ck-conjecture}.
It is radically different from conventional understanding \cite{BCKM}
that assumes break points of TTI and FDT are the same. 
As we discuss in section \ref{subsec:separation-breaking},
the separation of the break points of TTI and FDT will become 
apparent when time (length) dependences are explicitly 
considered.

\subsection{Quasi-Equilibrium Regime}
\label{subsec:c-quasieq}

\begin{figure}[]
\resizebox{\figwidthsmall}{!}{\includegraphics{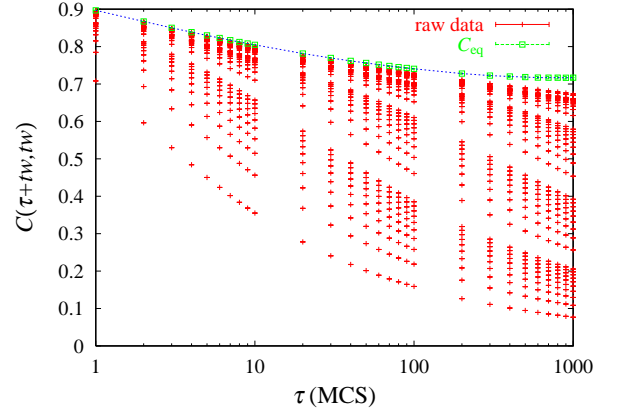}}
\caption{Spin autocorrelation function at $T/J=1.2$.
The top data points are the equilibrium limit obtained 
by the extrapolation method explained in the text.
}
\label{fig:dense12}
\end{figure}
\begin{figure}[]
\resizebox{\figwidth}{!}{\includegraphics{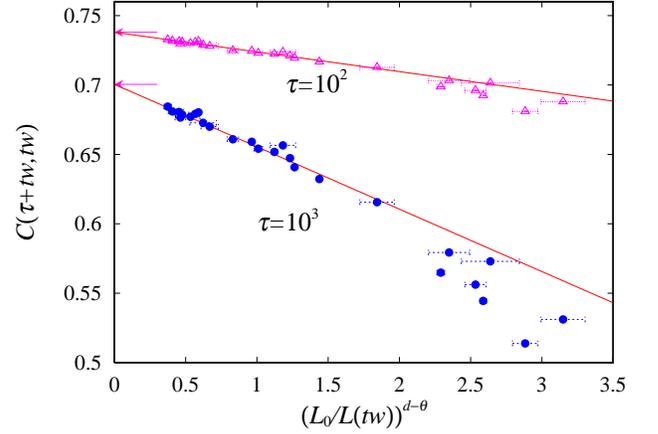}}
 \caption{
Typical examples of the extrapolation to the long $t_w$ limit with
fixed $\tau$. Two arrows represent the extrapolated
values of $C_{eq}(\tau)$ of $\tau=100$ and $1000$ at $T/J=1.2$.
The horizontal error bars are those of $L(t)$ presented in 
Fig. \protect{\ref{fig:xi-raw}}.
}
\label{fig:ext-tw}
\end{figure}

\begin{figure}[]
\resizebox{\figwidth}{!}{\includegraphics{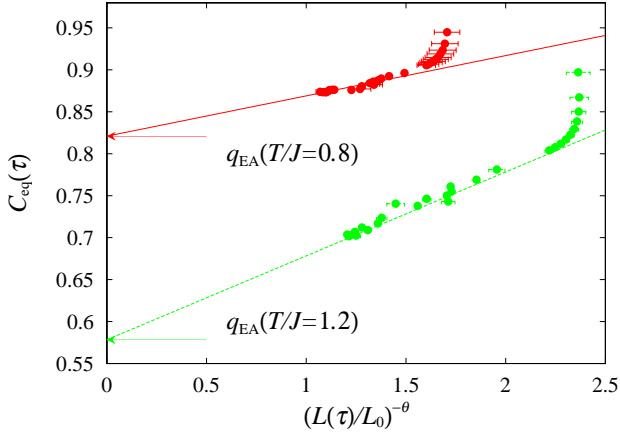}}
 \caption{Equilibrium spin autocorrelation function $C_{eq}(\tau)$
at $T/J=1.2$ (upper data)and $0.8$ (lower data).
 The horizontal error bars are those of $L(t)$ presented in 
Fig. \protect{\ref{fig:xi-raw}}.
The lines represent fittings according to the formula 
 (\protect{\ref{eq-c-eq}}).}  
 \label{fig:ceq}
\end{figure}

\begin{figure}[]
\resizebox{\figwidth}{!}{\includegraphics{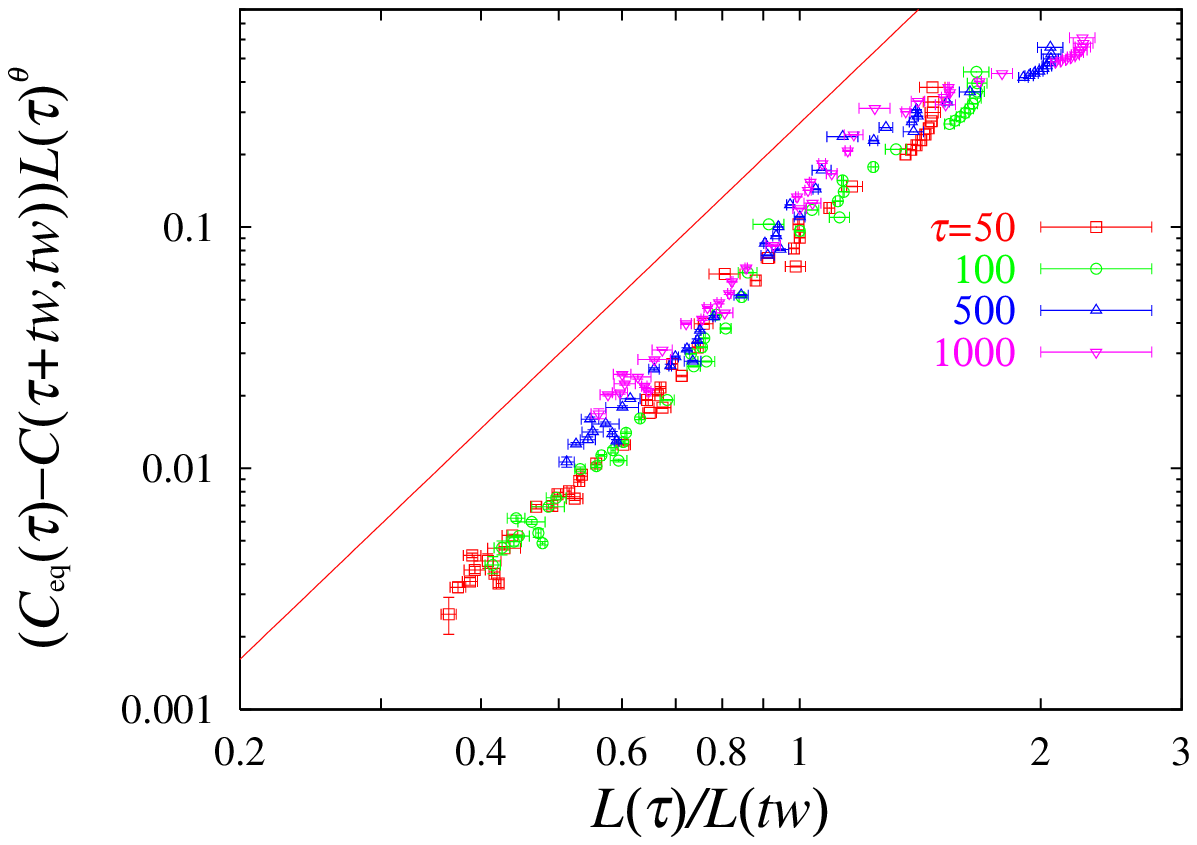}}
\resizebox{\figwidth}{!}{\includegraphics{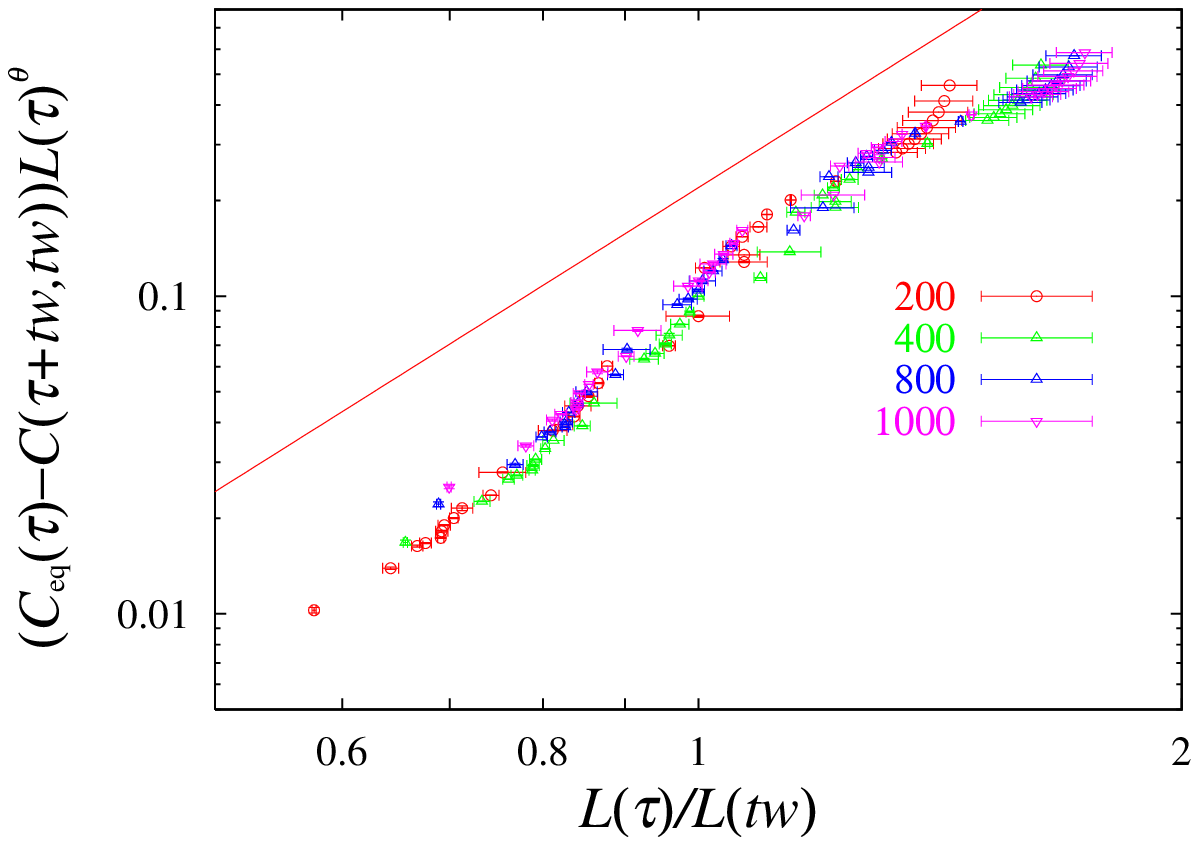}}
 \caption{Scaling plot of the non-equilibrium 
correction terms of $C(\tau+\tw,t_w)$ in the quasi-equilibrium regime
at $T/J=1.2$ (upper figure) and $0.8$ (lower figure).
 The horizontal error bars are those of $L(t)$ presented in 
Fig. \protect{\ref{fig:xi-raw}}.
The line has the expected slope $d-\theta$ for small $L(\tau)/L(t_w)$
 limit (see \protect{\eq{eq-c-quasieq}}). }
 \label{fig:coft-scal}
\end{figure}

Let us now examine the scaling properties in the quasi-equilibrium 
regime $x=L(\tau)/L(\tw) < 1$ discussed in section 
\ref{subsubsec:theory-quasi-eq}.
Since we have seen the FDT is well satisfied in the quasi-equilibrium
regime  $x  < 1$ (see Fig. \ref{fig:scale-c-m}), we will only use 
the data of the spin autocorrelation function. 
To focus on the quasi-equilibrium
regime, we took another dense data set of the autocorrelation function 
$C(\tau+\tw,\tw)$ (shown in Fig.\ref{fig:dense12})
of many waiting times $\tw=1,\ldots,30000$(MCS) and 
relatively short time separations $\tau < 1000$ (MCS).
Here we don't use the fitting of the growth law of $L(t)$ to
parameterize the time but
use directly the data of  $L(t)$ obtained in section
\ref{sec:growthlaw}.

The formula \eq{eq-c-quasieq} combined with \eq{eq-c-eq}
gives a protocol to extract the static EA order parameter
as the following. This analysis is already reported 
partly in  our previous work \cite{HYT}.

At first we take the equilibrium limit $t_w\rightarrow\infty$ of
$C(\tau+\tw,t_w)$ for each time separation $\tau$.
From \eq{eq-c-quasieq}, 
% which reads as, 
% \begin{eqnarray}
% && C(t=\tau+\tw,\tw)   = C_{\rm eq}(\tau)   \nonumber \\
% &-&  c'\rhoo m^{2} \frac{\kb T}{\Upsilon(L(\tau)/L_0)^{\theta}} 
% \left(\frac{L(\tau)}{L(t_{\rm w})}\right)^{d-\theta} 
%  +\cdots ,  \nonumber
% \end{eqnarray}
one finds that  the autocorrelation 
function becomes only  a linear function of $1/L^{d-\theta}(t_w)$
for a fixed $\tau$. Thus we plotted the data points of a given
$\tau$ of various $\tw$ against $1/L^{d-\theta}(t_w)$ and
fitted to a linear function in the large $L(t_w)$ regime. 
The equilibrium limit $C_{\rm eq}(\tau)$ is read off
directly from the fit.
Here we use the stiffness exponent $\theta=0.82$\cite{KH}.
For $L(\tw)$
we used the data obtained in the previous section \ref{sec:growthlaw}.
Typical fitting results are shown in Fig.~\ref{fig:ext-tw}. 
The linearity as a function of $1/L^{d-\theta}(t_w)$ 
supports the validity of the formula (\ref{eq-c-quasieq}). 
We repeated the same procedure for each $\tau$. 
The obtained equilibrium curve 
$C_{\rm eq}(\tau)$ was displayed in Fig.~\ref{fig:cinv-t}.

Next we take the large $\tau$ limit of the extracted $C_{\rm eq}(\tau)$
using \eq{eq-c-eq}.
%  which reads as,
% \bmat
% C_{\rm eq}(\tau) 
% =q_{\rm EA} +  c \rhoo m^{2}(\kb T/\Upsilon)(L(\tau)/L_{0})^{-\theta}.
% \emat
As shown in Fig.~\ref{fig:ceq},
$C_{\rm eq}(\tau)$ appears as a linear function of $1/L^{\theta}(\tau)$ 
which support \eq{eq-c-eq}.
A linear fit gives the EA order parameter $q_{\rm EA}$
as the limiting value. We obtained the EA
order parameter $q_{\rm EA}=0.58(2)$ at $T/J=1.2$
and $0.82(1)$ at $T/J=0.8$.  To our knowledge
this and our previous work \cite{HYT} is the fist 
which confirmed this fundamental scaling law \eq{eq-c-eq}.

Finally, we discuss the weak non-equilibrium correction term
to the equilibrium limit which is responsible for the weak
violation of TTI in the quasi-equilibrium regime.
The expression (\ref{eq-c-quasieq}) suggests that the weak
non-equilibrium correction term
$\Delta C(\tau+\tw,t_w)=C(\tau+\tw,t_w)-C_{\rm eq}(\tau)$  multiplied by 
$L^\theta(\tau)$ becomes only a function of $L(\tau)/L(t_w) ( < 1)$. 
As shown in Fig.~\ref{fig:coft-scal}, we confirm this scaling
form. For the limit of $L(\tau)/L(t_w)\ll 1$, the scaling function 
shows the expected power law behavior
$\left(L(\tau)/L(t_w)\right)^{d-\theta}$ being consistent with 
\eq{eq-c-quasieq}. 

\subsection{Crossover Regime}
\label{subsec:crossover}
\begin{figure}[]
\resizebox{\figwidth}{!}{\includegraphics{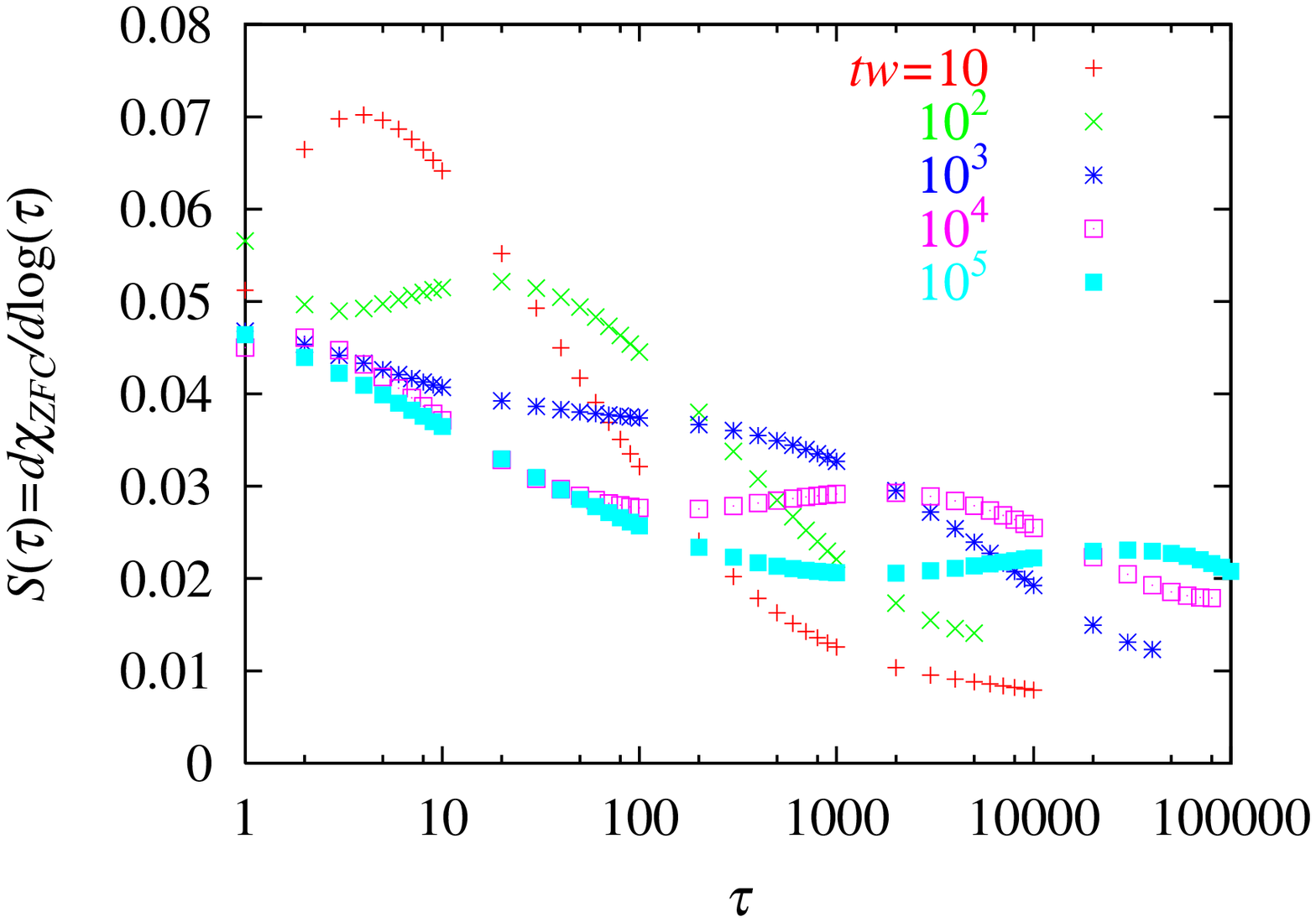}}
\resizebox{\figwidth}{!}{\includegraphics{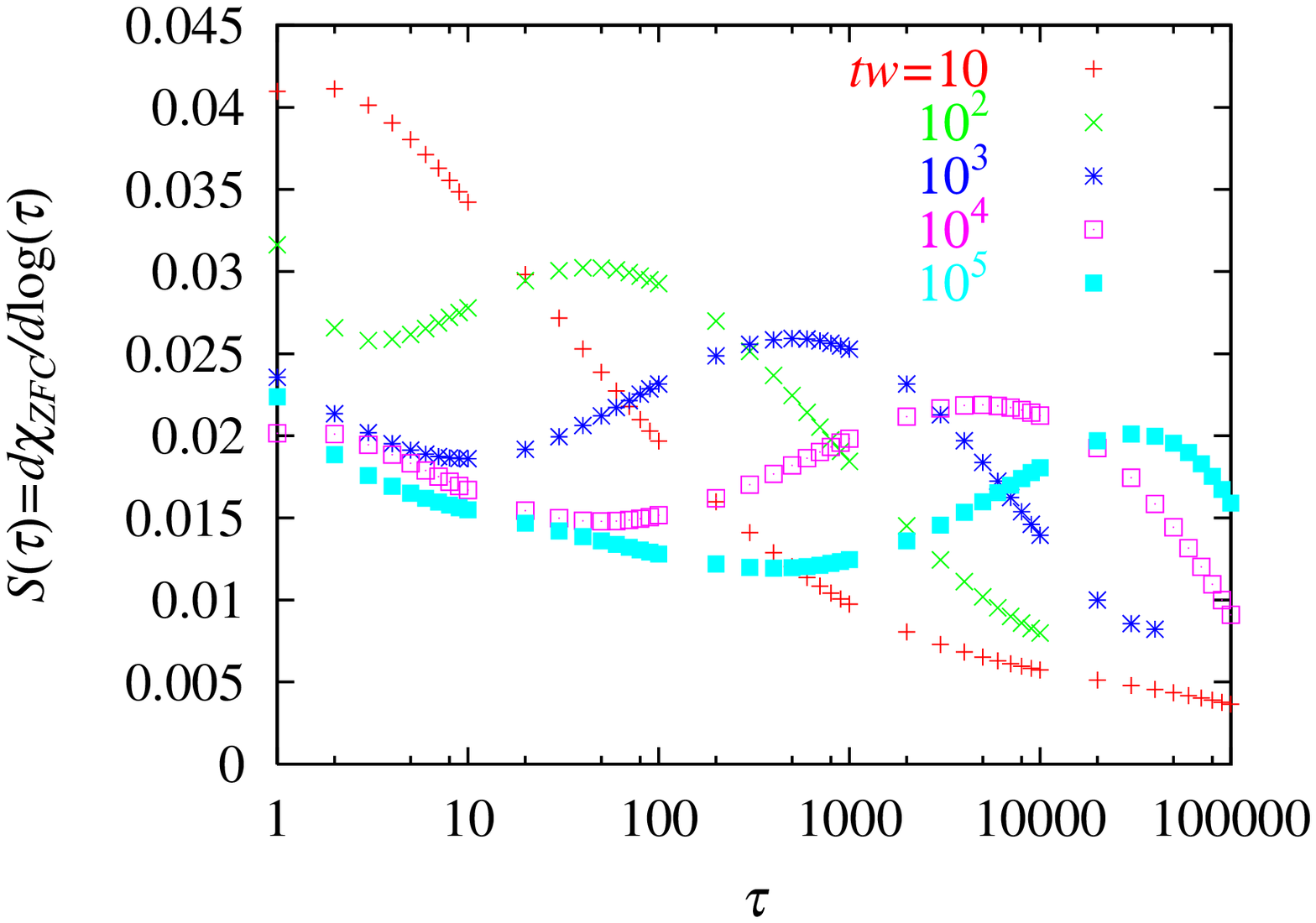}}
 \caption{Plot of the usual relaxation rate function
 $S(\tau,\tw)$ at $T/J=1.2$ (upper figure) and $0.8$ (lower figure).
}
 \label{fig:s}
\end{figure}

\begin{figure}[]

\resizebox{\figwidth}{!}{\includegraphics{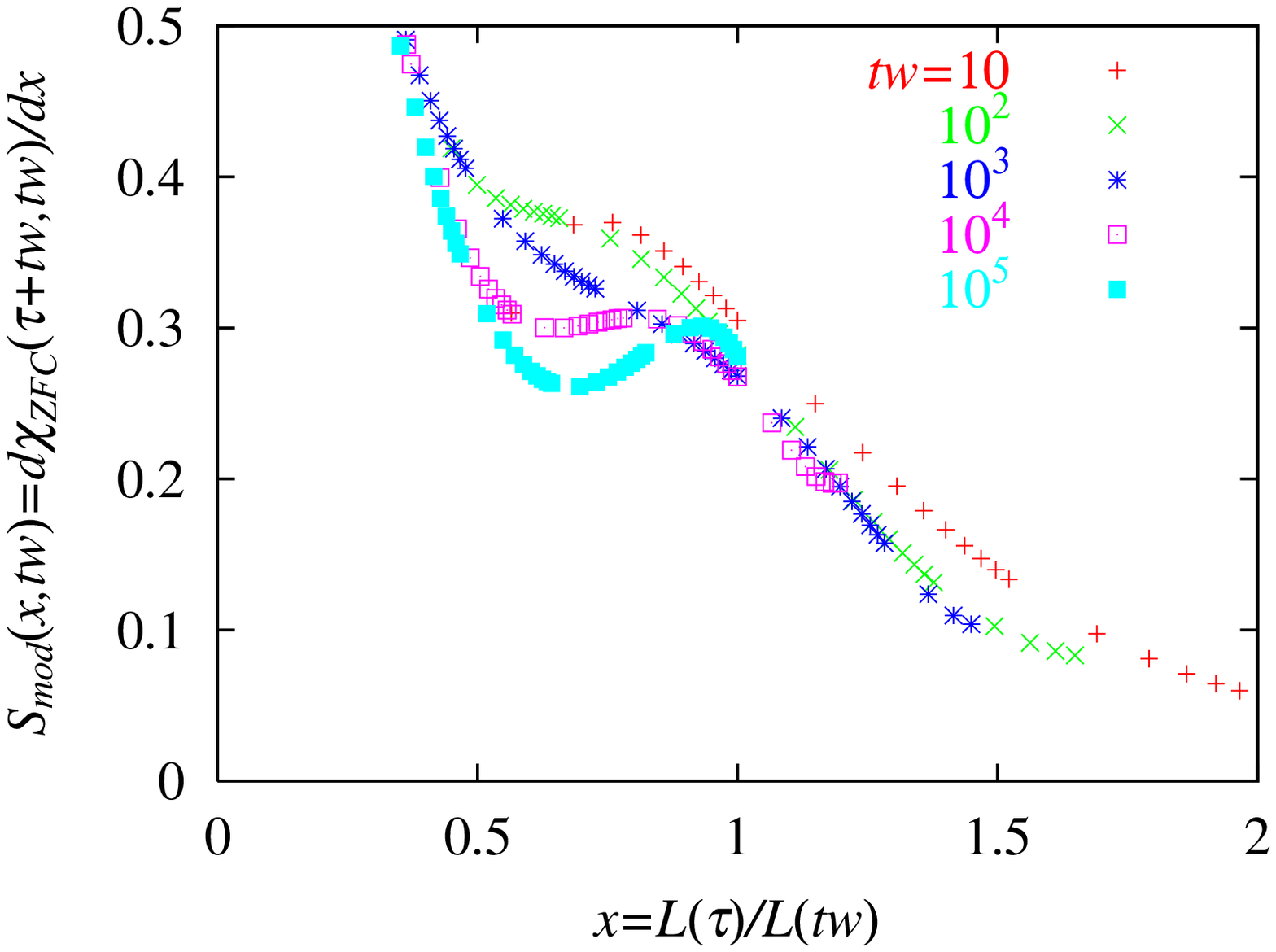}}
\resizebox{\figwidth}{!}{\includegraphics{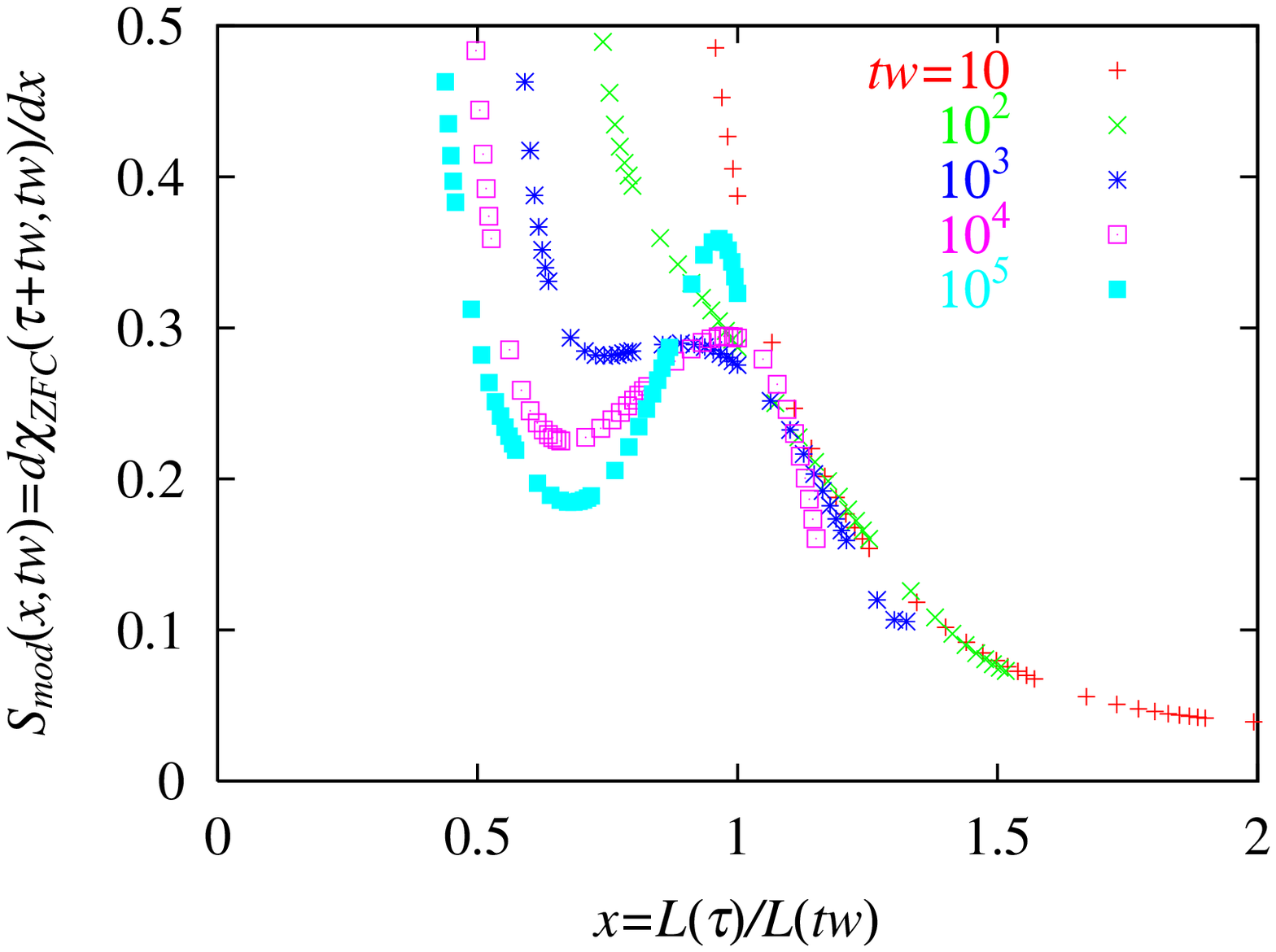}}
 \caption{Scaling plot of the modified relaxation rate function 
$S_{mod}(x,\tw)$  at $T/J=1.2$ (upper figure) and $0.8$ (lower figure).}
 \label{fig:s-mod}
\end{figure}

In the crossover regime $x=L(\tau)/L(\tw) \sim 1$, we expect
a vertical jump of the two time quantities at asymptotic
limit $L(\tw) \to \infty$. In Fig. \ref{fig:s} we display
the plot of the usual relaxation rate function $S(\tau,\tw)$ of
\eq{eq-def-s}, 
which shows the well known peak structure observed 
in experiments \cite{Uppsala-S} and a MC simulation \cite{MC-S}. 
This can be interpreted as the signal of the rapid changes 
in the crossover regime.
However, our scenario naturally suggested a more appropriate, 
{\it modified} relaxation rate function  $S_{\rm mod}(x, \tw)$ of
\eq{eq-def-s-mod}. 
In Fig. \ref{fig:s-mod} we show the plot of the modified relaxation rate
function. It clearly develops a sharp peak at around $x \sim 1$ 
with increasing $L(\tw)$ as expected.

\subsection{Scaling of Susceptibilities}
\label{subsec:scaling-sus}
\begin{figure}[h]
\resizebox{\figwidth}{!}{\includegraphics{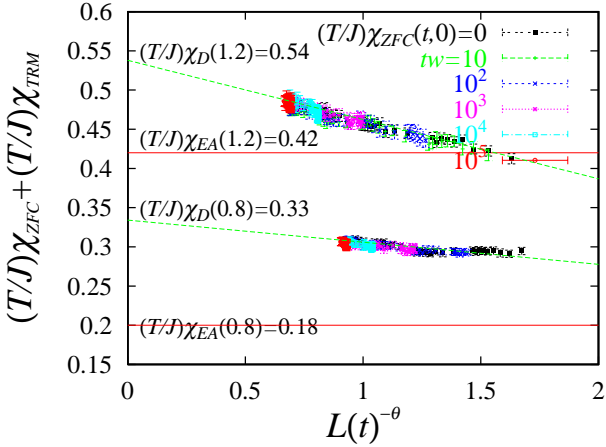}}
\caption{
Growth of the sum $(T/J)\chi_{\rm ZFC}(\tau+\tw,\tw)+(T/J)\chi_{\rm TRM}(\tau+\tw,\tw)$ with $t$. The data is the same as those used
in Fig. \ref{fig:check-linear}.
The fitting lines are due to \eq{eq-scaling-chizfc0}.
}
\label{fig:check-linear-lt}
\end{figure}

For the growth of the ZFC susceptibility with zero waiting time
we expect a scaling form \eq{eq-scaling-chizfc0} 
 which reads
 \bmat
  \chi_{\rm ZFC}(t,0) \sim \chi_{\rm D} -
 c''' m^{2}\frac{m^{2}}{\Upsilon(L(t)/L_{0})^{\theta}}
 \emat
 with $c'''$ being a numerical constant defined in \eq{eq-c'''}.

In Fig. \ref{fig:check-linear-lt} we show the sum 
$(T/J)\chi_{\rm ZFC}(\tau+\tw,\tw)+(T/J)\chi_{\rm
TRM}(\tau+\tw,\tw)$ plotted against $L(t=\tau+\tw)^{-\theta}$ using
$\theta=0.82$\cite{KH}.
We have checked that they satisfy the sum rule  \eq{eq:sum-rule}
and agrees with  $(T/J)\chi_{\rm ZFC}(t,0)$ 
(see Fig.\ref{fig:check-linear}).
As can be seen in the figure, the 
data are indeed consistent with the expected scaling form
being linear with $1/L(\tau)^{\theta}$ and pointing toward a constant
in the limit $L(\tau) \to \infty$. 
From the linear fit shown in the figure we find the values
of the dynamical susceptibility $(T/J)\chi_{\rm D}$ as
$0.54$ at $T/J=1.2$ and $0.33$ at $T/J=0.8$. 
The corresponding  dynamical oder parameter $q_{\rm D}$
can be determined via FDT $(T/J)\chi_{\rm D}=1-q_{\rm D}$ 
\eq{eq-def-chid} 
as $0.46$ at $T/J=1.2$ and $0.67$ at $T/J=0.8$. 

In the analysis of section \ref{subsec:c-quasieq}, we have obtained
the value of the equilibrium EA order parameter 
$q_{\rm EA}$ from which we readily find the equilibrium susceptibility
$(T/J)\chi_{\rm EA}=1-q_{\rm EA}$. 
Interestingly enough we find the data of $(T/J)\chi_{\rm D}(t,0)$ 
shown in the figure clearly goes over the static susceptibility 
$(T/J)\chi_{\rm EA}$ and the anticipated inequality 
$\chi_{\rm FC} =\chi_{\rm D} > \chi_{\rm EA}$ holds (see \eq{eq-chifc-chid}). 
This is one of the main results of the present numerical simulation.

\begin{figure}[h]
 \resizebox{\figwidth}{!}{\includegraphics{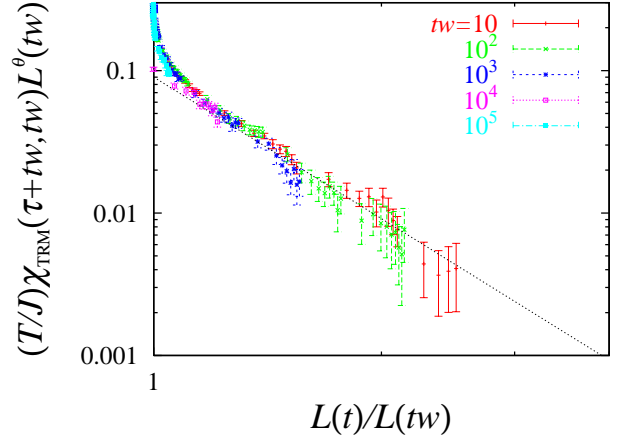}}
 \resizebox{\figwidth}{!}{\includegraphics{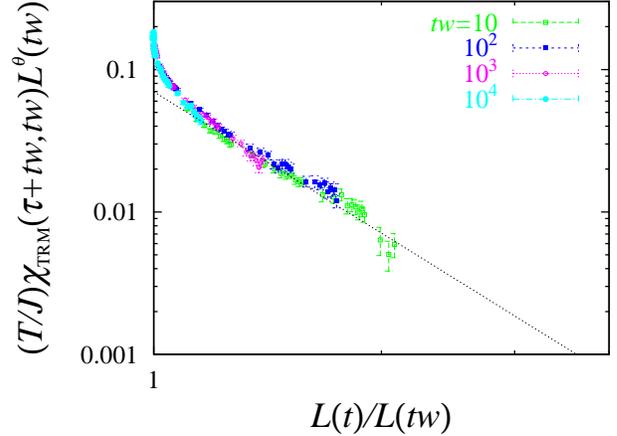}}
 \caption{Scaling plot of $(T/J)\chi_{\rm TRM}(\tau+\tw,\tw)$ 
at $T=1.2$ (upper figure) and $T=0.8$ (lower figure).
The dotted line  is a power law fit with exponent $\lambda=3.5$.
}
\label{fig:scale-trm}
\end{figure}

The sum rule \eq{eq:sum-rule} requires us to examine only 
either the TRM or ZFC susceptibility plus the growth of 
the ZFC susceptibility with zero waiting time  $\chi_{\rm ZFC}(t,0)$ 
which was obtained above. We have already 
analyzed the ZFC susceptibility in the quasi-equilibrium regime
and crossover regime in the preceding sections. 
In the following, we examine the TRM susceptibility in the 
aging regime which will complete our scaling analysis of 
the linear susceptibilities.

Let us examine the scaling ansatz  \eq{eq-chitrm-aging} which reads as,
\bmat
 \chi_{\rm TRM}(\tau+\tw,\tw) 
\sim  c_{\rm nst} \frac{m^{2}}{\Upsilon(L(\tw)/L_{0})^{\theta}}
\left(\frac{L(t)}{L(\tw)}\right)^{-\lambda}
\emat
for the TRM decay in the aging regime $L(\tau)/L(\tw) > 1$.
In Fig. \ref{fig:scale-trm}, we show the scaling plot of
the the decay of the TRM susceptibility $\chi_{\rm TRM}$ 
using the stiffness exponent $\theta=0.82$ \cite{KH}.
The results indeed agrees very well with the prediction 
\eq{eq-chitrm-aging} anticipated by Fisher and Huse 
\cite{FH2} and gives the non-equilibrium exponent 
$\lambda\sim 3.5$. The latter satisfies the 
bound \eq{eq-bound-lambda} with $d=4$.  To our knowledge
this is the fist time that this fundamental scaling law is confirmed.

\subsection{Separation of the Break points of TTI and FDT}
\label{subsec:separation-breaking}

\begin{figure}[t]
 \resizebox{\figwidth}{!}{\includegraphics{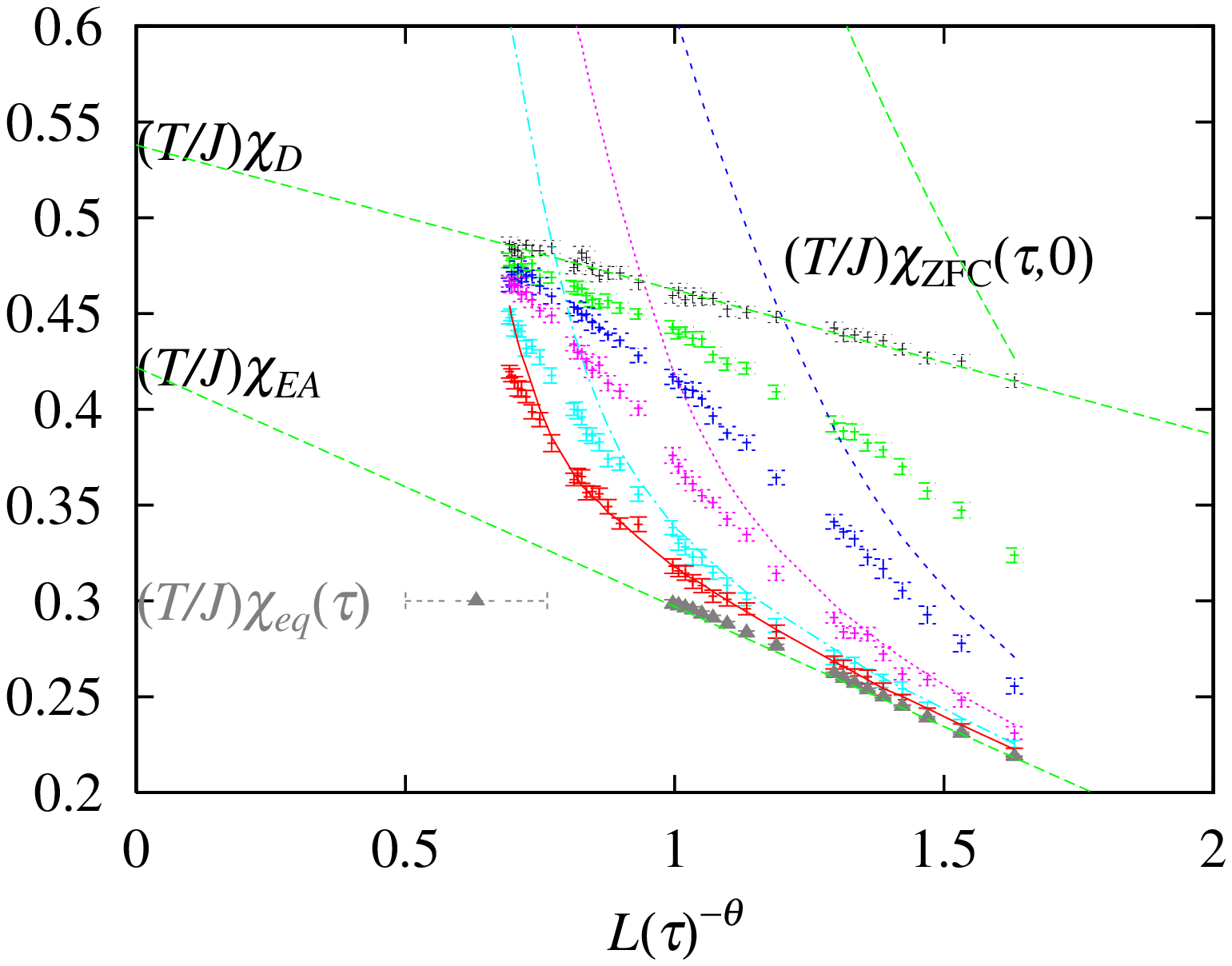}}
 \resizebox{\figwidth}{!}{\includegraphics{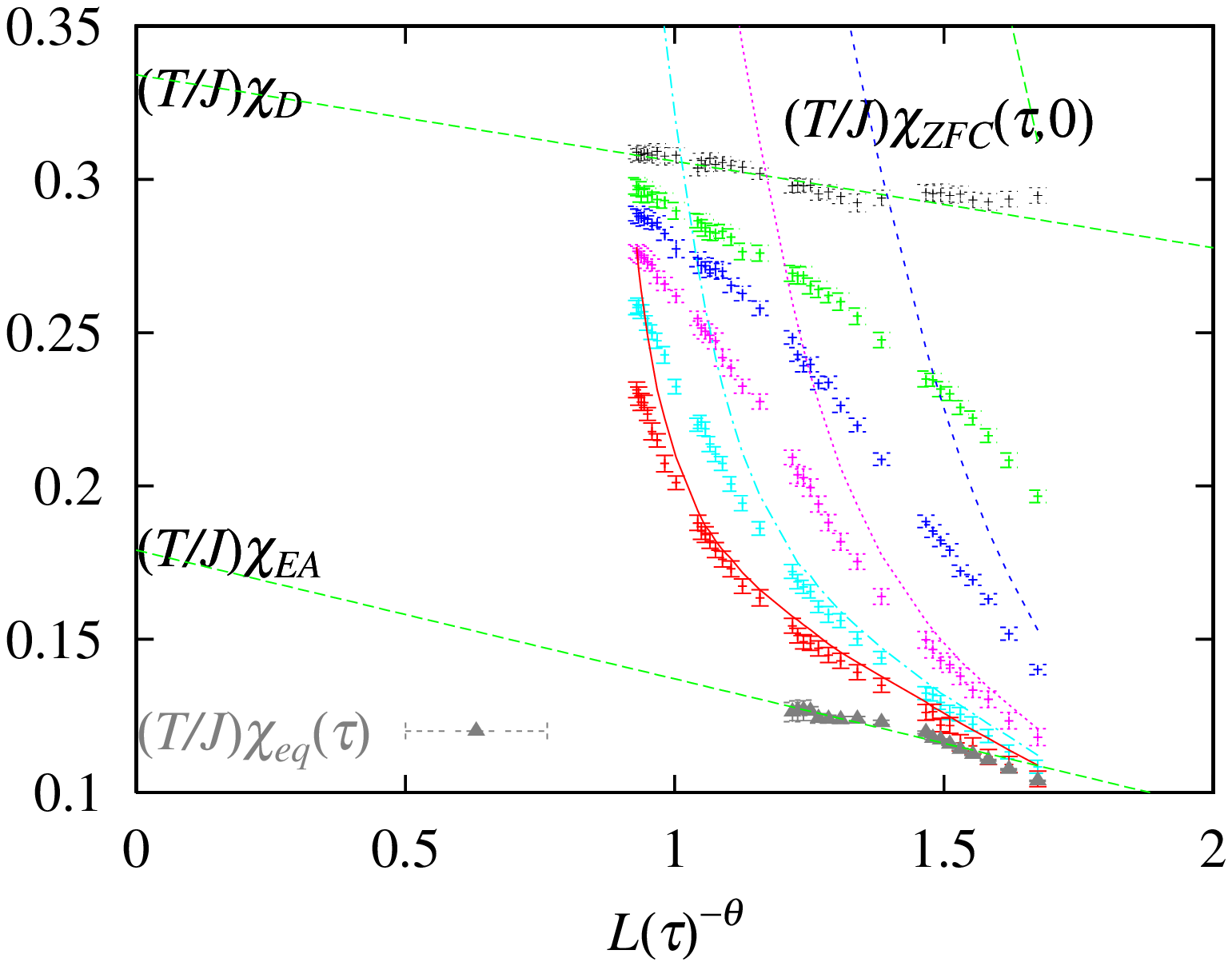}}
 \caption{ZFC linear susceptibilities vs $1/L^{\theta}(\tau)$
at $T/J=1.2$ and $T/J=0.8$.
The curves with small symbols are $\chi_{\rm ZFC}(\tau+\tw,\tw)$ with
$\tw=0,10,10^{2},10^{3},10^{4},10^{5}$ from the top to the bottom.
The curves with solid lines are $1-C(\tau+\tw,\tw)$ of 
$\tw=0,10,10^{2},10^{3},10^{4},10^{5}$ from the top to the bottom.
The data of $T\chi_{\rm eq}(\tau)$ (filled triangle) 
and its linear fit are shown at the bottom. 
The linear fit to $T\chi_{\rm ZFC}(\tau,0)$ is also shown at the top.
}
\label{fig:m-ltau}
\end{figure}

The results of the analysis presented so far supports 
well the existence of the anticipated asymptotic
regimes displayed in Fig. \ref{fig:c-conjecture} and
Fig. \ref{fig:ck-conjecture} which predicts separation of
the breaking of TTI and FDT.
In order to look more directly on the separation, we present in Fig
\ref{fig:m-ltau} the ZFC linear  
susceptibility $(T/J) \chi_{\rm ZFC}(t,\tw)$ and spin 
autocorrelation function $1-C(t,\tw)$  of various waiting 
times $\tw$  plotted against $1/L(\tau)^{\theta}$. 

In the figure, we included the equilibrium 
ZFC linear susceptibility $(T/J)\chi_{\rm eq}(\tau)$ for comparison.
The latter was obtained by the FDT
$(T/J)\chi_{\rm eq}(\tau)=1-C_{\rm eq}(\tau)$  using 
the equilibrium spin autocorrelation function $C_{\rm eq}(\tau)$
extracted previously in section \ref{subsec:c-quasieq}. 
Since $C_{\rm eq}(\tau)$ 
is a linear function of $1/L(\tau)^{\theta}$ 
as we confirmed in section \ref{subsec:c-quasieq} 
(see Fig. \ref{fig:ceq}), $(T/J)\chi_{\rm eq}(\tau)$ becomes
a straight line in the plot pointing toward
the equilibrium susceptibility $(T/J)\chi_{\rm EA}$.
The top curve is  $(T/J)\chi_{\rm ZFC}(\tau,0)$
which is at the other extreme: zero waiting time.
It becomes also a straight line in the plot
as we already saw in section \ref{subsec:scaling-sus}
(See Fig. \ref{fig:check-linear-lt}) pointing toward the
dynamical susceptibility $(T/J)\chi_{\rm D}$ which
is significantly larger than the equilibrium 
susceptibility  $(T/J)\chi_{\rm D}$.

Note that the slope of the susceptibility in the zero waiting time limit 
is smaller than that of the equilibrium limit.  This can be explained
as the following. From \eq{eq-chi-eq}, the slope of the
equilibrium limit is expected to be proportional to $c$ defined
in \eq{eq-def-const-c} while \eq{eq-scaling-chizfc0} suggests
the slope of the zero waiting time limit is proportional to
$c'''=A(1)-c_{\rm nst}$ given in\eq{eq-c'''}. Here $c_{\rm nst}$
is a positive constant and the value of $A(1)$ can be evaluated
by \eq{eq-def-a}. The latter implies $A(1) < c$ because 
the first term in \eq{eq-def-a} becomes identical to $c$ defined
in \eq{eq-def-const-c} while the second term is negative
because $\Upsilon/\Upsilon_{\rm eff}[y] \geq 1$. Thus we expect
$c > c'''$.

Now a surprising observation is that the break points 
of FDT moves further away from the equilibrium curve 
by increasing $\tw$. It appears very unlikely that
the break points of the FDT converge to the equilibrium susceptibility
$(T/J)\chi_{\rm EA}$. This feature strongly suggests
that the violation of TTI and FDT do not take place
simultaneously but separates asymptotically.
In our scenario presented in section \ref{subsec:theory-two-time}
the anomalously extended FDT regime is attributed to the 
soft droplets. It satisfies FDT but is absent in the ideal 
equilibrium where there are no frozen-in extended defects:
it is  a dynamical object.

\subsection{Integral Violation of FDT}
\label{subsec:integralFDT}

\begin{figure}[t]
\resizebox{\figwidth}{!}{\includegraphics{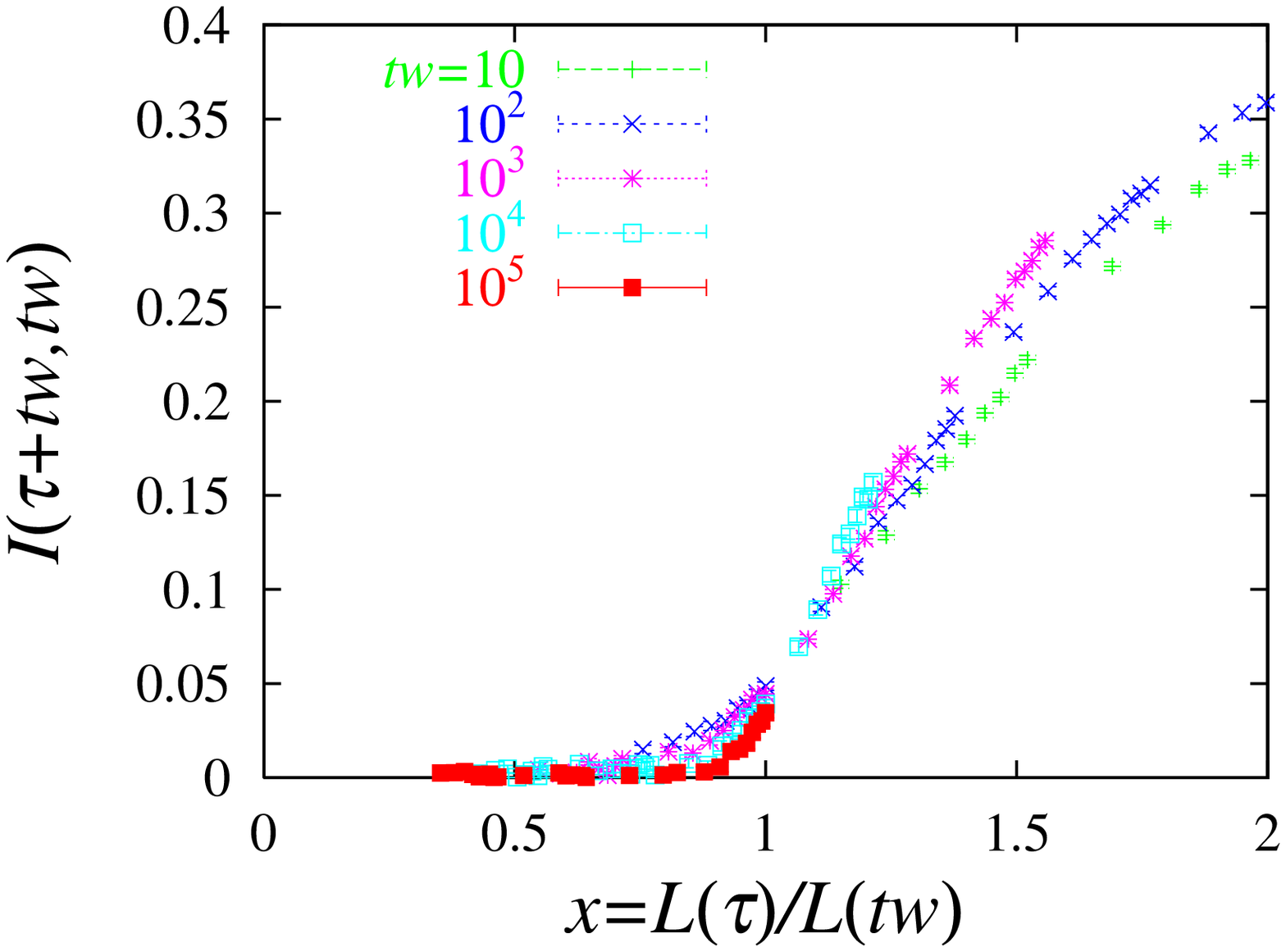}}
\resizebox{\figwidth}{!}{\includegraphics{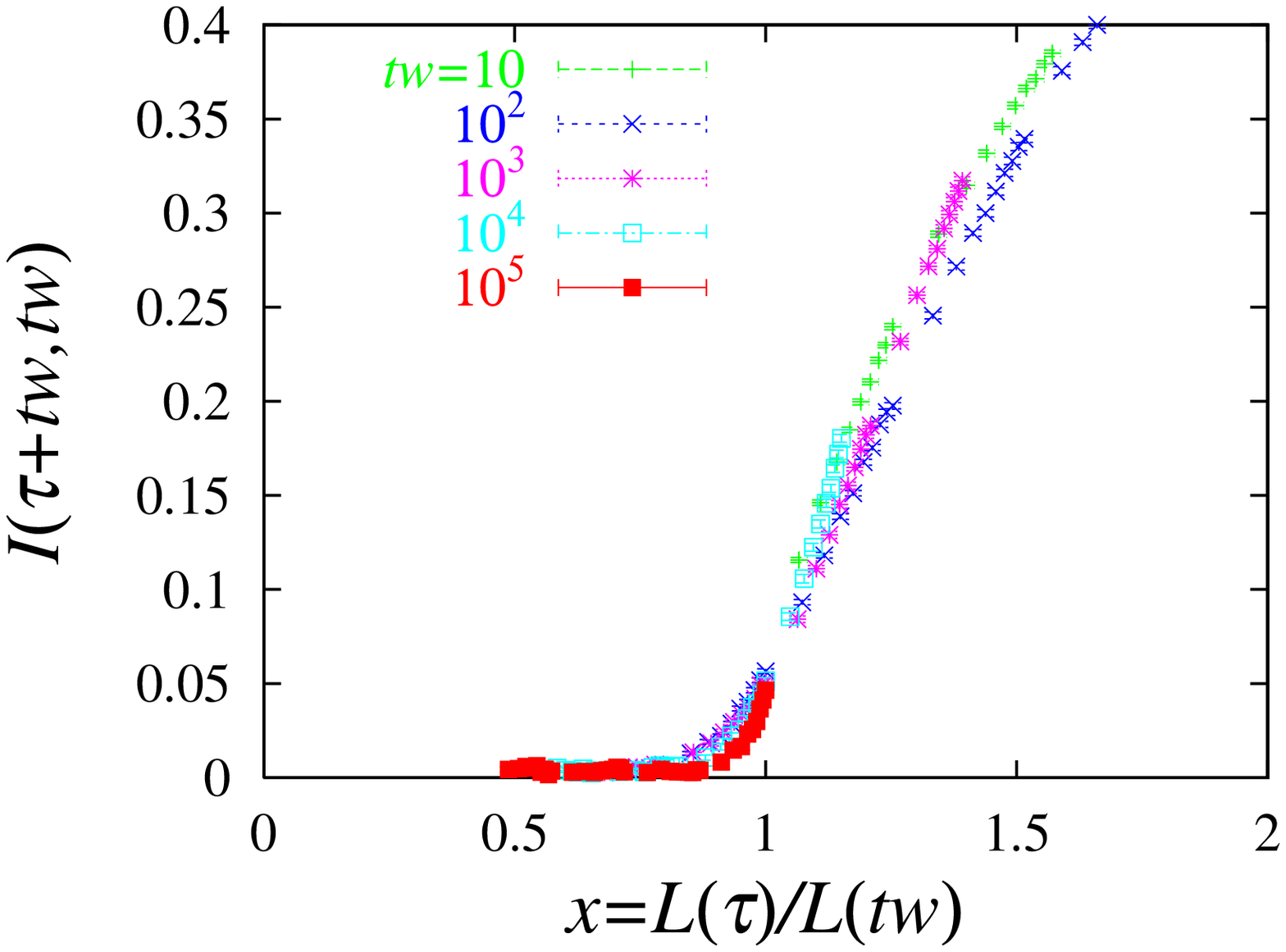}}
\caption{Integral violation of FDT at $T/J=1.2$ (upper figure)
and $T/J=0.8$ (lower figure) plotted against $L(\tau)/L(\tw)$}
\label{fig:fdt-lt}
\end{figure}

\begin{figure}[t]
\resizebox{\figwidth}{!}{\includegraphics{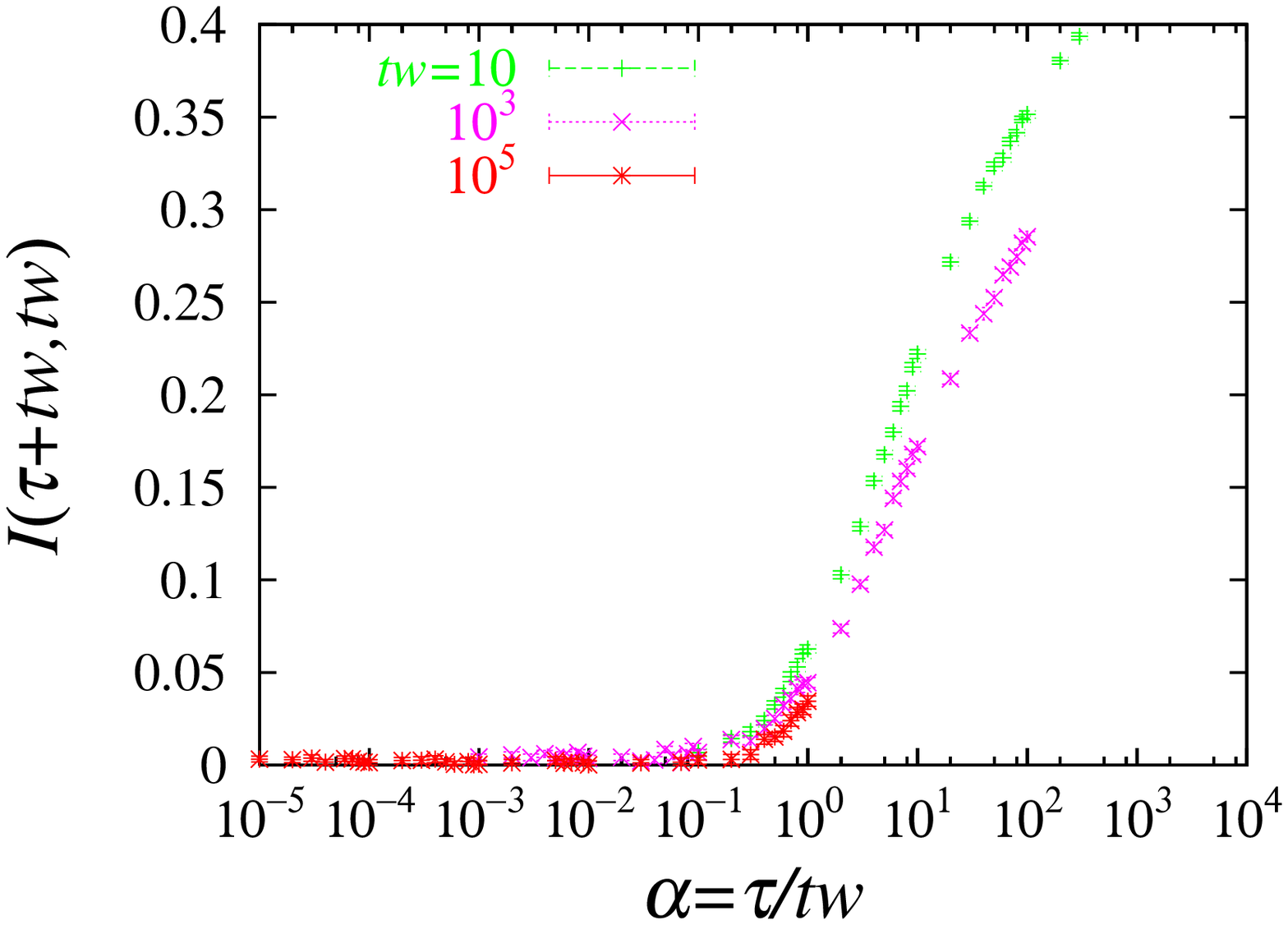}}
\resizebox{\figwidth}{!}{\includegraphics{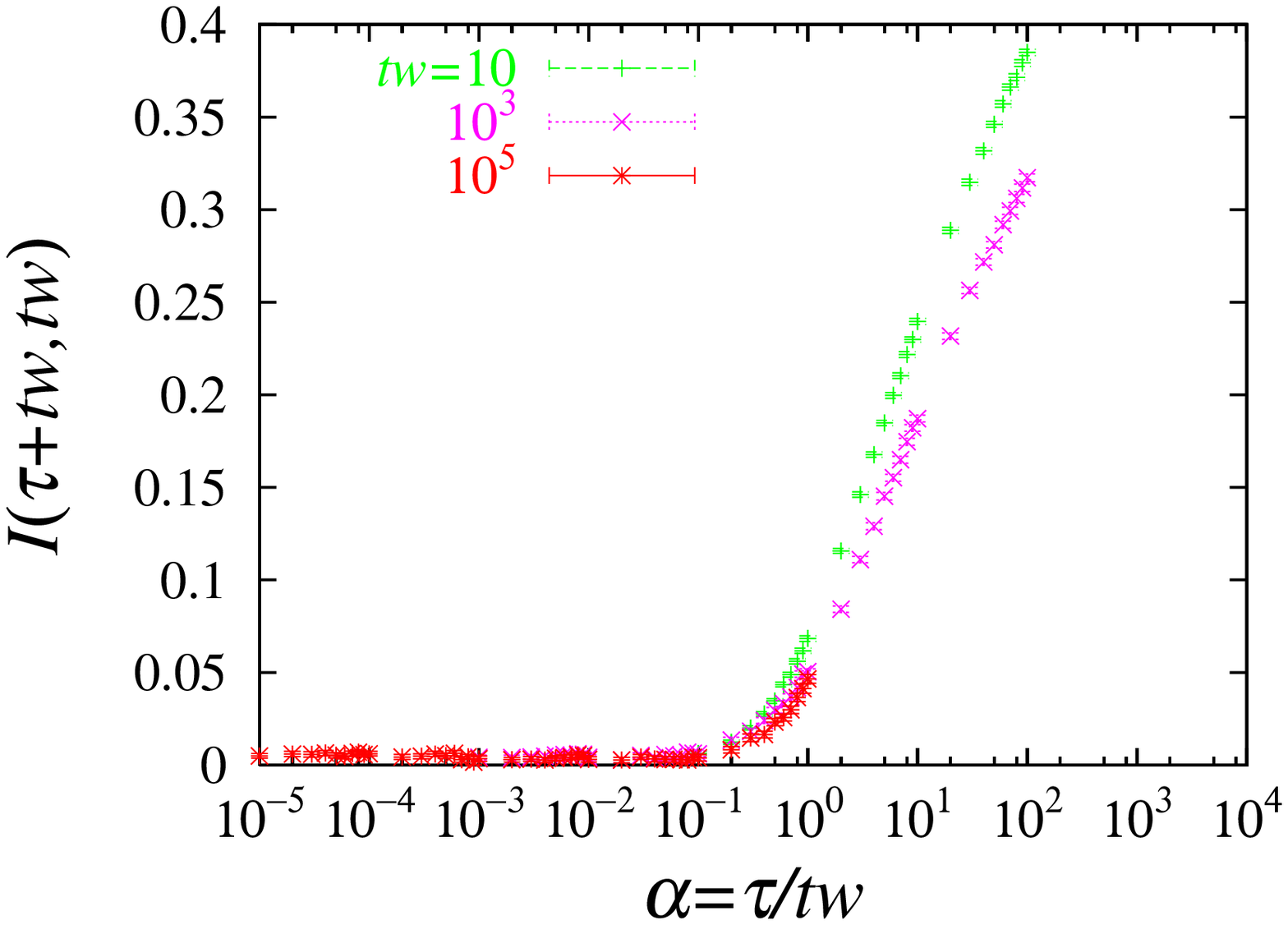}}
\caption{Integral violation of FDT at $T/J=1.2$ (upper figure)
and $T/J=0.8$ (lower figure) plotted against $\tau/\tw$.}
\label{fig:fdt}
\end{figure}

Let us finally examine the integral violation of the FDT defined in
\eq{eq-integral-violation-fdt}. 
In Fig.~\ref{fig:fdt-lt}, the integral violation $I(\tau+\tw,\tw)$
obtained using our data is plotted against $L(\tau)/L(\tw)$.
We see it decreases with increasing waiting time $\tw$
deep in the quasi-equilibrium regime  $x=L(\tau)/L(\tw) \ll 1$ 
and increases deep in the aging regime $x=L(\tau)/L(\tw) \gg 1$.

The intriguing problem is the validity of the FDT in the
crossover regime $x=L(\tau)/L(\tw) \sim 1$. 
In section \ref{subsubsec:theory-crossover} we conjectured that
it is satisfied in the crossover regime.
In Fig.~\ref{fig:fdt-lt}, one can see that the integral violation 
is indeed decreasing with increasing $\tw$ at $L(\tau)/L(\tw) \sim 1$
which appears compatible with our expectation. 

As noted in section  \ref{subsubsec:theory-crossover}, 
a possible scaling variable for the interior of the crossover regime 
would be $\alpha=\tau/\tw$. 
In the limit $\tw \to \infty$, 
possible limits classified by different $\alpha$ 
are all smashed into the crossover regime $x=L(\tau)/L(\tw)=1$.
In Fig.~\ref{fig:fdt}, we show the integral violation 
against $\alpha=\tau/\tw$. Indeed, it is decreasing function 
for any value of $\alpha=\tau/\tw$ including even the highest $\alpha$
within our data. These observations are compatible
with our conjecture that the FDT is asymptotically valid in the
crossover regime.

%Discussion
%\input{sec07-v6}

\section{Discussion and Summary}
\label{sec:discussion}

To summarize, we have presented the results of a detailed
MC simulation of a $4d$ EA Ising spin-glass model in the
present paper. We demonstrated that the results can be well understood 
within the extended droplet theory we proposed recently \cite{our-letter}.

We demonstrated the dynamical susceptibility $\chid$ larger
than the equilibrium susceptibility $\chiea$ exists in agreement
with the extended droplet theory. Within the latter scenario, 
the two different limits emerge as different large time 
(size) limits of the two-time (length) quantities.
This phenomenon is presumably intimately related with the well 
known experimental observation $\chifc > \chiea$ which has
been known for a long time since the very discovery of 
spin glasses \cite{Nagata79}. 
The question was, if it is a short time transient phenomena or not. 
The dynamical MFT was the first which clearly appreciated that the 
difference $\chifc- \chiea$ (called ``anomaly'' \cite{MFT}) exists 
in a well-defined large time limit. 
Our scaling theory provides systematic extrapolation 
scheme to obtain the two susceptibilities in a controlled way. 
It will be certainly interesting to perform such an analysis 
in experiments and numerical simulations. 

We found evidences of the abrupt change of the two-time quantities 
in the crossover regime $L(\tau) \sim L(\tw)$ which was 
anticipated in the extended droplet theory. This is directly 
related to the existence of the two different susceptibilities mentioned above.
In previous experimental studies, it was well known that the relaxation rate
$S(\tau,\tw)$ has a broad peak as a function of $\tau$ at around $\tau \sim \tw$.
This phenomenon is expected to be intimately related with the abrupt changes
in the crossover regime. We proposed a modified relaxation 
rate $S_{\rm mod}(x,\tw)$ with $x=L(\tau)/L(\tw)$. As expected, it was
found numerically to develop a peak at $x\sim 1$ which 
{\it sharpens with increasing $\tw$ when plotted against $x$}. 
Probably the latter will be more useful in future studies.

A very important feature of our analysis is the two-stroke strategy.
In numerical simulations one can obtain directly the data of 
dynamical length scale  $L(t)$ by which {\it time-dependent} quantities 
are immediately translated into {\it length-dependent} quantities.  
In addition, the information of the stiffness exponent $\theta$ and 
fractal dimension $d_{f}$ of droplets are known from independent 
studies of equilibrium properties. Thus we had virtually {\it zero} fitting 
parameter left for us in this stage to test various data collapse expected from
the scaling ansatz.

The analysis of the growth law itself can be done separately. 
We found that the growth law $L(t)$ measured at various temperatures
below $T_{\rm c}$ shows the anticipated 
crossover from short time (length) critical dynamics to asymptotic 
activated dynamics. In our scaling analysis, we used the critical exponents 
$\nu$, $z$ and the critical temperature $T_{\rm c}$ determined 
in previous studies on equilibrium critical properties  so that we have
{\it only one free-parameter} $\psi$ 
which was found to be $2.5-3$ in the present model. All the temperature
dependences are renormalized into the crossover length $L_{0}(T)$
and the corresponding crossover time scale 
$\tau_{0}(T) \propto L_{0}(T)^{z}$ by which
all data are collapsed onto a universal growth law function.
The latter becomes the expected power law for the critical dynamics 
at short times (length)  and a slower function at large times (length)
suggesting activated dynamics.

Similar analysis should be done on experimental studies \cite{HO} 
and numerical simulations of other systems, particularly for $3d$
models, 
as well. In experiments, however, this two-stroke strategy is not possible
and probably the best way is to start from simplest
quantities such as relaxation of the AC susceptibility \cite{Uppsala01}
to work out the parameters of the growth law.
The advantage of the experiment, on the other hand, is that 
one can explore  the order of $100$ lattice spacings while the present 
numerical simulations are limited to  $1 -10$ lattice spacings.

Concerning the effects of critical fluctuation, many recent studies 
have pointed out the importance to renormalize its effect. 
In three dimensions, previous results on the growth law by 
numerical simulations \cite{Kisker,Marinari98,KYT} as 
well as experiments\cite{Joh} are well fitted to a power law as 
$L(t)\sim t^{1/z(T)}$ where the exponent $1/z(T)$ is proportional 
to $T$, (but this $T$-linearity of $1/z(T)$ starts to break in the
temperature range around $T\sim 0.75T_c$\cite{KYT}). Probably it is a
sort of interpolation formula between 
$T_{c}$, where critical fluctuations are dominant, and $T \to 0$, 
where all time scales associated with thermally activated processes diverge.  
Indeed a recent experiment in $3d$ systems \cite{Uppsala01}
suggests the logarithmic growth law works well if critical fluctuation 
is properly considered. A recent numerical study of the growth law
\cite{BB02} has also found a signature of the crossover. 

It is useful to note that the empirical power law $L(t)\sim t^{1/z(T)}$, 
combined with the scaling laws in terms of $L(t)$, explains many 
of the empirical formulas proposed in previous experiments 
and numerical studies in a unified manner being consistent 
with our two-stroke approach.
For example a fitting formula $C_{\rm eq}(\tau) = q_{\rm EA} + 1/\tau^{\alpha}$
used in previous numerical studies (For example Ref.~\onlinecite{Parisi96})
can be understood as a variant of \eq{eq-c-eq} with  $\alpha=\theta/z(T)$.
The well-known ``sub-aging'' scaling for the TRM susceptibility
\cite{Saclay-TRM,Saclay}
$
\chi_{\rm TRM}(\tau+\tw,\tw) \sim F(\tau/\tw^{\mu})
$
with $\mu < 1$ can be understood as a variant of
\eq{eq-chitrm-aging} with $\mu=1-\theta/\lambda$.
The  ``$\omega t$ -scaling'' of the AC susceptibility \cite{Saclay}
can also be understood similarly as already noted in a numerical
study \cite{KYT2} and an experimental \cite{Uppsala01} study.
In addition, some apparent sub-aging feature of AC and  DC 
susceptibilities can be removed by considerations of finiteness 
of cooling rates in real experiments \cite{Uppsala01,BB02,UCLA}.

The fundamental assumption of our  scenario is that
the effective stiffness constant $\Ueff$ of the free-energy gap
$\FtypLR$ of droplets is a function of the ratio $x=L/R$ of
the two length scales, namely the length scale of the droplet 
itself $L$ and of the extended defect $R$ which surrounds 
the droplet. Most importantly we assumed
the vanishing of the effective stiffness constant as $x \to 1$.
This allowed the emergence of the two different order parameters
$\qea$  and $\qd$  and the associated susceptibilities $\chiea$
and $\chid$. It is desirable to clarify the scaling of the stiffness
constant explicitly in the whole range of $ 0 < x < 1$.

We expect our conjecture is consistent with the results of recent studies 
of low energy excitations in spin-glass models \cite{LEE}.
In spin glasses of finite sizes $R$, it is likely that 
the existence of boundaries will intrinsically induce certain defects
as compared with infinite systems \cite{NS98,M99,KYT}. 
Then we expect droplet excitations as large as the system 
size itself $L \sim R$ is anomalously soft.
Such an anomaly is indeed found in the series of recent studies \cite{LEE}. 
Although there new exponent $\theta'=0$ was conjectured \cite{LEE}, 
we consider it is better to attribute this to the zero stiffness constant 
$\Ueff[1]=0$ as in (\ref{eqn:effU-large}). The stiffness exponent
$\theta > 0$, on the other hand, is associated with a defect in $\Gamma$
as we adopt in (\ref{eqn:effU-general}).
The anomalously low energy and large scale excitations at $L \sim R$
explains the apparently non-trivial overlap distribution
function $P(q)$ found in numerous numerical studies of finite 
size systems \cite{low-temp-MC,Rome-review} which appear
very similar to the prediction of the equilibrium mean-field 
theory \cite{MFT-eq}.  
Although the meaning of the apparently non-trivial (and probably
non-self averaging) $P(q)$ in equilibrium is not obvious \cite{NS01},
we consider the contribution of the anomalous excited 
states to the macroscopic magnetic susceptibility in the present 
dynamical situation, {\it which  is the realistic situation},
is very important. As we discussed in section \ref{subsubsec:comarison},
our scenario implies the {\it average} 
overlap $\bar{q}=\int_{0}^{1}dq q P(q)$ measured in equilibrium 
of finite size systems $R$ (with built-in defects)   is equivalent to
the dynamical order parameter  
$\qd$. We found the latter is related to the Field Cooled susceptibility
$\chifc$ as $\chifc=\chid=(1-\qd)/\kb T$ (See \eq{eq-def-chid}
and \eq{eq-chifc-chid}).
It would remind one of a folklore found in some literatures that the 
difference of $\chifc$ and $\chizfc$ is {\it somewhat} related to
difference of $\bar{q}$ and $\qea$ which is a consequence of 
the non-trivial $P(q)$ found in 
Parisi's replica symmetry broken solution of the mean-field 
model \cite{MFT-eq}.
However, we still have to recall the intrinsically dynamical 
nature of the situation:
the domain walls, which allow the anomalous softening of droplets,
are dynamical objects and they should be absent in ideal equilibrium
where $P(q)$ will have only one delta peak at $q=\qea$ as predicted
by the original droplet theory \cite{FH1}.

Concerning the dynamical MFT, a serious problem in practice is 
that correction terms to the asymptotic limit is not known.
They should obviously exist in the $C-\chi$ relation 
since the curves in Fig. \ref{fig:ck} systematically moves with
increasing $\tw$. 
Such a feature exist in the data of a recent
experiment of simultaneous magnetic noise/response measurement\cite{HO}
and previous numerical studies 
\cite{FR95,Parisi96,Marinari98,RPR01}.

Our numerical results indeed suggests the separation of the breaking of 
TTI and FDT (see Fig. \ref{fig:m-ltau}). 
Such a feature has not been realized in previous 
studies  \cite{FR95,Parisi96,Marinari98,RPR01,HO}.
Within the scaling theory, the  correction terms to the expected asymptotic
limit (Fig. \ref{fig:c-conjecture} and Fig. \ref{fig:ck-conjecture})
themselves are predicted to have salient universal scaling properties 
which are amenable to be examined in practice. Our numerical results were
well explained by the latter scaling ansatz including the decay
of the TRM susceptibility  \eq{eq-chitrm-aging} which itself was 
predicted by the original droplet theory more than a decade ago.

\section*{Acknowledgements}
This work is supported by a Grant-in-Aid for Scientific Research
Program(\#12640367), and that for the Encouragement of Young
Scientists(\#13740233)
from the Ministry of Education, Culture, Sports, Science 
and Technology of Japan.  
The present simulations have been performed on Fujitsu VPP-500/40 at the
Supercomputer Center, Institute for Solid State Physics, the University
of Tokyo. 

%References
%\input{ref-v6}


\begin{thebibliography}{99}
\bibitem{note}
Unfortunately the ambiguity of the width $\ln a$ is unavoidable 
within the independent droplet model.
\end{thebibliography}
\end{document}